\documentclass{sn-jnl}% Math and Physical Sciences Reference Style
\usepackage{comment}
\usepackage{graphicx}
\usepackage{subfigure}
\usepackage{hyperref}
\usepackage{xcolor}
\usepackage[section]{placeins}
\jyear{2022}%

\theoremstyle{thmstyleone}%
\newcommand{\bR}{{\mathbb{ R}}}
\theoremstyle{thmstyletwo}%

\theoremstyle{thmstylethree}%
\usepackage{natbib}
\setcitestyle{authoryear}
\begin{document}

\title[Nonlocal Mechanistic Models in Ecology]{Nonlocal Mechanistic Models in Ecology: Numerical Methods and Parameter Inferencing}

%%=============================================================%%
%% Prefix	-> \pfx{Dr}
%% GivenName	-> \fnm{Joergen W.}
%% Particle	-> \spfx{van der} -> surname prefix
%% FamilyName	-> \sur{Ploeg}
%% Suffix	-> \sfx{IV}
%% NatureName	-> \tanm{Poet Laureate} -> Title after name
%% Degrees	-> \dgr{MSc, PhD}
%% \author*[1,2]{\pfx{Dr} \fnm{Joergen W.} \spfx{van der} \sur{Ploeg} \sfx{IV} \tanm{Poet Laureate} 
%%                 \dgr{MSc, PhD}}\email{iauthor@gmail.com}
%%=============================================================%%

\author*[1]{\fnm{Erin} \sur{Ellefsen}}\email{ellefer@earlham.edu}

\author[2]{\fnm{Nancy} \sur{Rodr\'iguez} }\email{rodrign@colorado.edu}
%\equalcont{These authors contributed equally to this work.}

%\author[1,2]{\fnm{Third} \sur{Author}}\email{iiiauthor@gmail.com}
%\equalcont{These authors contributed equally to this work.}

\affil*[1]{\orgdiv{Department of Mathematics}, \orgname{Earlham College}%, \orgaddress{\street{}, \city{Richmond}, \postcode{}, \state{}, \country{}}
}

\affil[2]{\orgdiv{Department of Applied Mathematics}, \orgname{University of Colorado Boulder}%, \orgaddress{\street{}, \city{}, \postcode{}, \state{}, \country{}}
}

%\affil[3]{\orgdiv{Department}, \orgname{Organization}, \orgaddress{\street{Street}, %\city{City}, \postcode{610101}, \state{State}, \country{Country}}}

%%==================================%%
%% sample for unstructured abstract %%
%%==================================%%

\abstract{Animals use various processes to inform themselves about their environment and make decisions about how to move and form their territory. 
In some cases, populations inform themselves of competing groups through observations at distances, scent markings, or memories of locations where an individual has encountered competing populations. 
As the process of gathering this information is inherently nonlocal, mechanistic models that include nonlocal terms have been proposed to investigate the movement of species.  
Naturally, these models present analytical and computational challenges. In this work we study a multi-species model with nonlocal advection.  We introduce an efficient numerical scheme using spectral methods to compute solutions of a
nonlocal reaction-advection-diffusion system for a large number of interacting species. 
Moreover, we investigate the effects that the parameters and interaction potentials have on the population densities.  
Finally, we propose a method using maximum likelihood estimation to determine the most important factors driving species' movements and test this method using synthetic data.}


%%\pacs[JEL Classification]{D8, H51}

%%\pacs[MSC Classification]{35A01, 65L10, 65L12, 65L20, 65L70}

\maketitle

\section{Introduction}\label{sec1}
 Learning about the local environment is essential to a species' survival. Interactions within a social group, competition with conspecifics, foraging, and avoiding predators are a few examples of processes that inform how a species moves in its environment, and these are essential in understanding and predicting territories of species.  
For example, wolves move using information about the scent of their pack, scents of conspecifics, pack density, and the location of wolf pups
%, foraging to find food, and dispersal to mark territory 
\citep{White1996}.
%Species take into account several factors while forming their territories, some of which include the environment, population density, and the territories of predators, prey, and conspecifics. For example, %%% EXAMPLE
Understanding the mechanisms that drive species movement and how they inform territory decisions is a focus in spatial ecology, and these issues have become increasingly important to understand as climate change alters the environment \citep{Molnar2011,Stenseth2002}, and thus alters the territory of species \citep{Parmesan2003}. 

Mathematical models can use the mechanisms that inform territory decisions to predict how territories change or redistribute in response to the environment, a change in the territory of conspecifics, or a change in the territory of another species. 
Mechanistic models can be used to study how species live and move in their environment by utilizing the underlying theories of how individuals move. In some ways, these models can be more beneficial than their statistical counterparts because they can be used to test hypotheses of mechanisms driving animal movement, have strong predictive power, and can be verified using field measurements \citep{ Moorcroft1999}. Thus, mechanistic models have been used to describe foraging \citep{Grunbaum1996, Grunbaum1998}, aggregation \citep{Gueron1994}, and home ranges \citep{Lewis1993, Powell2004} . Mechanistic home range models incorporate movement processes, influence of conspecifics, environmental pressures, and space-use \citep{MoorcroftLewis2006, Moorcroft2008, Bateman2015}. In \citep{Holgate1971} and \citep{Okubo1980}, mechanistic home range models with diffusion and attraction to a home center, such as a den site, were used to describe territories. Since then, mechanistic models have incorporated various processes to produce patterns more accurate to observed territories, such as scent marking \citep{Moorcroft1999} and habitat selection \citep{MoorcroftLewisCrabtree2006}.
Local reaction-advection-diffusion (RAD) models have been used with success to predict the territories of species such as wolves, \citep{White1996}, coyotes, \citep{Moorcroft1999}, and meerkats, \citep{Bateman2015}. 
%Modeling multi-species or multi-group territory has been of much interest. In fact, there have been links between local mechanistic models of social groups and data.  
In \citep{Bateman2015}, meerkat location data and environmental data are incorporated into a local mechanistic model describing territory formation of social groups of meerkats via parameter estimation. 

However, the process of  gathering information about a local environment is inherently nonlocal; using observations, sounds, and smells to gain information about an environment happens at a distance \citep{Carrillo2015,Lutscher2019}. In addition, it has been shown that nonlocal advection is essential for keeping swarms coherent \citep{Mogilner1999}, and for the well-posedness of solutions to mechanistic models of home-range and territory formation \citep{Briscoe2002,Potts2016}. 
Moreover, field observations provide evidence that meerkats avoid locations where they have previously encountered members from an opposing group, further justifying the importance of nonlocal interactions in certain species \citep{Bateman2015, Clayton2001}. 
Thus, models with nonlocal advection have been of much interest -- see \citep{Briscoe2002,Topaz2006,Delgadino2020} and references within. 
 We are interested in a nonlocal mechanistic model that can be used to describe territory formation of social groups of a species, removing the necessity of artificial dynamics. 
 However, with nonlocal advection, classical theory no longer applies \citep{Briscoe2002, Carrillo2019}. Therefore, analyzing the model is difficult and solving it numerically is computationally expensive \citep{Bernoff2015}.  
 In order to investigate the behavior of solutions in two-dimensions with several interacting species and also connect the model with data, we must address these challenges.
 
In this work, we first introduce an efficient numerical scheme to solve a nonlocal reaction-advection-diffusion equation for a large number of interacting species.  
%We work on a bounded, periodic grid and compute the convolutions and derivatives in Fourier space. 
We investigate the effects that the interaction potential, the environment, the diffusion strength, and the number of groups have on the behavior of the equilibrium solutions as well as the computation time. 
Interestingly, we find that in some cases, changing the slope of a interaction potential has more of an effect on a population's territory than changing the variance. We also see that the balance between the diffusion and aggregation strength has a large influence on the computation times. 
Finally, using synthetic data we test a data fitting method to determine which forces are the primary drivers in the movement of social groups using maximum likelihood estimation and stochastic gradient descent.

 \textit{Outline:} In Section \ref{sec:background} we present some background and introduce the mechanistic model, which is the subject of this study.  
 In Section \ref{sec:numerics_method}, we describe the numerical scheme for a general nonlocal reaction-advection-diffusion equation in two-dimensions. Then, we investigate the effect the parameters in the model have on the behavior of the solutions and the computational expense of adding more groups to the system.  Finally, we minimize the negative log-likelihood function to incorporate synthetic data into the model in Section \ref{sec:MLE} and conclude in Section \ref{sec:disc}. 
 
\section{Background and Model}\label{sec:background}

This work was motivated by the analysis done by Bateman and collaborators in \cite{Bateman2015}.
Mechanistic models of meerkat territory formation were connected with observed data in a local reaction-advection-diffusion (RAD) framework. The authors use various local RAD models to describe territories of social groups of meerkats. 
%They use maximum likelihood estimation to incorporate meerkat location data into these various models and to determine which version of these models is most accurate to the observed data. 
The core model considered is the following: 
\begin{align}\label{eq:Bateman}
\partial_t u_i(x,t) &= \nabla^2[u_i(x,t)D(x,t)] - \nabla \cdot[u_i(x,t)C_i(x,t)],
\end{align}
for $x \in \Omega \subseteq \mathbb{R}^d$, $t>0$, where $u_i$ represents the density of group $i$ with $i = 1...N$, $D$ is the spatial diffusion rate, and $C_i$ is the velocity of group $i$'s directed movement. Each group advects in the direction of $\hat{v}_i$, a unit vector pointing in the direction of group $i$'s home center. There are two core versions of this model: the groups either move away from direct interactions with conspecifics or away from scent markings by conspecifics. Within these core models, the authors can take into account both desirable and undesirable habitat features, social groups splitting, and territories shifting. For three different distinct time periods, the authors incorporate data into the various versions of equation \eqref{eq:Bateman} and determine which models are most accurate to the data during each period. The driving motivation for our work is the need to include a home center in this model. In fact, home centers are not observed in meerkat populations, and it has been noted that a mechanistic model should be able to recreate observed behaviors without fixing a home center \citep{Borger2008}.  Our aim is to introduce such a model as well as efficient numerical schemes which will aid in the exploration of the qualitative behavior of the system.  

\subsection{Nonlocal Mechanistic Models}\label{sec:NLMechanistic}
In order to consider gathering and use of information via nonlocal mechanisms, we study the model: 
\begin{align}\label{eq:System}
\partial_t u_i(x,t) &= \eta \Delta u_i^2(x,t) - \nabla \cdot \left\{ u \nabla\left( K_1 * u_i  -\hspace{-5pt} \sum_{j = 1, j\neq i}^N \hspace{-5pt}K_2*u_j + U(x,t)\right)\right\} 
\end{align}
for $x \in \Omega \subset \bR^d, t > 0.$ As in the local model, $u_i$ represents density of population $i$ with $i = 1, 2, ..., N$. In this study we impose periodic boundary conditions. The parameter $\eta$ in system \eqref{eq:System} represents the intra-group dispersal rate, which models an overcrowding effect; the group dispersal rate is higher with a larger population density.   The convolution terms describe the intra-group long-range aggregation and inter-group long-range repulsion, governed by the potentials $K_1$ and $K_2$, respectively.
%Note that t
The long-range aggregation term moves the group $u_i$ with a nonlocal velocity $-\nabla K_1 *u_i$, which helps maintain the group's coherence. Moreover, the long-range inter-group repulsion term moves the population $u_i$ away from other groups via the velocity field $\sum_{j=1,j\neq i}^N \nabla K_2 * u_j$ and serves as a segregation term. 
 An associated energy of the system motivates the use of the same kernel in both the aggregation and segregation terms, $K = K_1 = K_2$. In such case, the system can be seen as the gradient-flow of the following energy, 
 \begin{align*}
 E[u_i](t)\hspace{-2pt} := \hspace{-5pt} \int_\Omega \hspace{-3pt}\left[\eta \sum_{i=1}^N \hspace{-2pt} A(u_i) \hspace{-1pt} - \hspace{-1pt} \frac{1}{2}\sum_{i=1}^N (K*u_i)u_i \hspace{-1pt} + \hspace{-1pt} \hspace{-10pt} \sum_{i,j=1, i\neq j}^N \hspace{-9pt} (K*u_i)u_j  \hspace{-1pt}+ \hspace{-1pt} \sum_{i=1}^N U(x,t)u_i \hspace{-1pt} \right] \hspace{-2pt} dx.
 \end{align*}
Equilibrium solutions of the system can be found by minimizing this energy. However, if having such a structure to the system is not necessary, we can use two different kernels to describe the inter-group repulsion and intra-group aggregation. We handle this case numerically and explore this with a few examples in Section \ref{sec:diffPotentials}.
The term $U$ describes how favorable the environment is. In the case of meerkat territory development, meerkats have been known to avoid dense areas of sourgrass. Thus, $U$ could describe the density of sour grass at location $x$ and time $t$. While $U$ can be time-dependent, in this work, we only consider environments that are constant in time. We consider some of the environmental traits used in \cite{Bateman2015}, such as sand-type, interface between sand-types, and elevation, which we describe in more detail in Section \ref{sec:environmentanddata}.

A version of system \eqref{eq:System}  was used to study animal ecosystems with multiple groups \citep{Potts2019}, with the assumption that each population can detect opposing populations within a local neighborhood through direct observations or interactions, communication through scent marking, or memory of past interactions with opposing populations. Thus, in system \eqref{eq:System} we are able to take into account the processes modeled in system \eqref{eq:Bateman}, while eliminating the need for an artifical home center. 
Another version of system \eqref{eq:System} was introduced in \citep{Rodriguez2016} to represent social segregation. The $N=1$ case has received much attention, \citep{Bedrossian2011, Bertozzi2010, Li2010, Bernoff2015}, as well as versions in the $N=2$ case, \citep{Evers2017, Carrillo2017}, and this model has been used to study phenomena such as animal territories \citep{Potts2016}, predator-prey dynamics \citep{Fagioli2021}, and human gangs \citep{Barbaro2021}. It is further studied in \cite{Ellefsen2021}, where the authors investigate finding equilibrium solutions efficiently with multiple groups in two dimensions, an essential piece to connecting the model with data. In \cite{Giunta2021}, the authors investigate an efficient way to solve a multi-species model using spectral methods in one-dimension and two interacting species. 

Our contribution is the development of an efficient numerical method that allows us to solve system \eqref{eq:System} in two dimensions for a large number of groups. We perform numerical simulations for up to thirteen groups efficiently
%In order to connect the model with data, an efficient way to solve for the population densities in two-dimensions with many interacting species is imperative. In \citep{Ellefsen2021}, equilibrium solutions were found by minimizing an associated energy of the nonlocal system and a local approximation. These results showed that the energy landscape is very complicated, the local and nonlocal models did not always find the same minimizers. Thus, we follow a similar approach to \citep{Giunta2021}, using the Fast Fourier Transform to compute derivatives and convolutions in frequency space. However, we solve the system in two-dimensions with up to thirteen interacting-groups.  
which allows us to incorporate synthetic data into the model through parameter estimation via maximum likelihood estimation, as done in the local case \citep{Bateman2015}.

\section{Numerical Method and Solutions}\label{sec:numerics_method}
 In this section, we introduce the numerical methods to solve system \eqref{eq:System} and investigate how the parameters, potentials, and terms in the model affect the behavior of equilibrium solutions.
  
\subsection{Discretization Scheme} \label{sec:discretizationscheme}
We outline the process to numerically find equilibrium solutions of system \eqref{eq:System}.
We take advantage of working with a bounded, periodic grid and use the Fast Fourier Transform to move into frequency space, compute derivatives and convolutions, and use the Inverse Fast Fourier Transform to move back to physical space. This process gives us a discrete representation of the right hand side of our system, which allows us to discretize in time and numerically solve the system.

Consider the Fourier Transform, 
\begin{align*}
\hat{u}(k) = \int_{-\infty}^{\infty} u(x) e^{-2\pi i k x}dx, \hspace{.2cm} k \in \mathbb{R}.
\end{align*}
Because we are solving a nonlocal system of PDEs, we must address computing convolutions and derivatives efficiently. Therefore, there we employ two convenient properties of the Fourier Transform in our computations:
\begin{align}
\widehat{f'}(k) &=ik\widehat{f}(k)  \label{eq:deriv} \\
\widehat{K*f}(k) & = \widehat{K}(k)\widehat{f}(k) \label{eq:conv}.
\end{align}
As we are working on a bounded, periodic grid, we can compute derivatives and convolutions in frequency space using the Discrete Fourier Transform (DFT),
\begin{align}
\hat{v}_k &= h\sum_{j = 1}^{M} v_j e^{- i k x_j}, \hspace{.2cm} k \in -\frac{M}{2}+1,...,\frac{M}{2}, \label{eq:DFT}
\end{align}
which allows us to pass into frequency space. The Inverse Discrete Fourier Transform (IDFT), 
\begin{align}
v_j = \frac{1}{2\pi}\sum_{k=-M/2+1}^{M/2} \hat{v}_k e^{ i k x_j}, \hspace{.2cm} j \in 1, ..., M, \label{eq:IDFT}
\end{align}
allows us to return to physical space. 
We utilize equation \eqref{eq:DFT} to move into  the Fourier domain, we use properties \eqref{eq:deriv} and \eqref{eq:conv} to compute derivatives and convolutions in Fourier space, and use \eqref{eq:IDFT} to move back to physical space. In fact, we make use of the Fast Fourier Transform (FFT) and Inverse Fast Fourier Transform (IFFT) in order to move in and out of frequency space efficiently, reducing number of operations from $M^2$ to $M \log M$. 

We first compute the right hand side of system \eqref{eq:System} on the bounded, periodic grid for $t = n\Delta t$. 
%% how to compute fourier space
Let $u_{mj}^{i,n}$ be the value of $u$ at time $ n \Delta t$ for group $i \in \{1,...,N\}$ at grid point $(m,j)$ for $m,j \in \{ 1,...,M\}$ , and let $\hat{u}_{k\ell}^{i,n}$  represent the DFT for $k, \ell \in \{-M/2+1, ..., M/2\}$. We begin by using the FFT to move into Fourier space to determine $\hat{u}_{k\ell}^{i,n}$, $\hat{K}$, and $\hat{U}$. We compute the nonlocal terms in system \eqref{eq:System} by multiplying $\hat{u}_{k\ell}^{i,n}$ and $\hat{K}$ for each group $i$, utilizing the property in equation \eqref{eq:conv}.
We remain in Fourier space to compute the gradient of the nonlocal terms and $U$, using the property in equation \eqref{eq:deriv}. 
%We multiply by $im$ to compute $\partial_x u_{m,j}^{i,n}$  and by $ij$ to compute $\partial_y u_{m,j}^{i,n}$. 
We use the IFFT to return the physical domain and multiply these terms by $u$. We move back into frequency space in order to compute derivatives, using property \eqref{eq:deriv}. Finally, we take the IFFT to move back to physical space. 
This gives us a discrete representation of the right hand side for group $i$ at time $t = n\Delta t$, 
\begin{align}\label{eq:discrete}
\partial_t u^{i,n} = f(u^{i,n}), i = 1, ..., N,
\end{align}
where $f$ is the function described above. 
%%solve equilibrium problem
We can solve for equilibirum solutions directly by setting $\partial_t u^{i,n} = 0$  in \eqref{eq:discrete} for $i = 1 ... N$. To investigate dynamics of the system as well, we can time step until we reach equilibrium. One can use equation \eqref{eq:discrete} and an ODE solver to find $u^{i,n+1}$. In particular, we use a Runge Kutta method in MatLab.

\begin{comment}
We can also solve with $\partial_t u^{i,n} = 0$ and time step from the solution found to reduce error. We find that when fitting data, time stepping allows from the initial conditions determined by the data produces the best results.
%%time step
We discretize in time as follows:
\begin{align*}
\frac{u_{mj}^{i,n+1}  - u_{mj}^{i,n}}{\Delta t } &= \frac{1}{2}(f(u_{mj}^{i,n}) + f(u_{mj}^{i,n+1})),
\end{align*} 
and solve for $u^{i,n+1}$. 
\end{comment}

After solving directly for the equilibrium solution or between each time step, we mitigate error by following the same process as done in \cite{Bateman2015}. First, we set all negative entries for $u^{i,n}_{m,j}$ equal to zero. Second, we approximate the mass of each group, 
\begin{align*}
M_i^n &= \sum_{m,j=1}^M \frac{1}{4}( u^{i,n}_{m,j}  + u^{i,n}_{m,j+1} + u^{i,n}_{m+1,j} + u^{i,n}_{m+1,j+1})\Delta x \Delta y,
\end{align*}
and divide $u^{i,n}$ by $M_i^n$ to maintain a mass of $1$. Note that when $m$ or $j = M$, we use periodicity in $u$. 
We determine error at time $n \Delta t$ by computing the the $L^2$ norm of $f(u^{i,n})$ for for $i=1 ... N$. For an equilibrium solution, this value should be zero. We continue time-stepping until this value is below a chosen tolerance.

\subsection{Non-smooth and Non-periodic Environments and Location Data}\label{sec:environmentanddata} 

\begin{figure}[h!]
	\centering
	\subfigure[Sand Environmental Data]{\includegraphics[width = .45\linewidth]{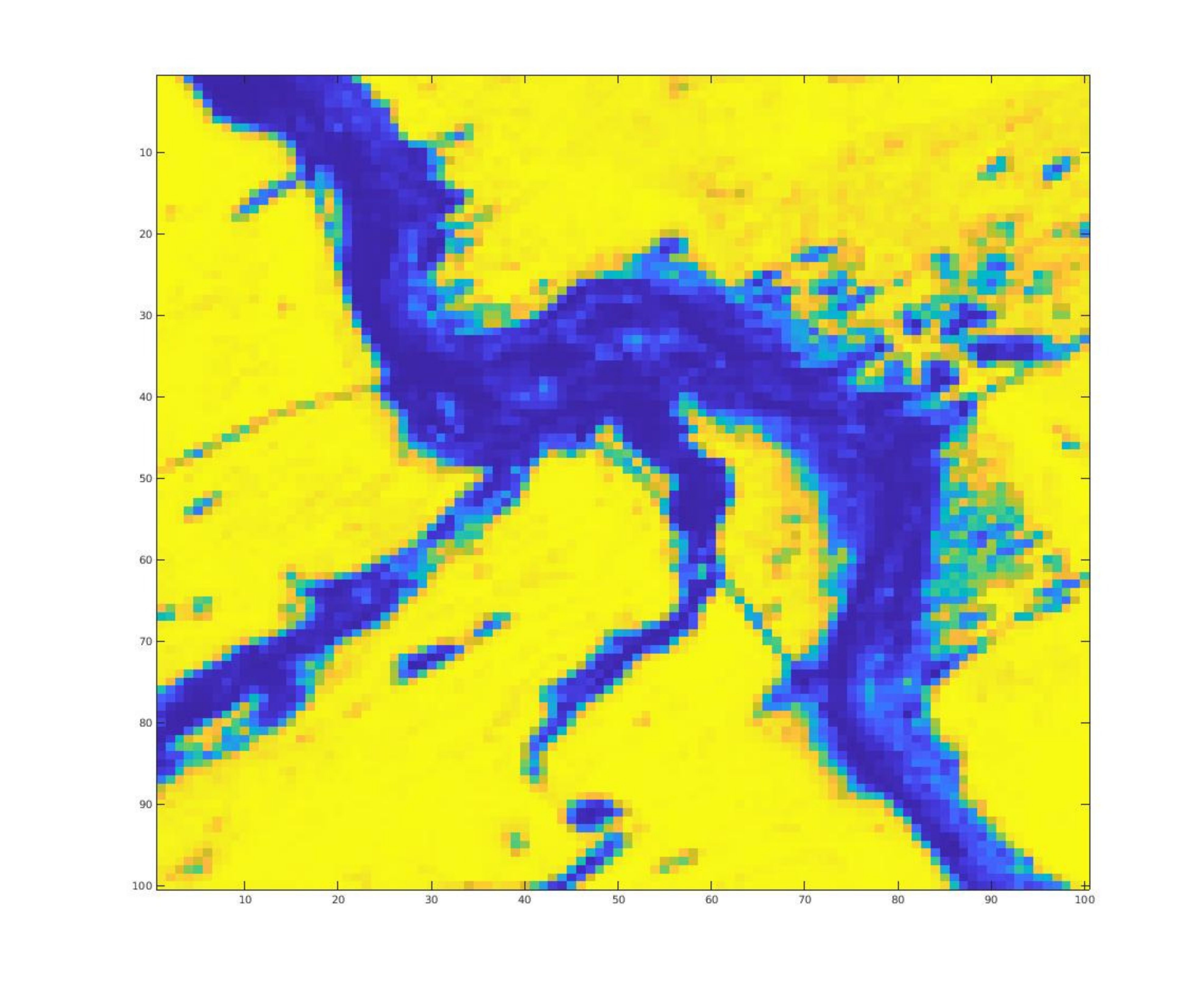} \label{fig:SAND1}}
	\subfigure[Mollified Sand Environmental Data with extended domain]{\includegraphics[width = .45\linewidth, height=1.6in]{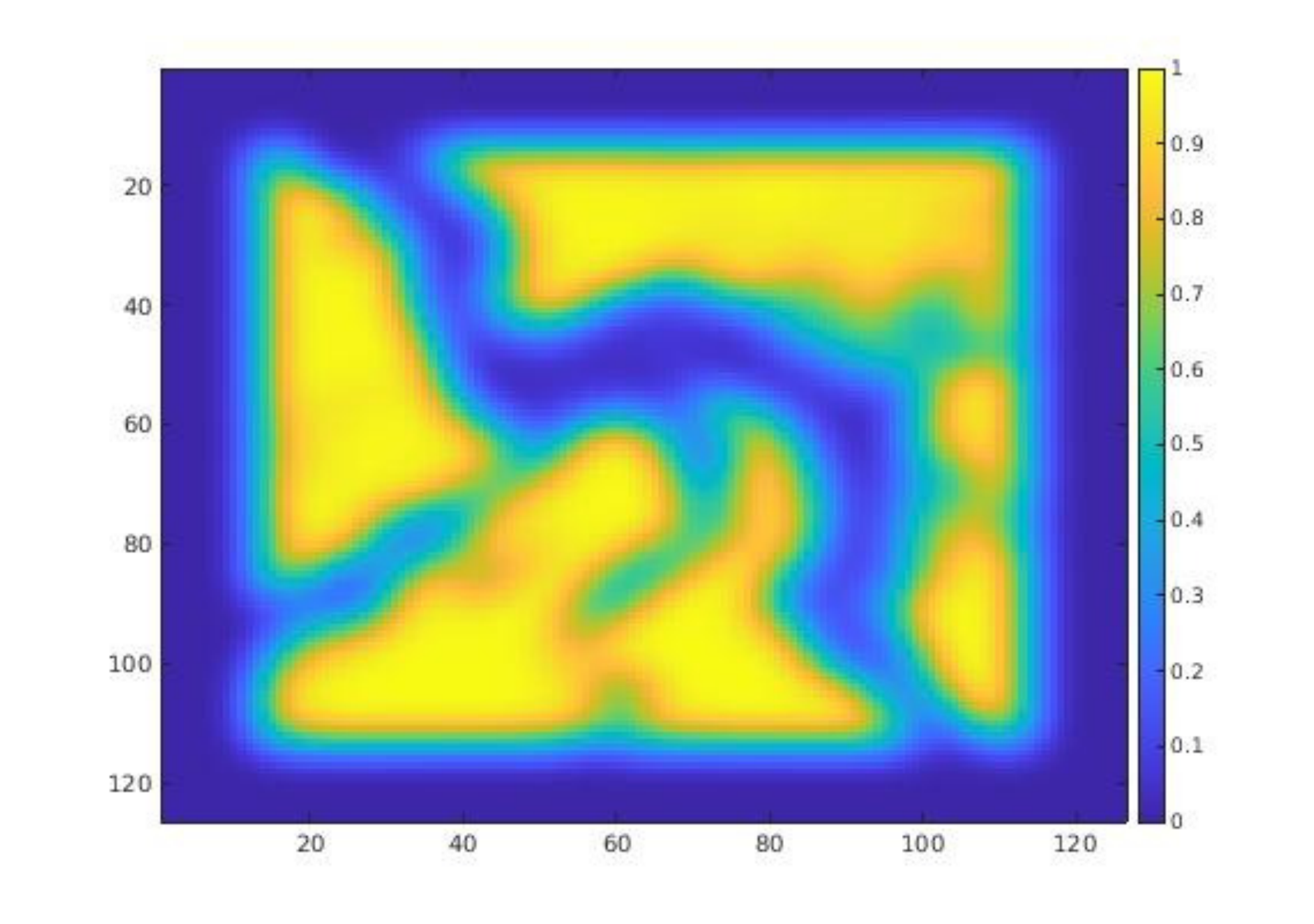} \label{fig:SAND}}\\
	\subfigure[Edge Environmental Data]{\includegraphics[width = .45\linewidth]{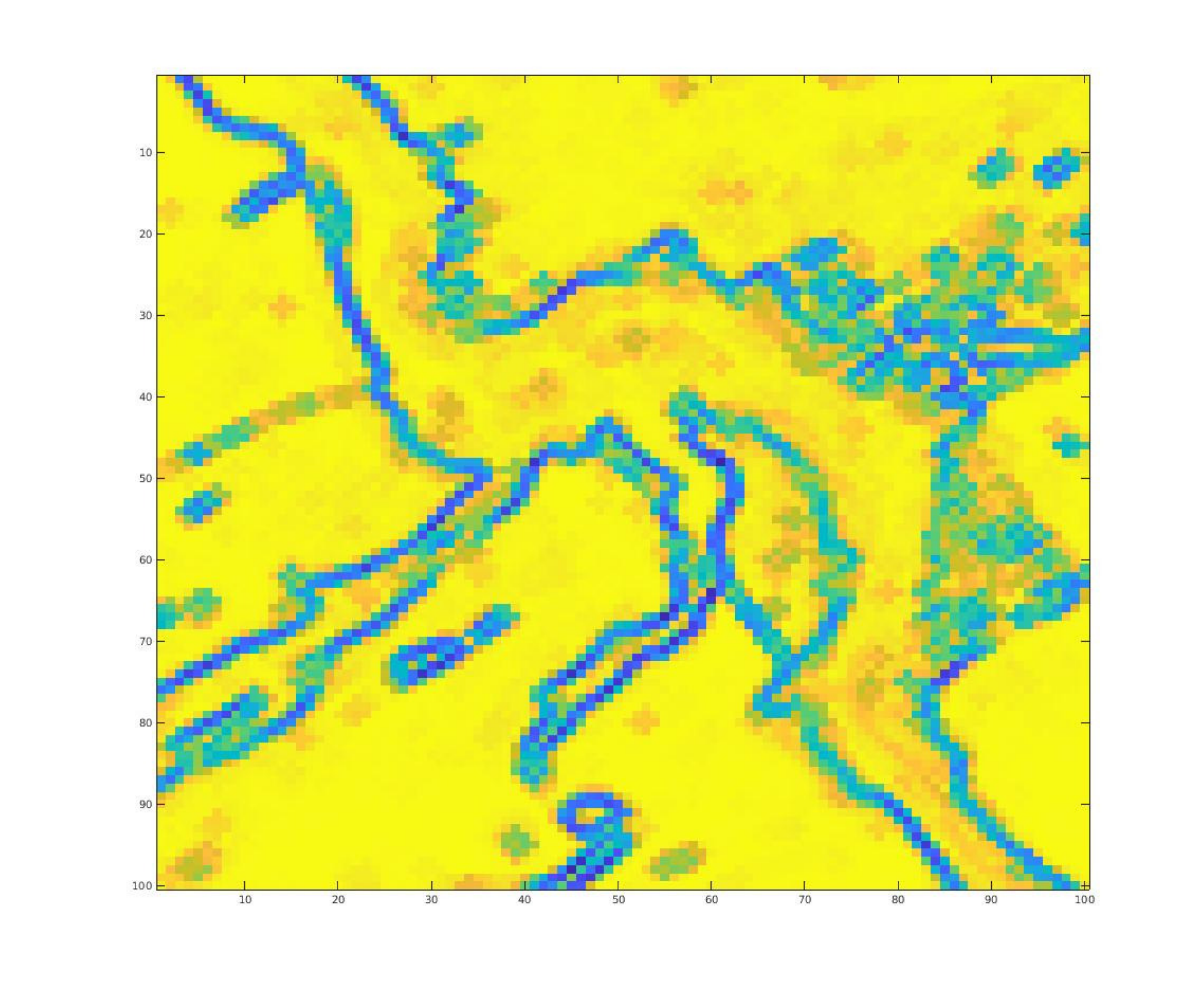} \label{fig:EDGE1}}
	\subfigure[Mollified Edge Environmental Data with extended domain]{\includegraphics[width = .45\linewidth, height = 1.6in]{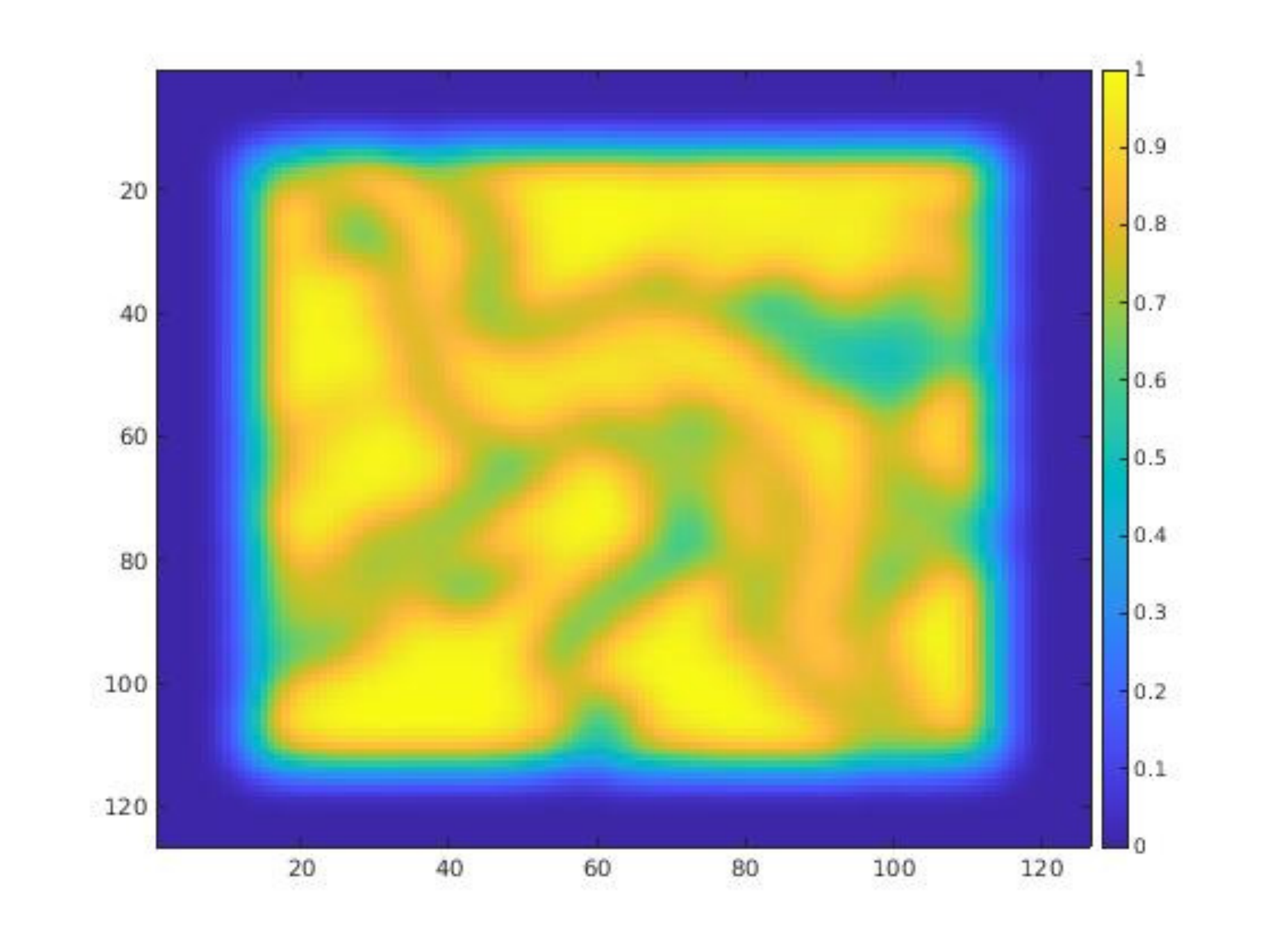} \label{fig:EDGE}}
	\caption{Environmental data and mollified environmental data on an extended domain in the same geographical area in South Africa for two environment types on a scale of 0 (unattractive environment)  to 1 (attractive environment)}
	\label{fig:Environment}
\end{figure}

\begin{figure}[h!]
	\centering 
	\subfigure[Location Data]{\includegraphics[width=.45\linewidth]{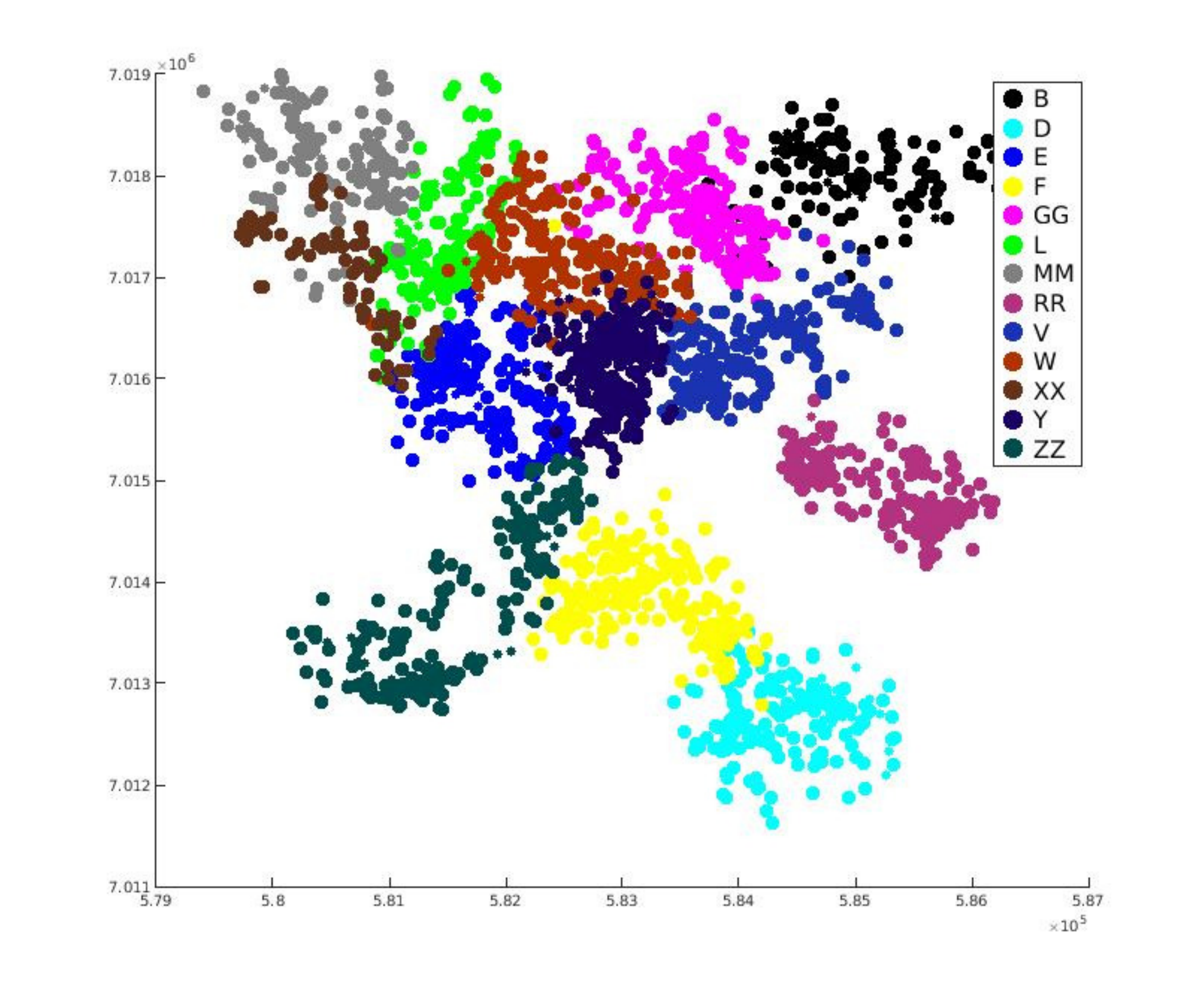}\label{fig:data}}
	\subfigure[Kernel Density Estimation of data]{\includegraphics[width = .45\linewidth]{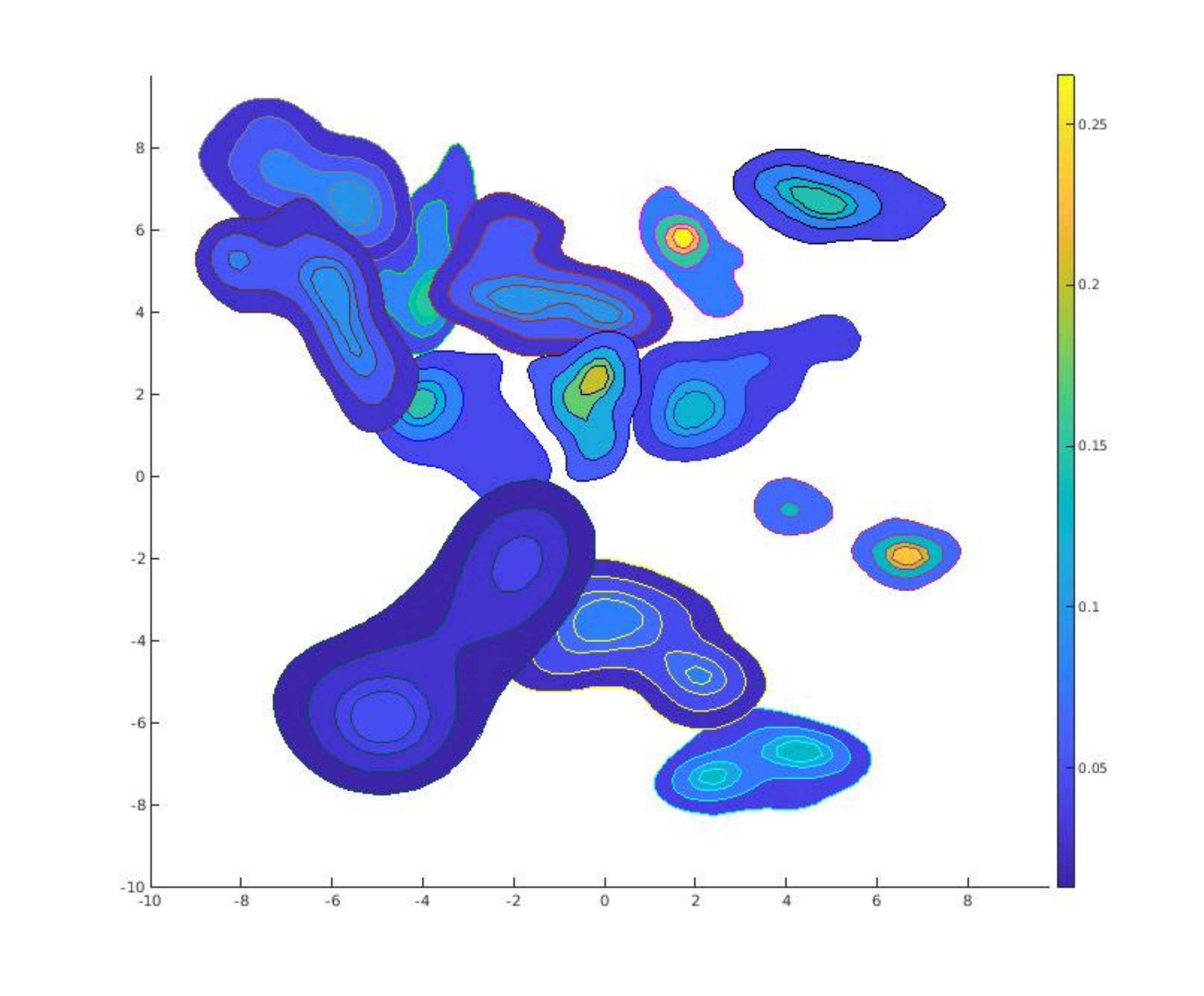} \label{fig:KDE}}
	\caption{(Meerkat location data on the same geographical area as Figure \ref{fig:Environment}, where each data point is a location where a meerkat from that group was observed and each color is a different group, and the kernel density estimation of the location data} 
	\label{fig:DataKDE}
\end{figure}

In application, environmental data is usually not smooth. In this section, we discuss how to deal with this issue. For an illustration, we work with the environmental and location data for meerkats. 
% In following simulations, we consider environment data and meerkat location data. 
The detailed account of acquisition of the habitat data can be found in \cite{Bateman2015}. The environmental data for sand-type and interface between sand types are shown in Figure \ref{fig:SAND1} and Figure \ref{fig:EDGE1}, respectively. The data are normalized to fall between zero and one. For the sand-type data, Figure \ref{fig:SAND1}, zero corresponds to clay sand, and one corresponds to ferrous sand. For the edge data, or the interface between sand-types seen in Figure \ref{fig:EDGE1}, zero corresponds to the interface and one corresponds to a distinct sand-type. To deal with the non-smooth nature of the data, we convolve the data with a mollifier so that we can consider a smoother $U$. These mollified environment functions are shown in Figure \ref{fig:SAND} and Figure \ref{fig:EDGE}. It is also the case that the environment data and meerkat habitats are not periodic, while we are assuming periodic boundary conditions. Therefore, to reduce boundary effects, we extend our domain with $U(\textbf{x}) = 0$, where $\textbf{x}$ is in the extended part of the domain. This is illustrated by the blue border in Figure \ref{fig:SAND} and \ref{fig:EDGE}.

  We also consider meerkat location data from the Kalahari Meerkat Project. In \cite{Bateman2015}, three time periods of data were considered: a steady period where territories are relatively static, a period where a social group split into two new social groups, and a period where territory locations shifted. The data include thirteen groups of meerkats, and each data point describes a time and location where a member of a group was observed. Figure \ref{fig:data} illustrates the location data from the steady period. To use initial conditions informed by this data, we perform the kernel density estimation of the given data for each group, shown in Figure \ref{fig:KDE}. Note that Figure \ref{fig:KDE} is shown without the extended domain. With the extended domain, the initial values for each group are further removed from the boundaries, reducing the effect of the periodic boundary conditions.

\subsection{Exploring Computation Times} \label{sec:computationtime}
\begin{table}[h!]
	\centering
\begin{tabular}{|c|c|c c c c c|c c c c c|}
	\hline
	& & \multicolumn{5}{|c|}{$a = .25$} & \multicolumn{5}{|c|}{$a = 1$}\\
	\hline
	\multirow{6}{4em}{$b = .5$}& $N$ /$\eta$  &1&3&5&7&9&1&3&5&7&9\\
	\hline
	&1& 6 & 25 & 14 & 16 & 20 & 15 & 8 & 11 & 12 & 13\\
	&3&  82& 31 & 58 & 56 & 69 &  83& 19 &24 & 27 & 28 \\
	&5& 362 & 107 & 120 & 108 & 133  & 61   & 30 & 40& 47 & 49 \\
	&7& 1038 & 270 & 216 & 169 & 206  &126 & 66& 64& 71 & 72\\
	%&9&  &  &  &   &   &  & & & & \\
	\hline
	\multirow{6}{4em}{$b = 1$} & $N$ /$\eta$  &1&3&5&7&9&1&3&5&7&9\\
	\hline
	&1& 14 & 4 & 10 & 12 & 14& 13& 4&8&10 & 11 \\
	&3&  108& 36 & 26 & 27 & 29 & 205& 16& 19 & 21 & 24 \\ 
	&5& 386 & 132 & 81 & 65 & 58 &1243 & 94& 61& 50& 48 \\
	&7& 874& 248 & 141 & 114 & 91 & -- & 164&  117& 92& 79\\
	%&9&  &  &  &  &  &  & & & &  \\
	\hline
\end{tabular}

\caption{Computation time (s) to find equilibrium solutions to system \eqref{eq:System} with varying environment strength ($a$), potential strength ($b$), group number ($N$), and diffusivity parameter ($\eta$), and error tolerance = .02.}
\label{T:Computation}
\end{table}

\begin{figure}[h!]
\centering
	\subfigure[$N=1$,$\eta=1$]{\includegraphics[width = .17\linewidth]{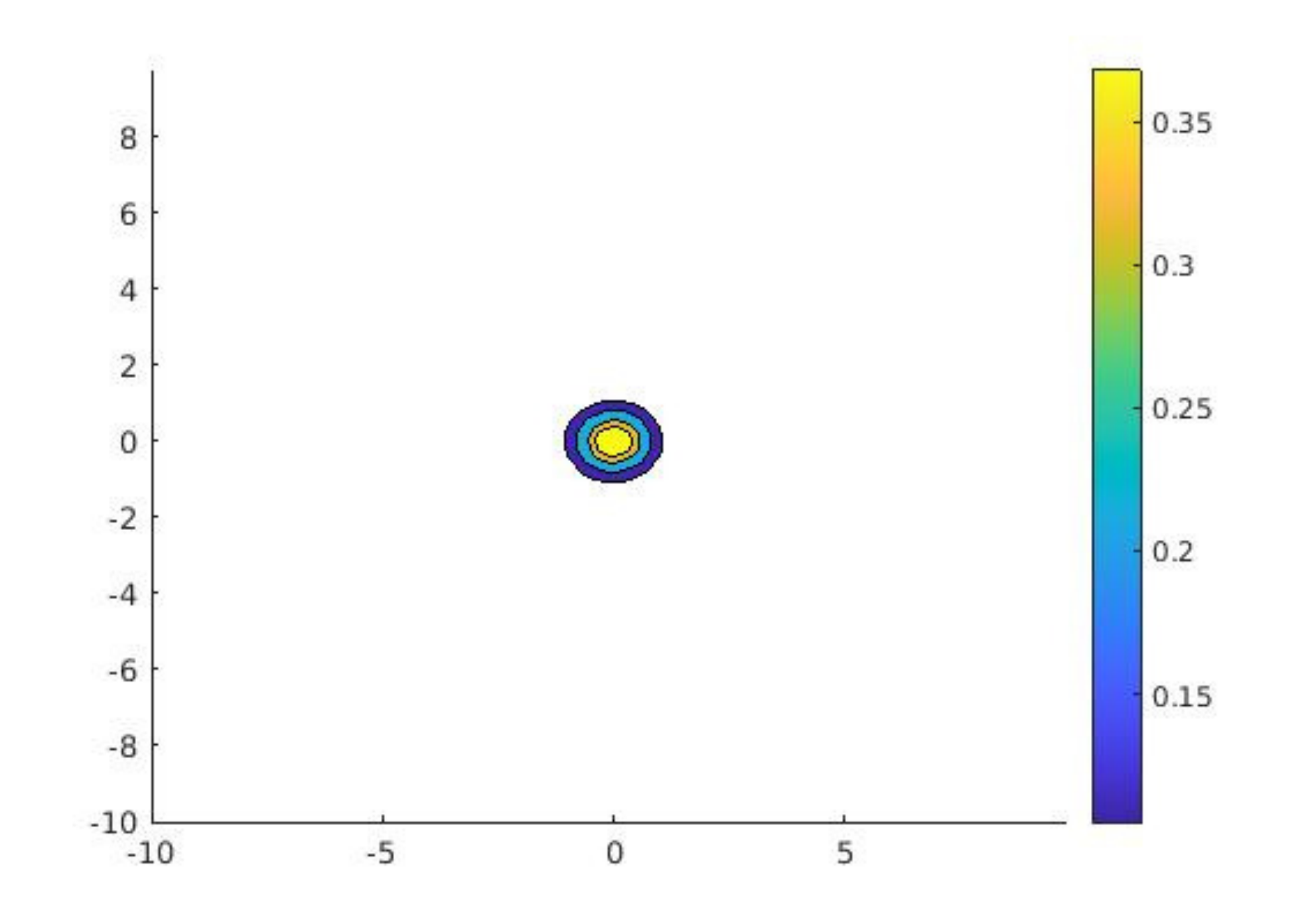}}
	\subfigure[$N=1$,$\eta=3$]{\includegraphics[width = .17\linewidth]{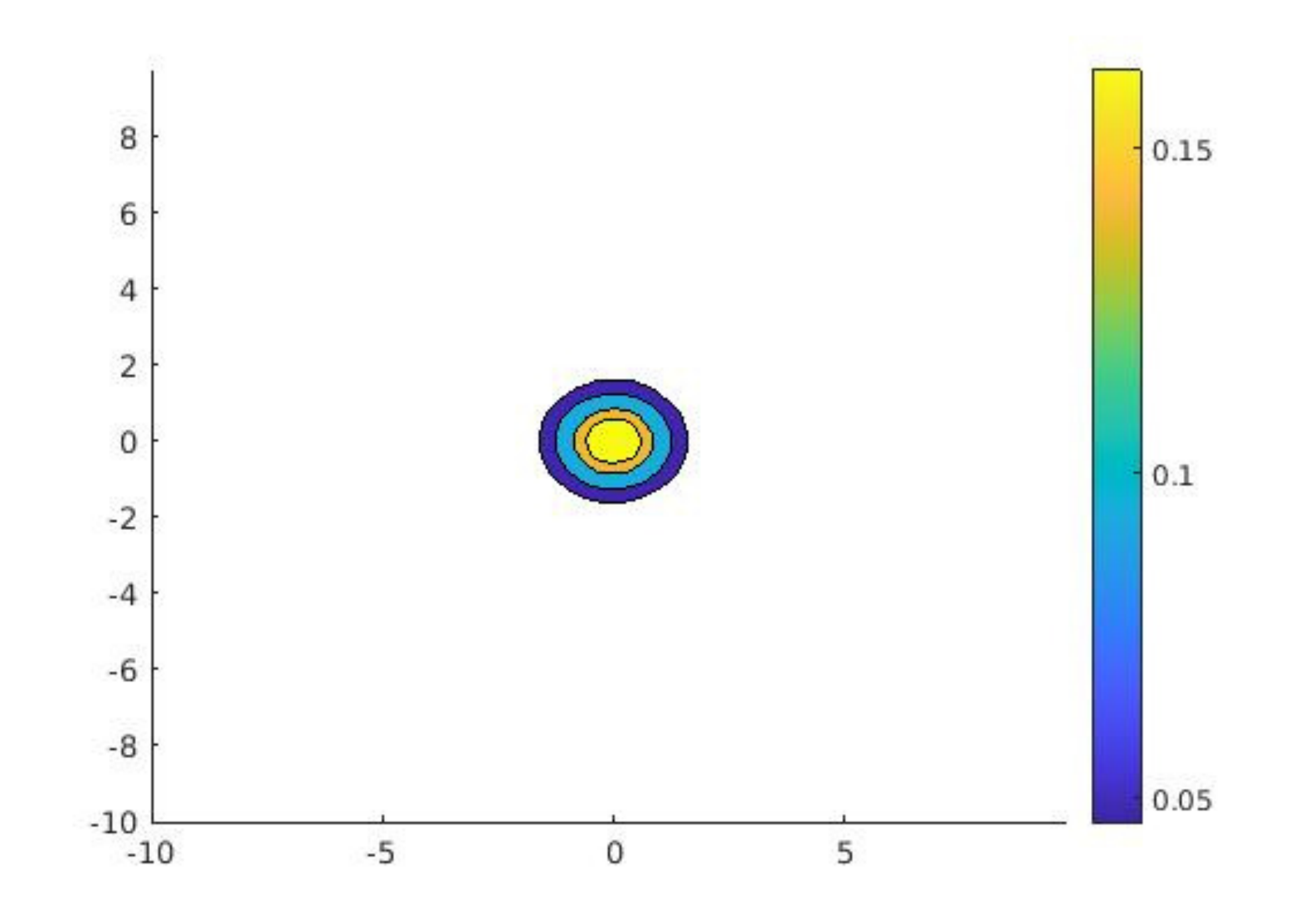}}
	\subfigure[$N=1$,$\eta=5$]{\includegraphics[width = .17\linewidth]{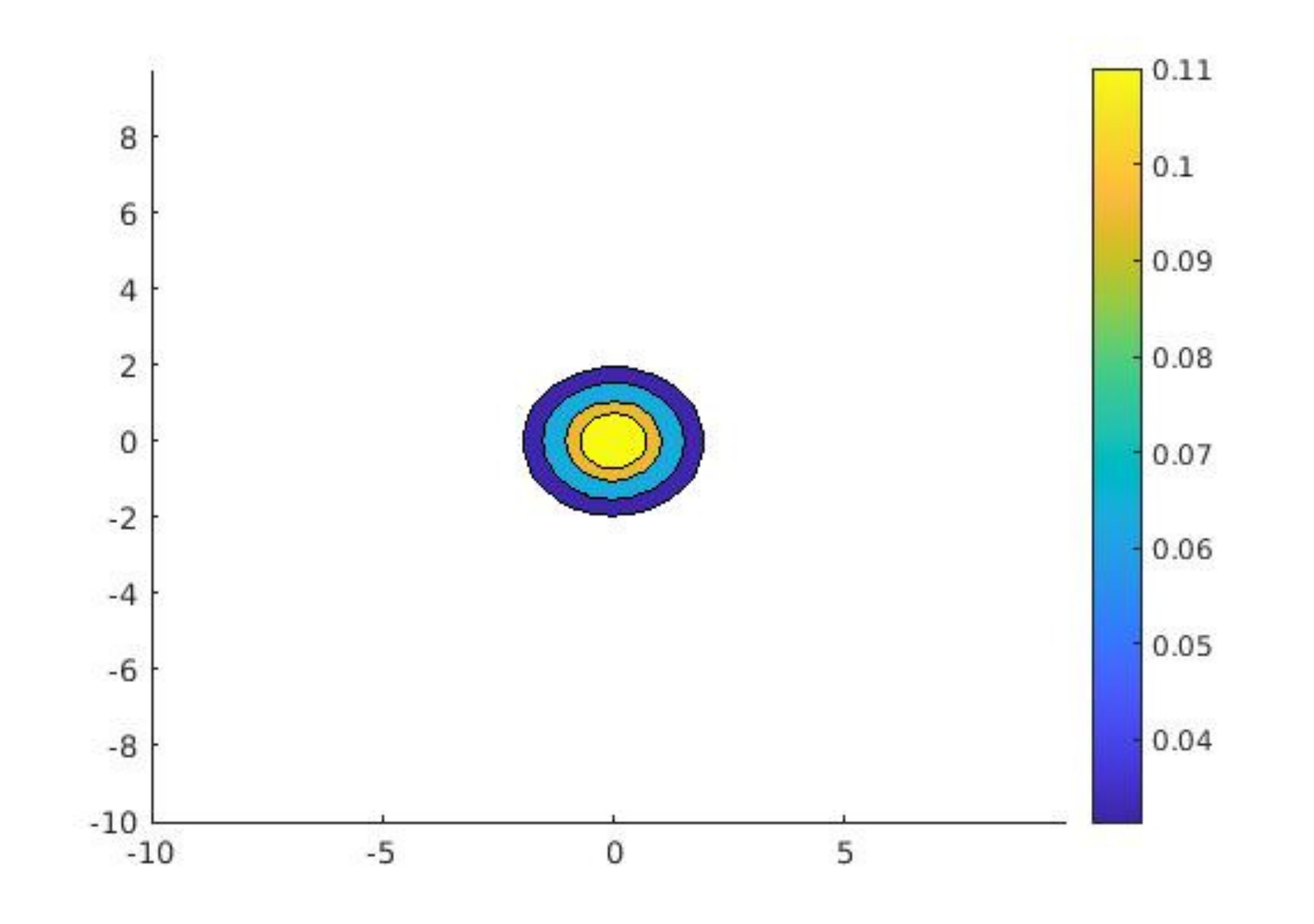}}
	\subfigure[$N=1$,$\eta=7$]{\includegraphics[width = .17\linewidth]{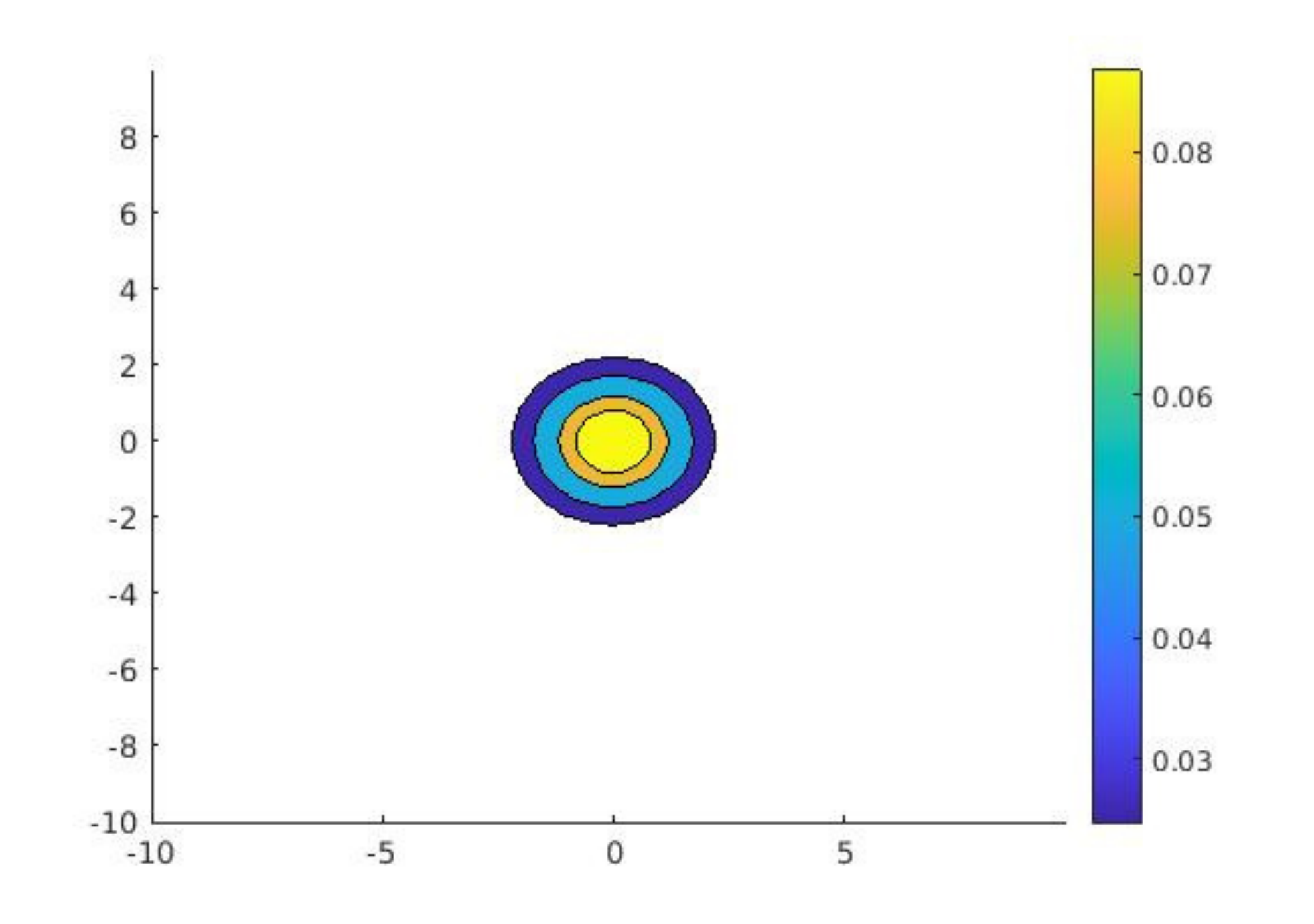}}
	\subfigure[$N=1$,$\eta=9$]{\includegraphics[width = .17\linewidth]{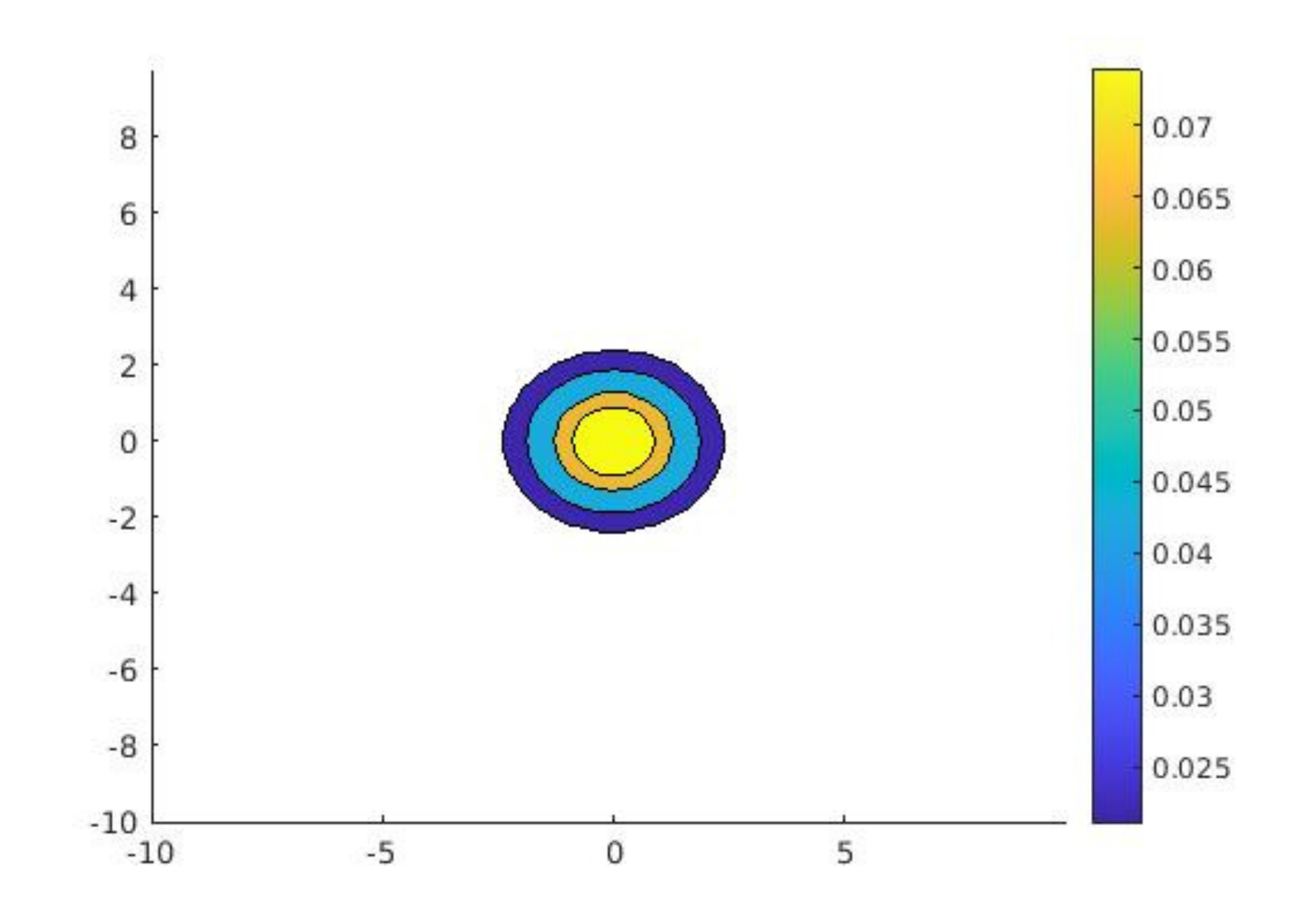}}\\
	\subfigure[$N=3$,$\eta=1$]{\includegraphics[width = .17\linewidth]{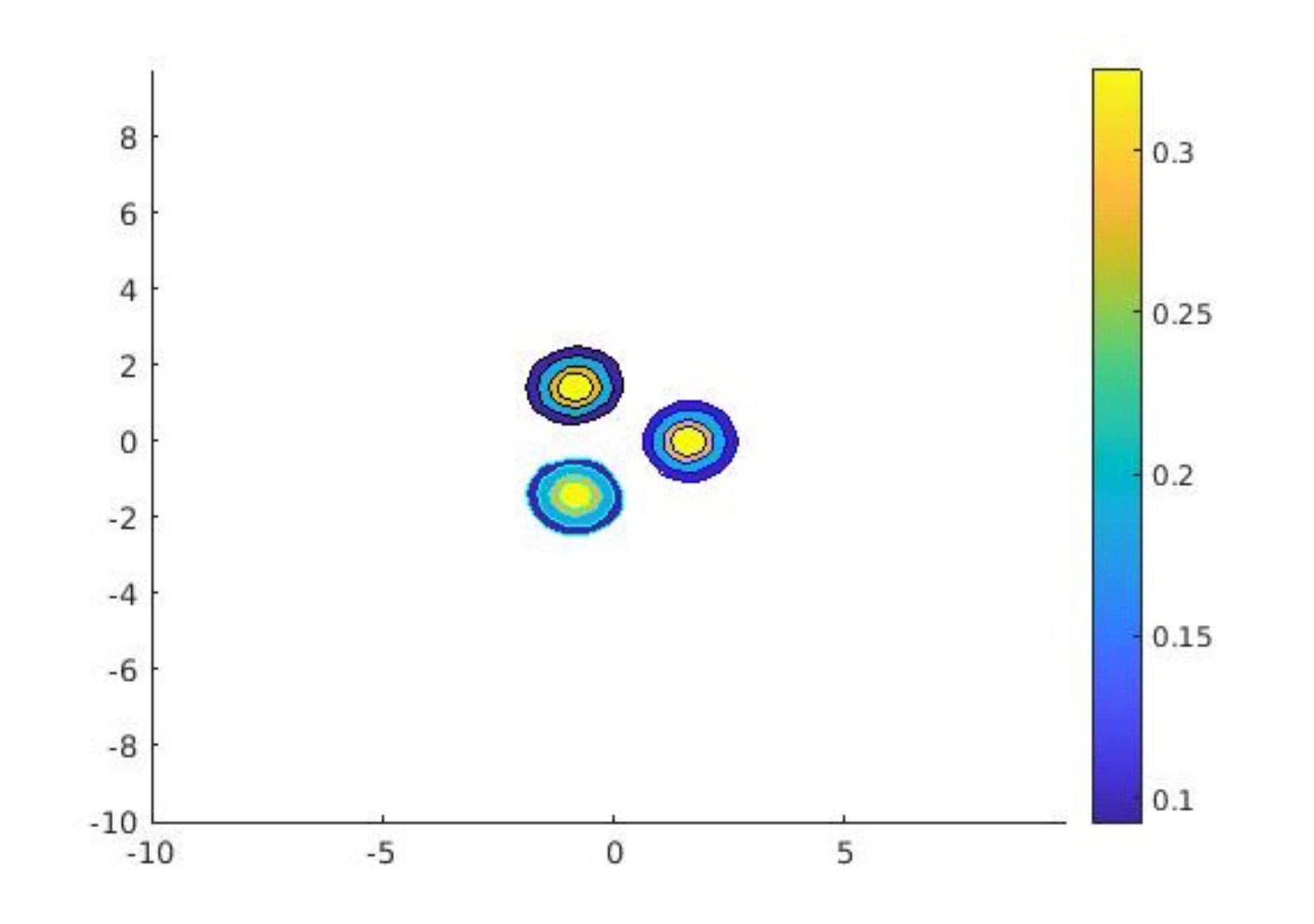}}
	\subfigure[$N=3$,$\eta=3$]{\includegraphics[width = .17\linewidth]{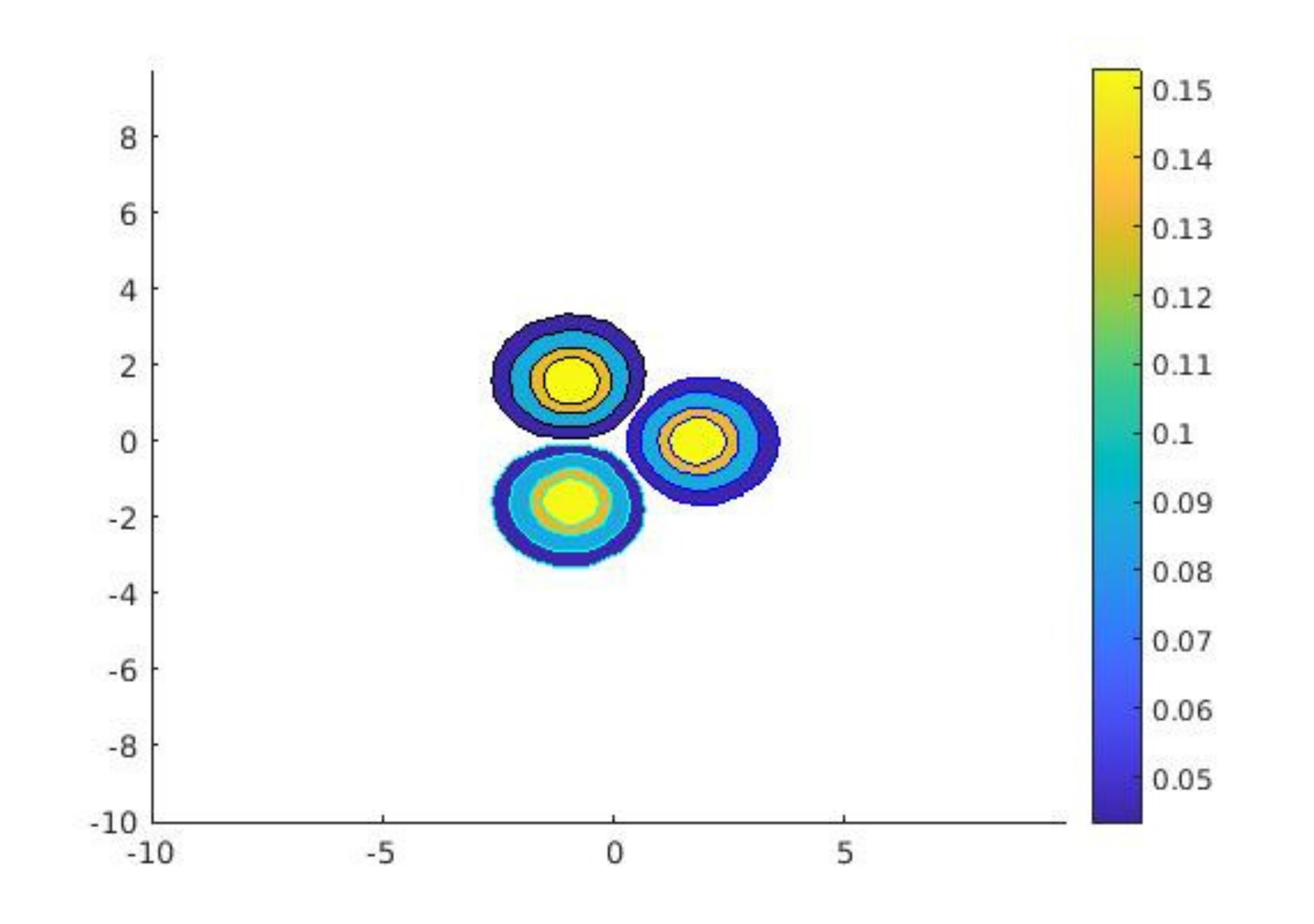}}
	\subfigure[$N=3$,$\eta=5$]{\includegraphics[width = .17\linewidth]{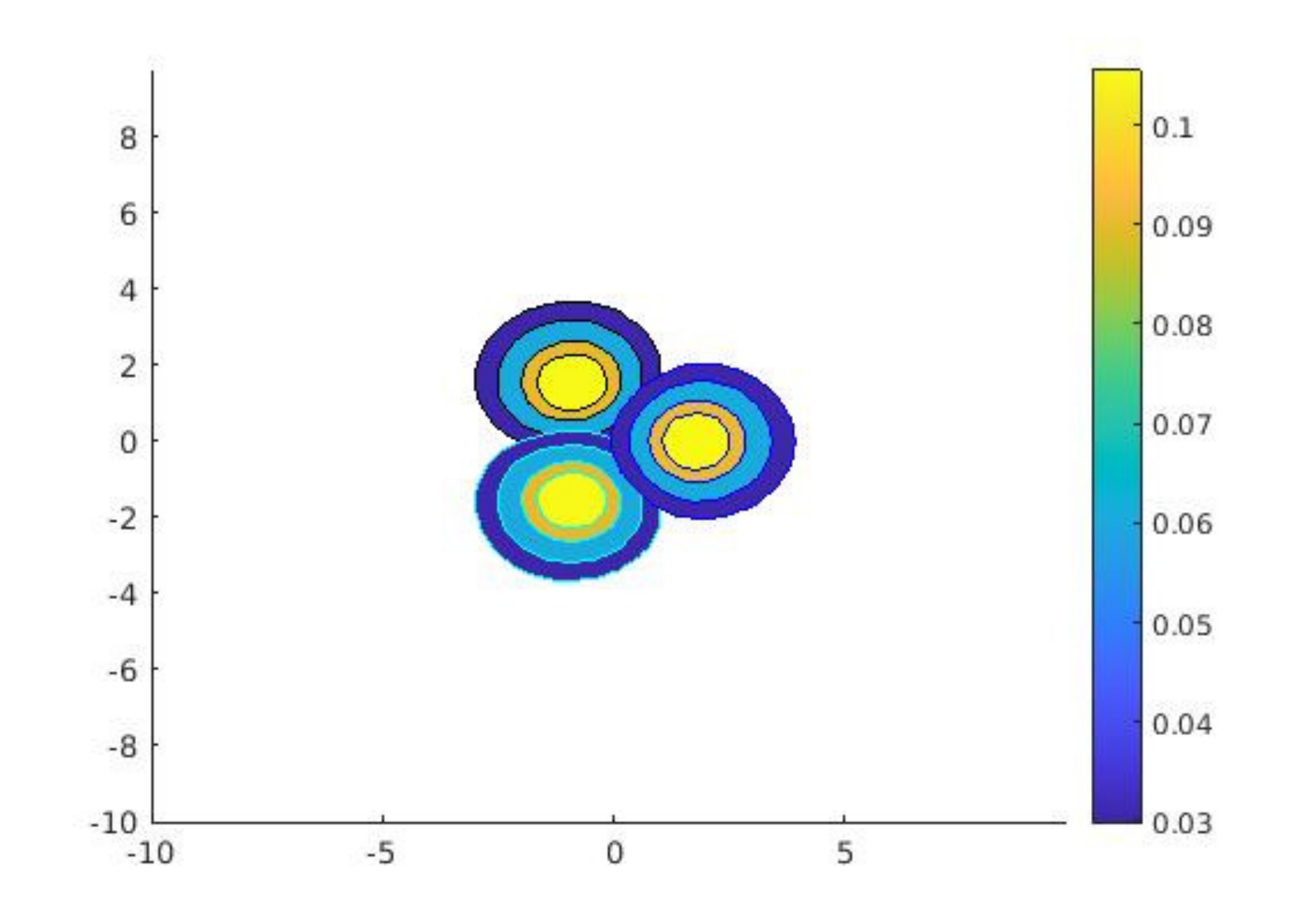}}
	\subfigure[$N=3$,$\eta=7$]{\includegraphics[width = .17\linewidth]{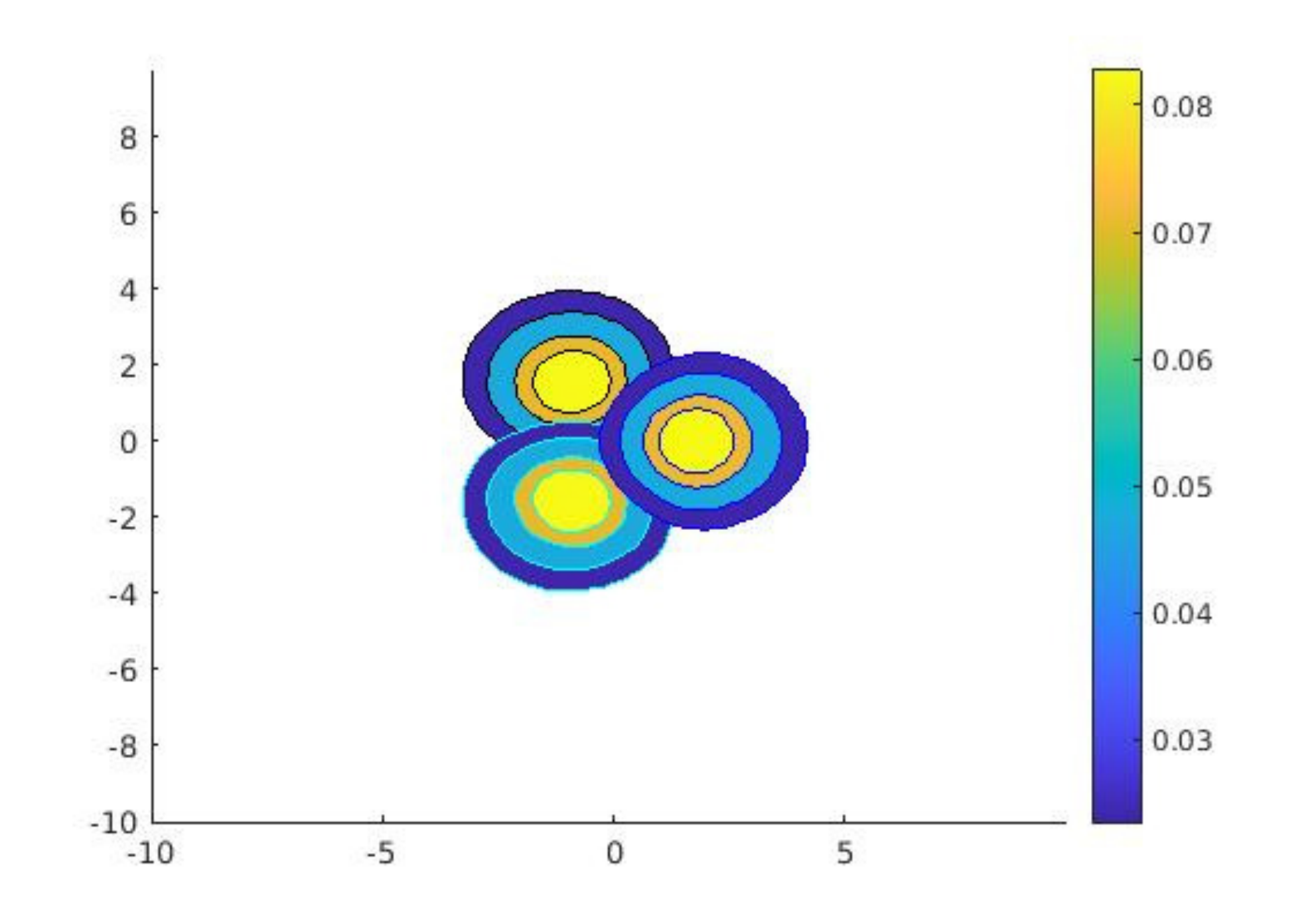}}
	\subfigure[$N=3$,$\eta=9$]{\includegraphics[width = .17\linewidth]{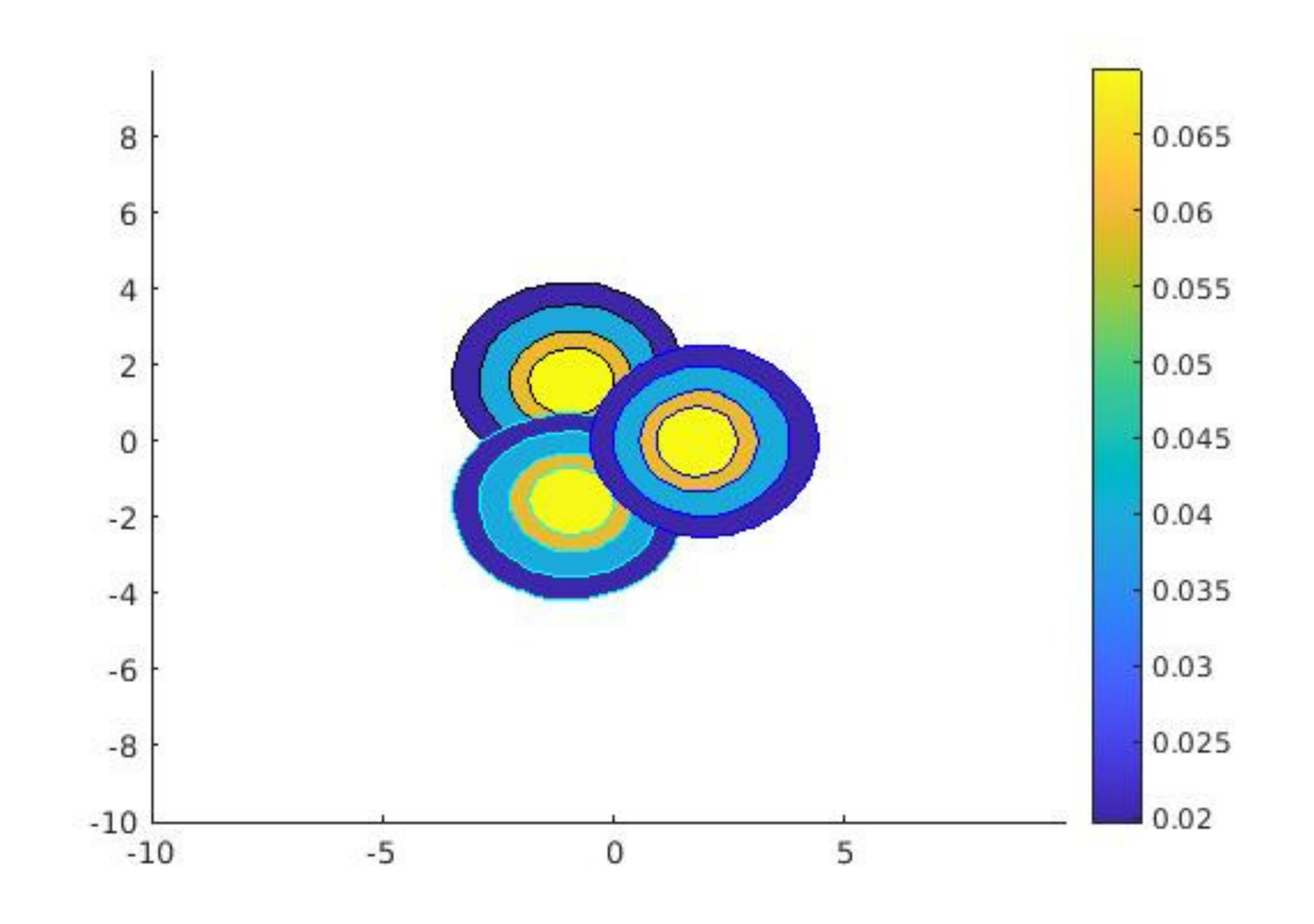}}\\
	\subfigure[$N=5$,$\eta=1$]{\includegraphics[width = .17\linewidth]{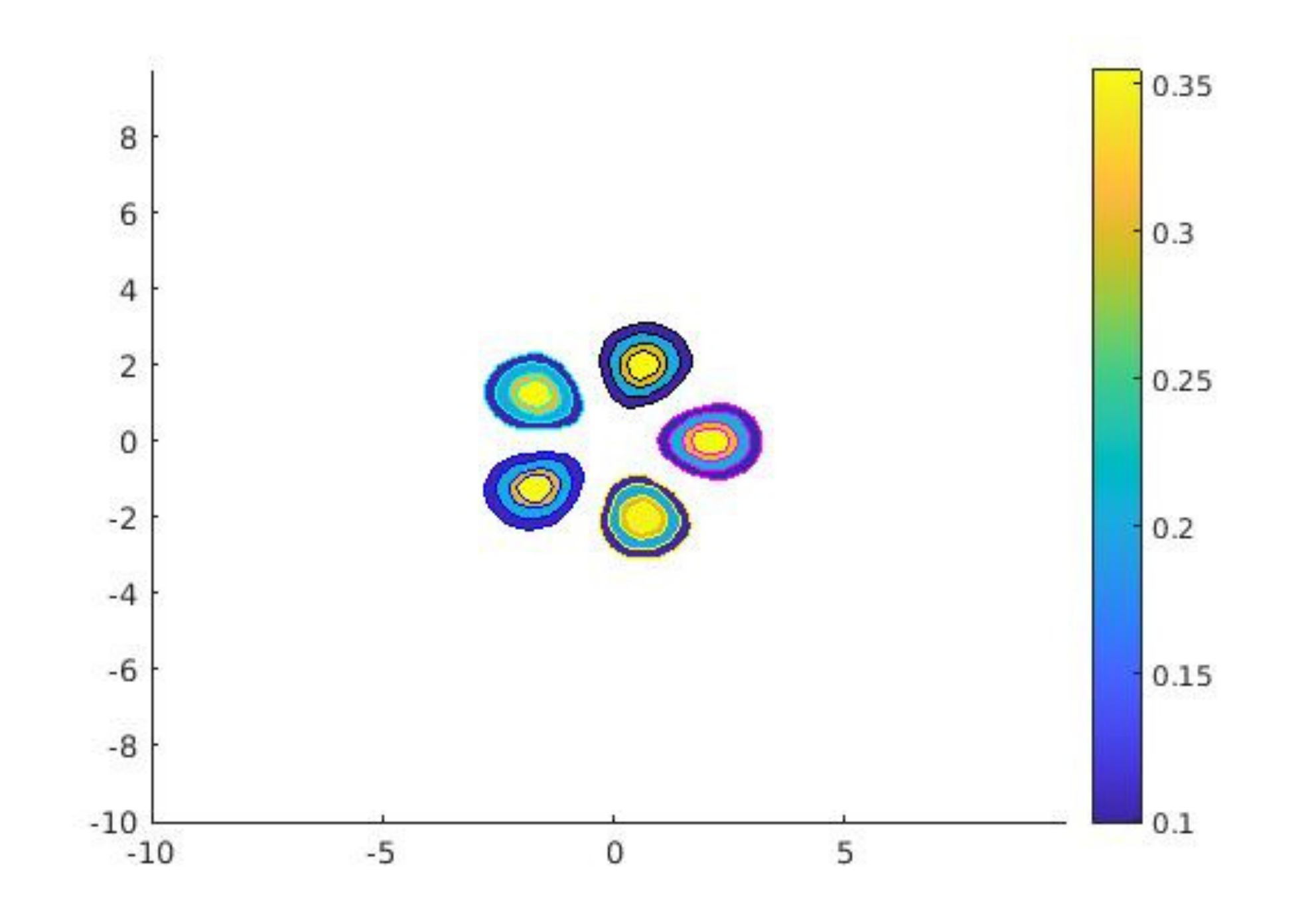}}
	\subfigure[$N=5$,$\eta=3$]{\includegraphics[width = .17\linewidth]{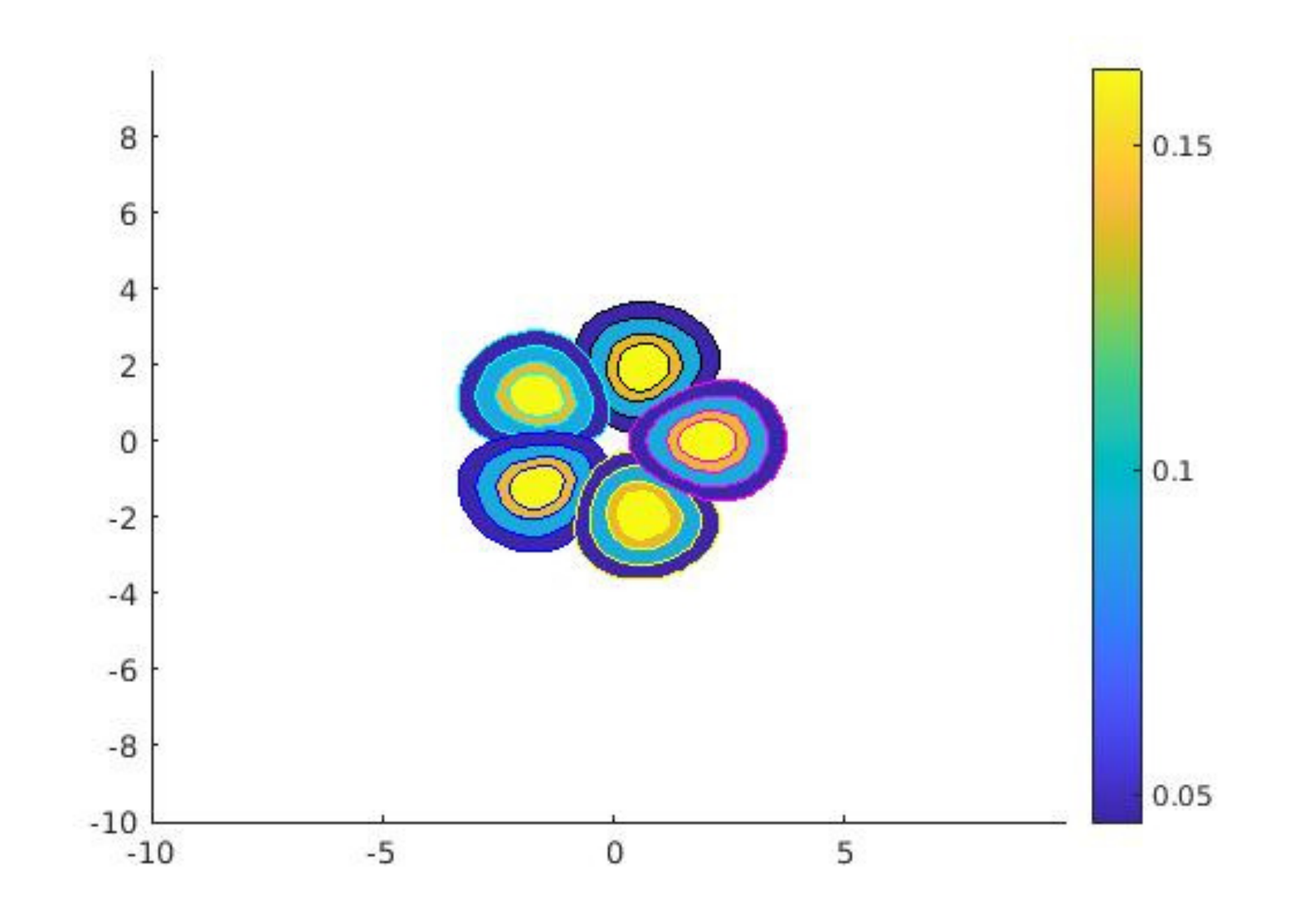}}
	\subfigure[$N=5$,$\eta=5$]{\includegraphics[width = .17\linewidth]{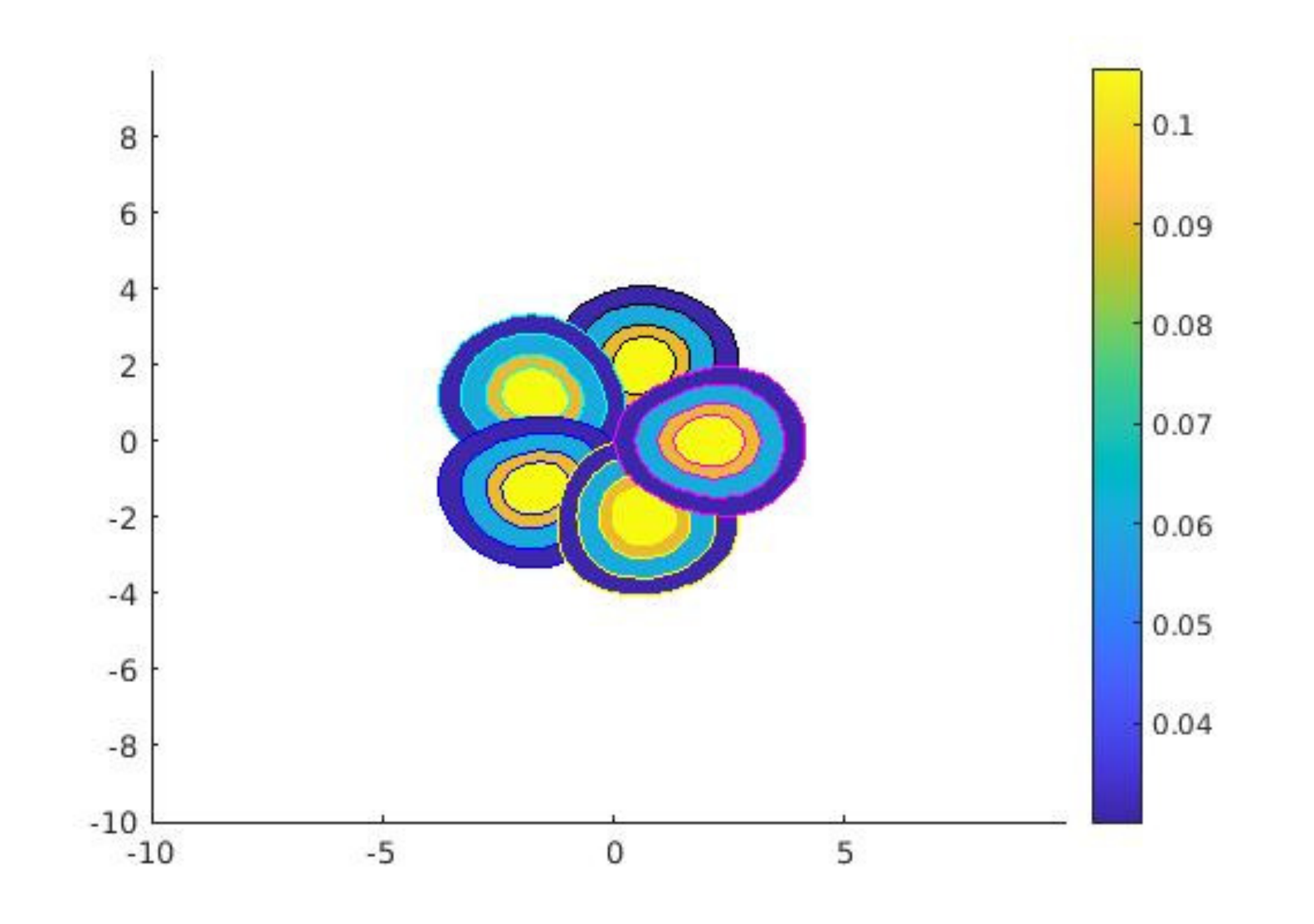}}
	\subfigure[$N=5$,$\eta=7$]{\includegraphics[width = .17\linewidth]{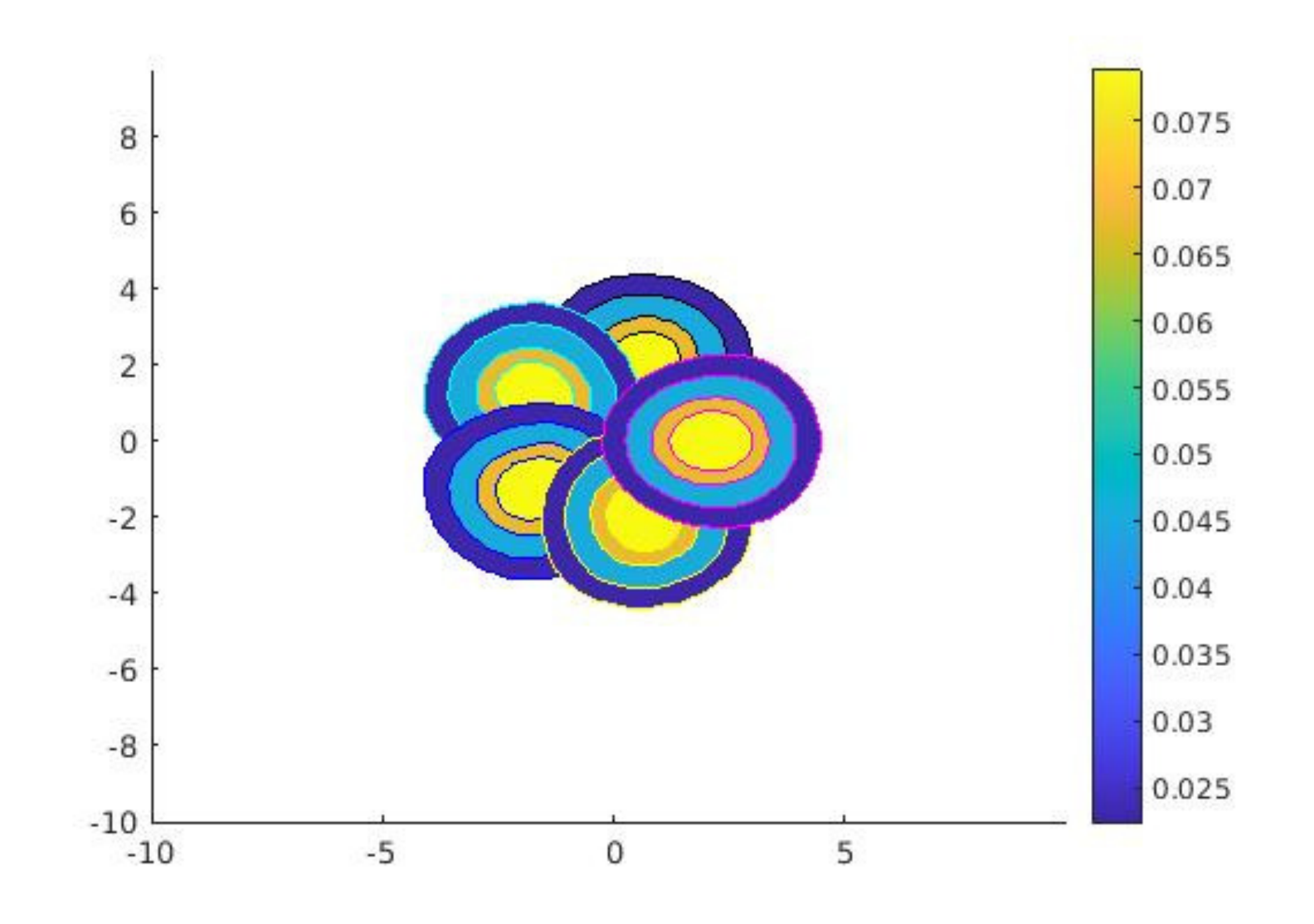}}
	\subfigure[$N=5$,$\eta=9$]{\includegraphics[width = .17\linewidth]{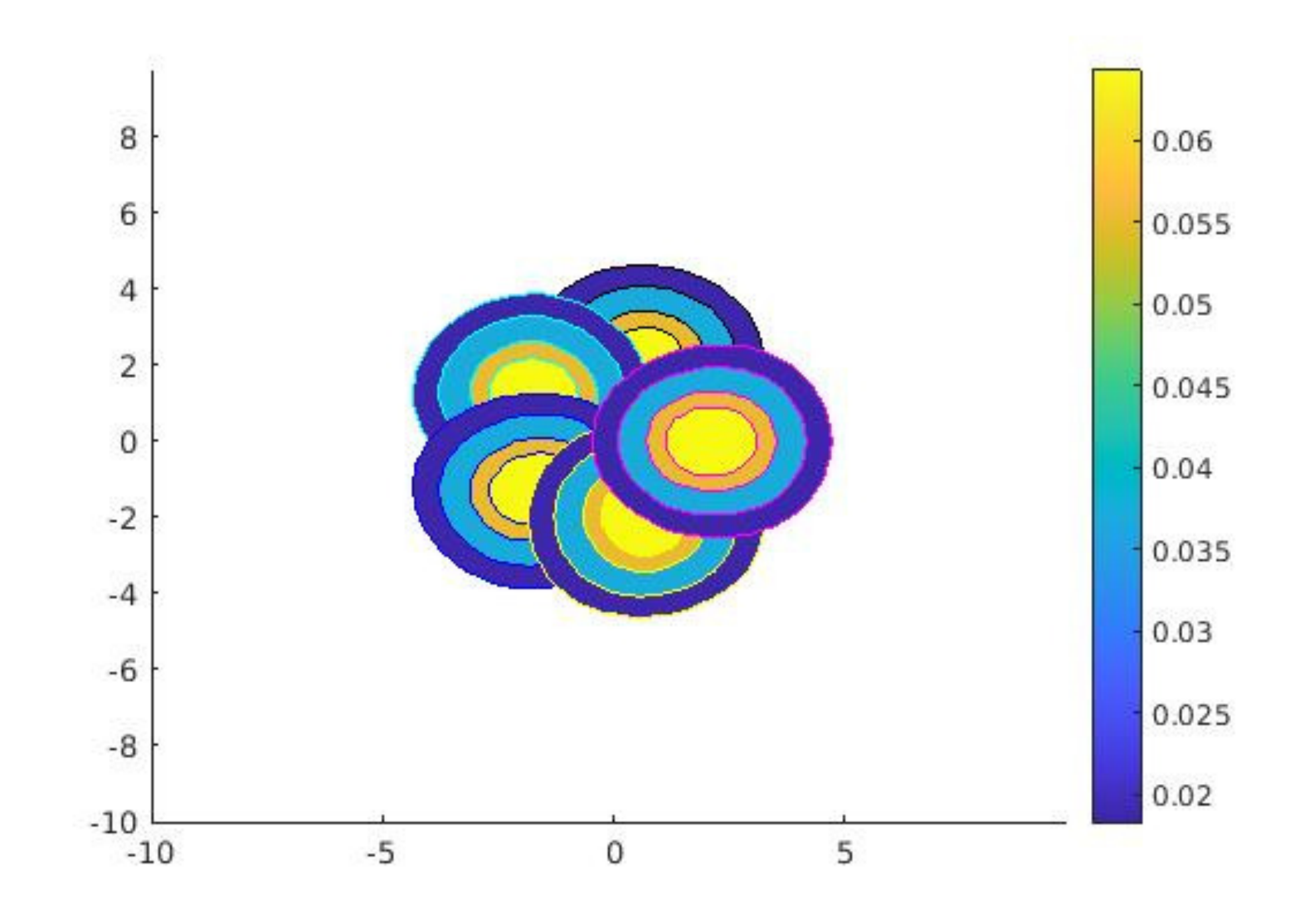}}\\
	\subfigure[$N=7$,$\eta=1$]{\includegraphics[width = .17\linewidth]{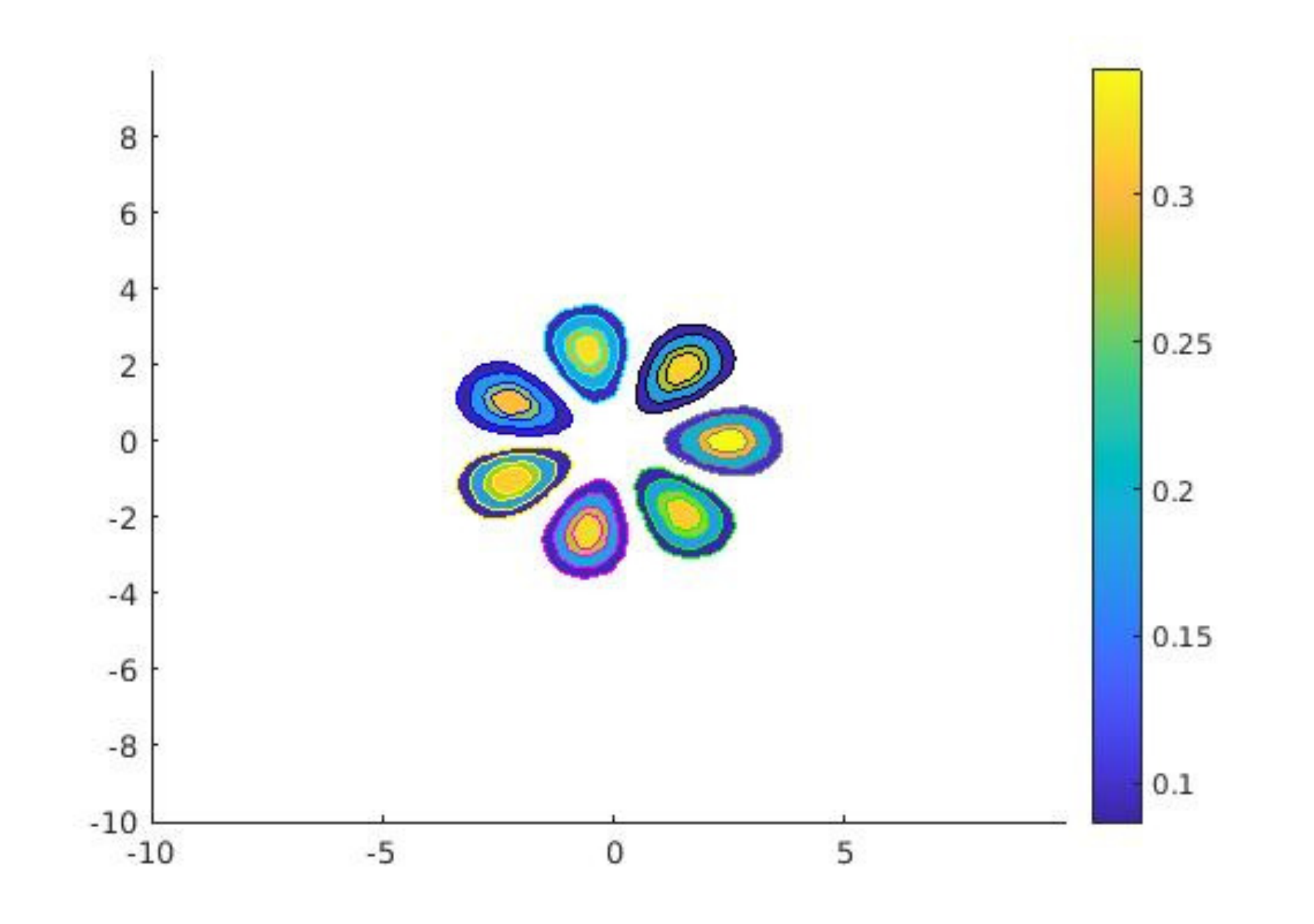}}
	\subfigure[$N=7$,$\eta=3$]{\includegraphics[width = .17\linewidth]{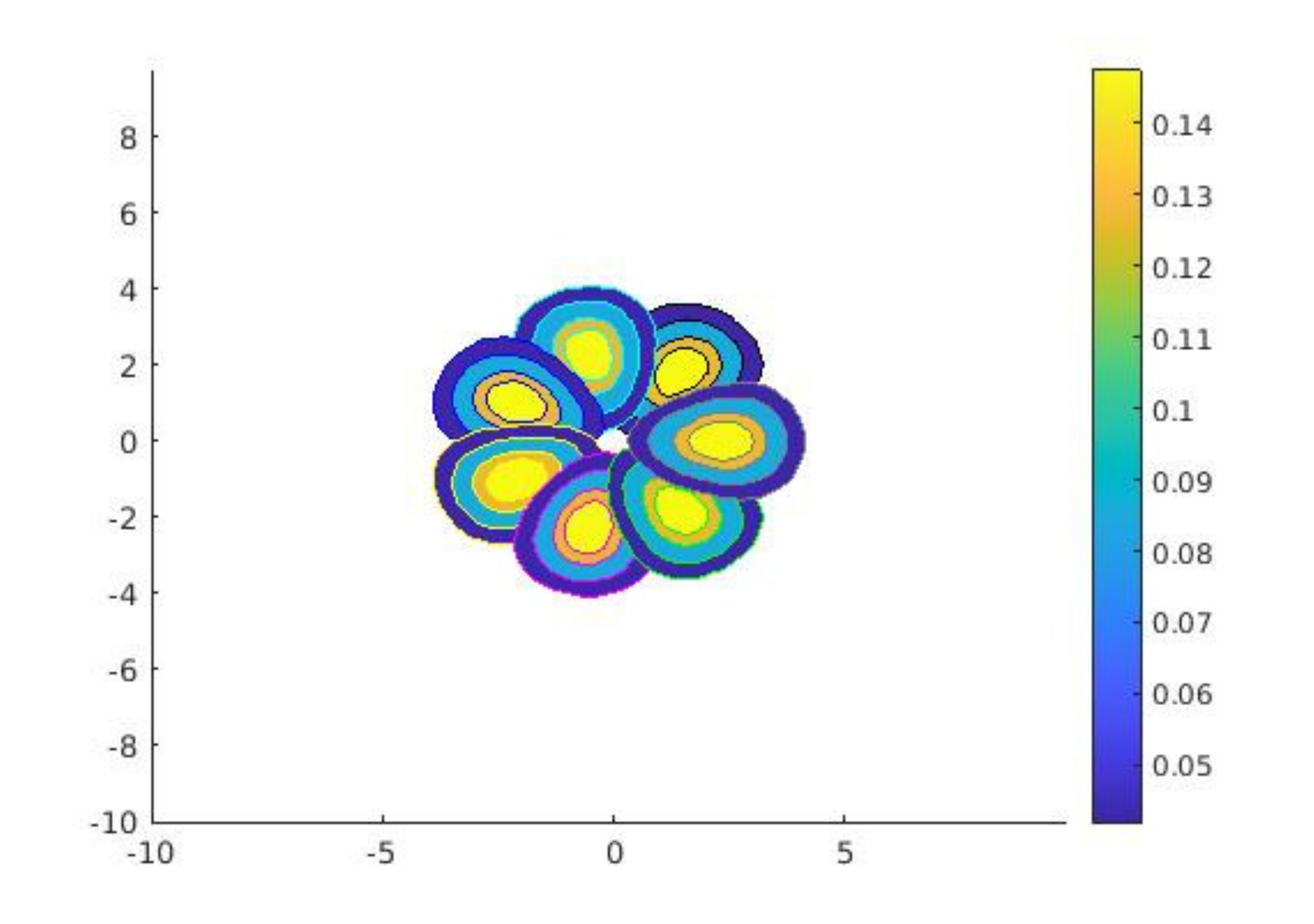}}
	\subfigure[$N=7$,$\eta=5$]{\includegraphics[width = .17\linewidth]{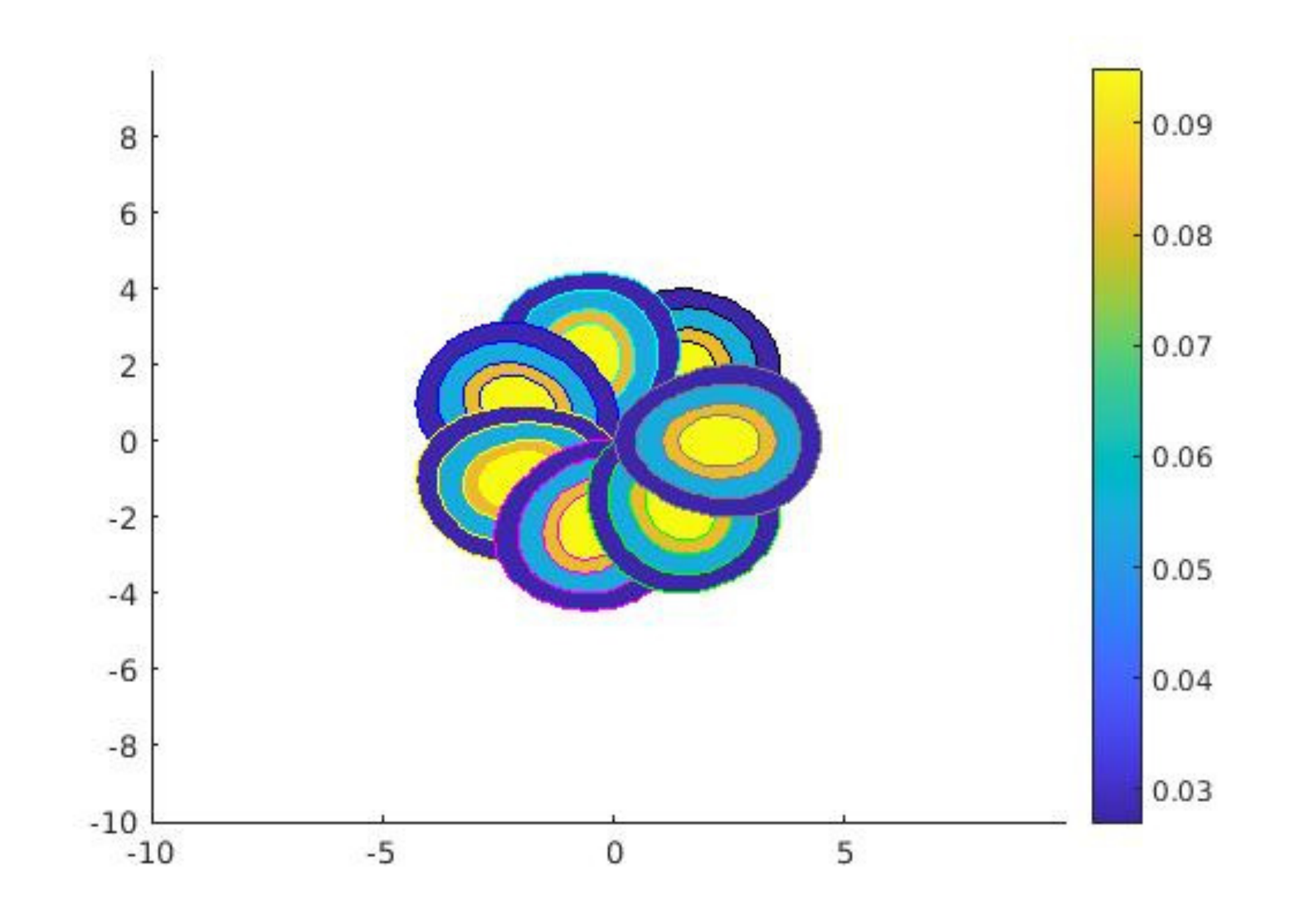}}
	\subfigure[$N=7$,$\eta=7$]{\includegraphics[width = .17\linewidth]{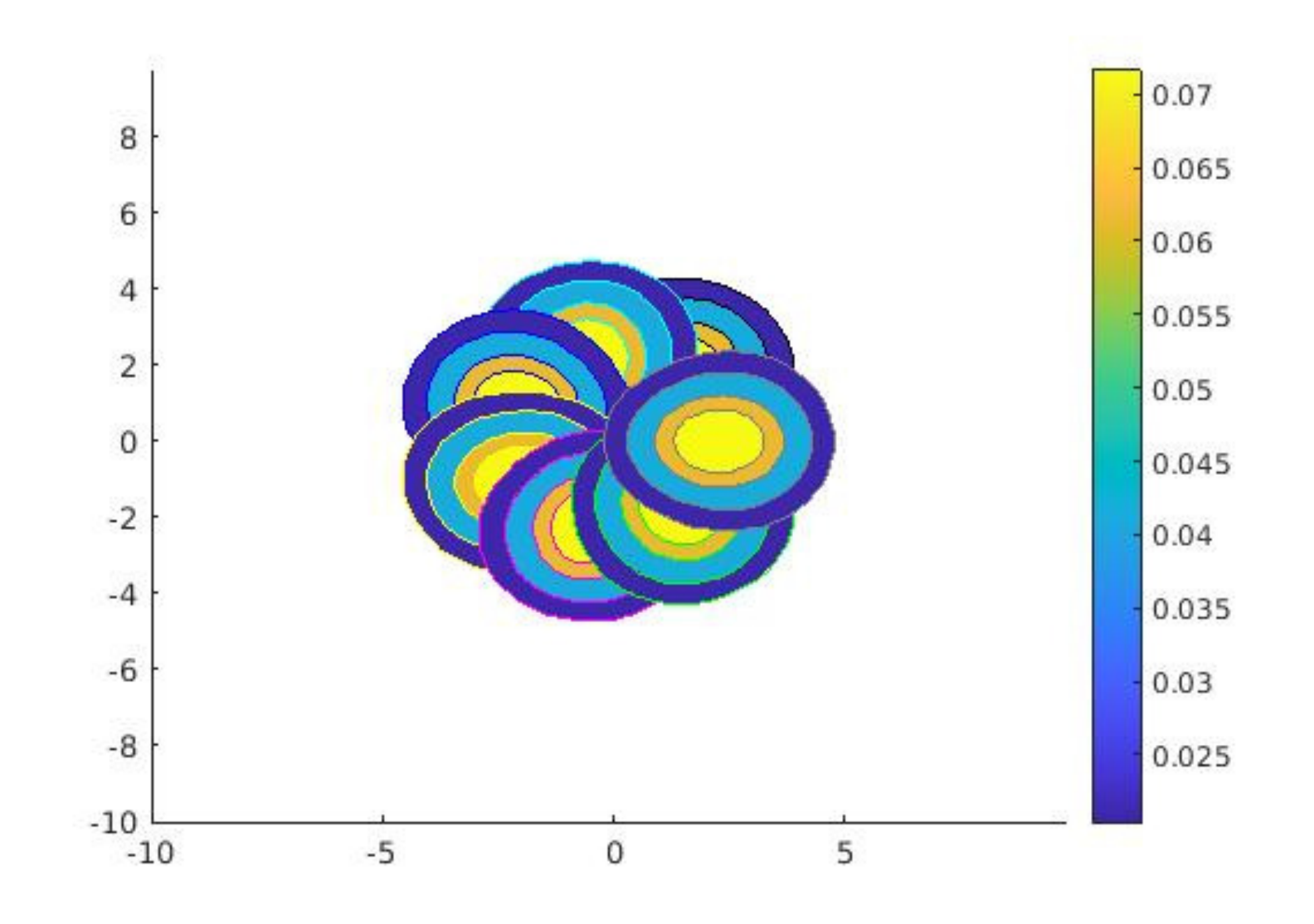}}
	\subfigure[$N=7$,$\eta=9$]{\includegraphics[width = .17\linewidth]{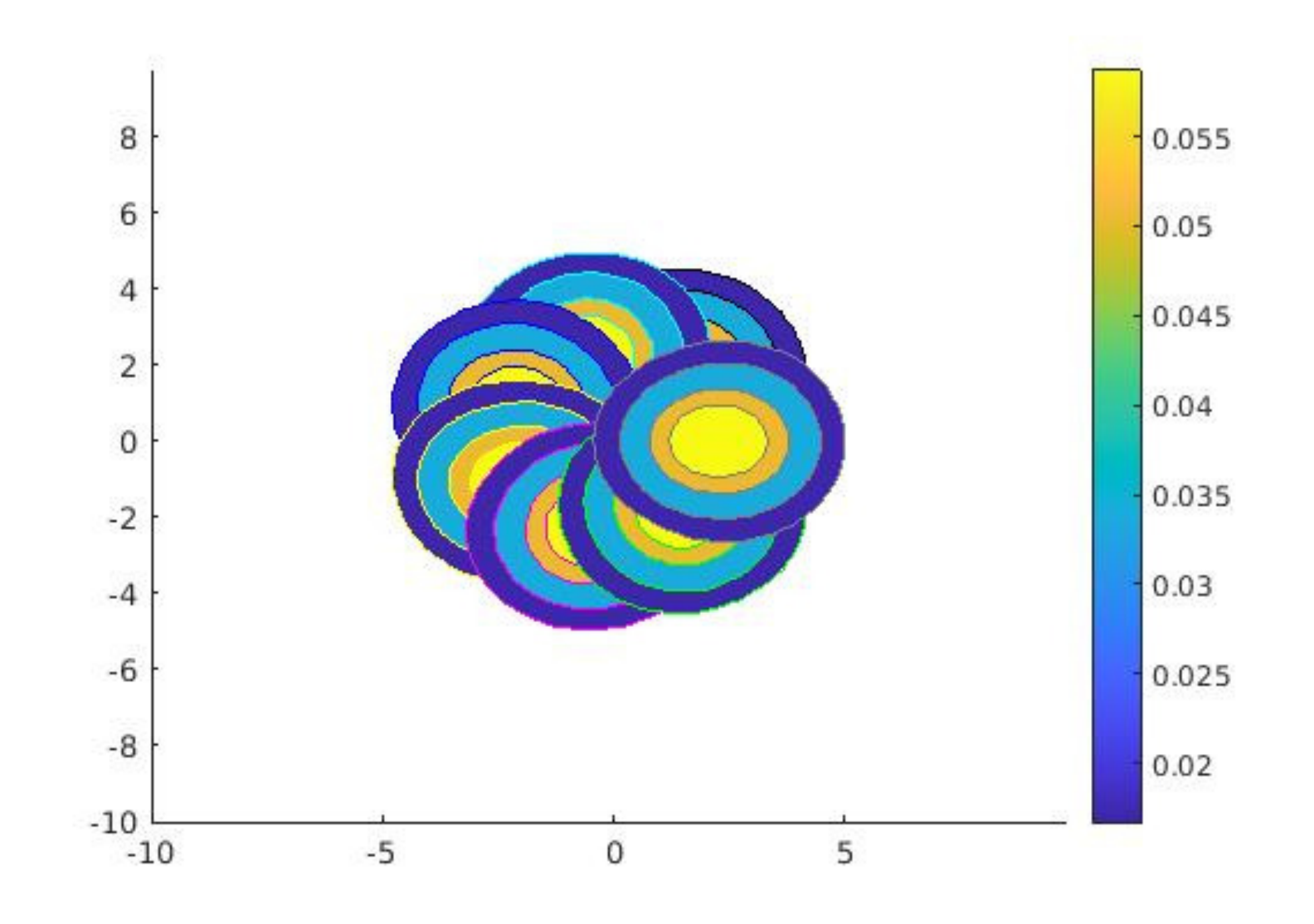}}\\
	\caption{Equilibrium solutions to system \eqref{eq:System} from the top right quadrant of Table \ref{T:Computation}, with $a=1$, $b=.5$, $K$ Laplace, varied $N$, and varied $\eta$  }
	\label{fig:Computation}
\end{figure}
As discussed in Section \ref{sec:environmentanddata}, we have location data for thirteen groups of meerkats. In order to incorporate data into system \eqref{eq:System}, we need to find equilibrium solutions for a large number of groups and a large range of parameter values. For these reasons, we find equilibrium solutions to the system with the number of groups ranging from one to nine and for a variety of parameter values. We pay particular attention to how the computation time scales as we increase the number of groups as well as how parameter regimes affect computation times.

In the simulations, we vary parameters $\eta$, $a$, and $b$, where $\eta$ is the strength of diffusivity, and $a$ and $b$ are constants multiplied by $U$ and $K$, respectively. We refer to these as the environment and interaction potential strength. 
%% what environmental potneitals do we use
 The environment potential we use in these trials is a Gaussian, $U(x,y) = Ae^{(-.1x^2-.1y^2)}$, where $A$ is a normalizing parameter so that $U$ has a maximum value of one. Therefore, the most attractive environment has a value of one, and a less attractive environment, or neutral environment, has a value of zero. 
%% What $K$ do we use 
 The interaction potential used is the Laplace potential, $K({\bf x})= Be^{-\sqrt{\frac{3}{2}}\vert\textbf{x}\vert}$, where $B$ is found so $K$ integrates to one over the domain.
 These simulations were done with the same initial condition for each trial, $u^{i,0} = e^{-(x- x^i_0)^2 - (y-y^i_0)^2}$. For $N=1$, $(x_0,y_0) = (0,0)$. For $N>1$, $(x_0^i,y_0^i)$ are spaced symmetrically about a circle centered at the origin with radius 2. The simulations were run until the error for each equation in system \eqref{eq:System}, described in Section \ref{sec:discretizationscheme}, was under a chosen tolerance, $tol = .02$.
 We are computing solutions over a 100x100 grid. This is due to the fact that this is the resolution of the data we intend to incorporate into our model.
 
 Table \ref{T:Computation} shows the time it takes to compute equilibrium solutions for different values of $\eta$, $a$, and $b$ for $N=1,3,5,7$ groups.
		It is clear that the computational expense increases as we increase the number of groups. However, it is worth noting that the computational expense is dramatically less than in \cite{Ellefsen2021}, where the time to compute an equilibrium solution in two dimensions with $N=1$ was on the order of hours. It is also the case that some regimes of $\eta$ and $b$ are more efficient than others for higher values of $N$. 
		%As previously discussed, there is interplay between $\eta$ and $b$. 
		When $b$ is larger and there is a stronger sense of aggregation within groups, a higher $\eta$ balances this out. It can be seen in Table \ref{T:Computation} that lower $\eta$ values take longer to compute when $b=1$ than when $b=.5$. In regimes where $\eta$ and $b$ are well-balanced, it seems the expense of adding groups is constant, while in the regimes where $\eta$ is small relative to $b$, it is more expensive to add groups. Thus, the balance between $\eta$ and $b$ is not only important in the behavior of solutions, but also in the computational expense to solve the system. Figure \ref{fig:Computation} shows equilibrium solutions when $b=.5$ and $a = 1$, the top right quadrant of Table \ref{T:Computation}. The first column, where groups are more aggregated, takes the longest to compute when compared to other simulations with the same number of groups, while the second and third columns are the fastest, but similar to the remaining four.

%%%
\subsection{The Effect of Potentials and Parameters on Solutions}\label{sec:exp}
In this section, we investigate how parameters and interaction potentials affect territory formation.  This allows us to determine how changing the shape of the interaction potential and the environment potential changes the behavior of solutions, which can help inform how one may choose potentials for specific situations.  We solve for the equilibrium solutions using the methods described in Section \ref{sec:discretizationscheme} while varying the parameters and potentials used. 

%We vary the shape of both the environment potential and the interaction potential.
We consider both a Laplace potential for the interaction potential as described in Section \ref{sec:computationtime}, as well as a Gaussian potential, $K(x,y) = Be^{\left(-\frac{x^2}{2\sigma^2}-\frac{y^2}{2\sigma^2}\right)}$, where we change the variance, $\sigma^2$, and $B$ is chosen so that $K$ integrates to one over the domain. We consider both a Gaussian potential for the environment as described in Section \ref{sec:computationtime} as well as a Laplace potential, $U({\bf x})= Ae^{-\sqrt{\frac{3}{2}}\vert\textbf{x}\vert}$, where $A$ is a normalizing parameter found so the maximum of $U$ over the domain is one. Therefore, environment is on a scale of one (most attractive) to zero (least attractive). As in the previous simulations, $a$ and $b$ are parameters multiplied by $U$ and $K$, respectively, affecting the strength of the environment and the interaction potential. In the plots of our simulations, we show two perspectives of each equilibrium solution in a column: a contour plot with contour lines drawn at 1/4, 1/2, 3/4, and 7/8 of the maximum population density, and a three dimensional plot of the equilibrium solution with each group represented by different color and the environment represented in black.

\subsubsection{Equilibrium solutions for a single group}\label{sec:onegroup}

\begin{figure}[h!]
	\centering
	\subfigure[$u(\textbf{x},\infty)$, $\eta = 3$]{\includegraphics[width=.27\linewidth]{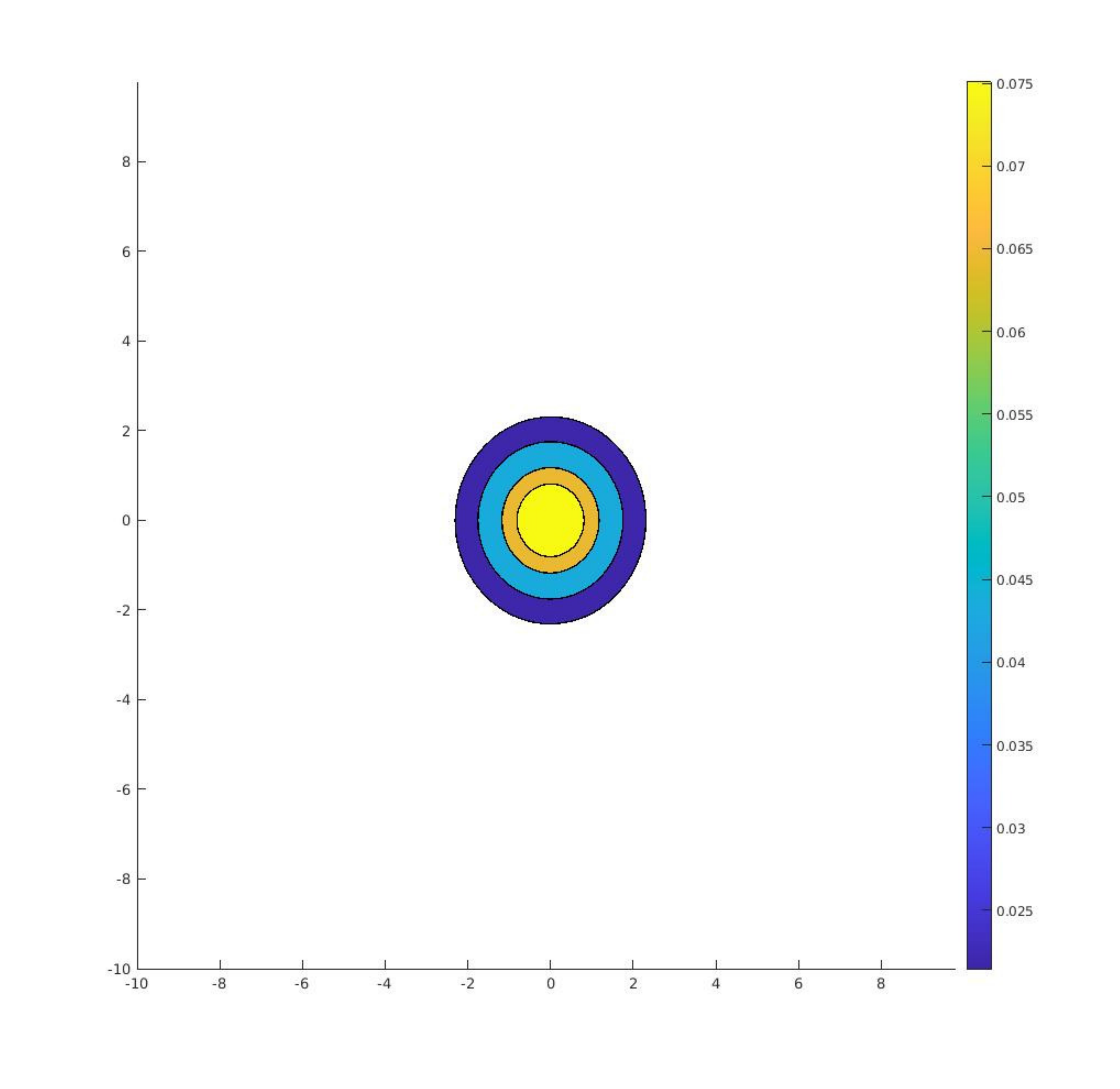} \label{fig:eta3} } 
	\subfigure[$u(\textbf{x},\infty)$, $\eta = 5$]{\includegraphics[width=.27\linewidth]{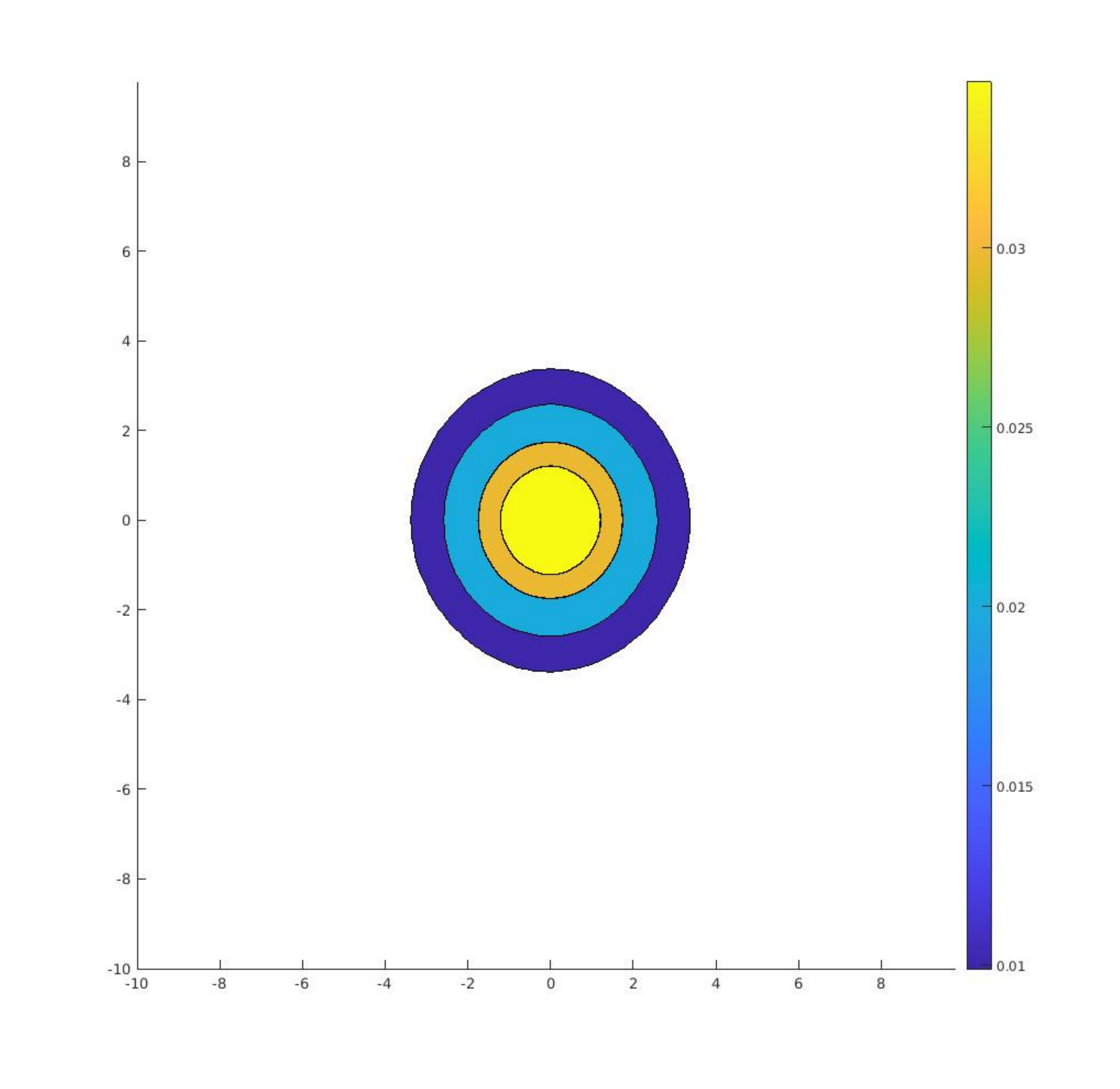}\label{fig:eta5} } 
	\subfigure[$u(\textbf{x},\infty)$, $\eta = 7$]{\includegraphics[width=.27\linewidth]{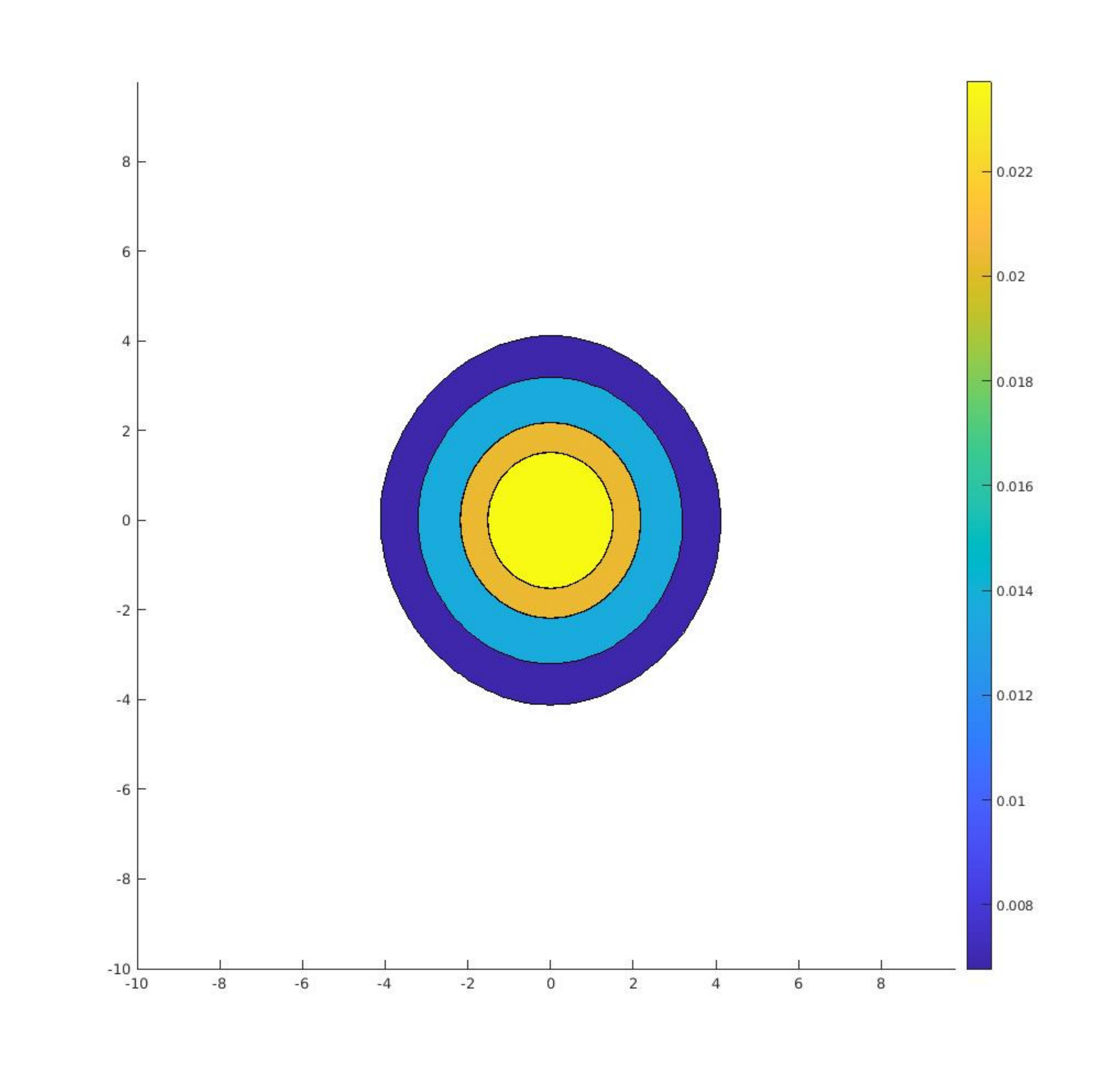}\label{fig:eta7}}\\
	\subfigure[$u(\textbf{x},\infty)$, $U(\textbf{x})$, $\eta = 3$]{\includegraphics[width=.27\linewidth]{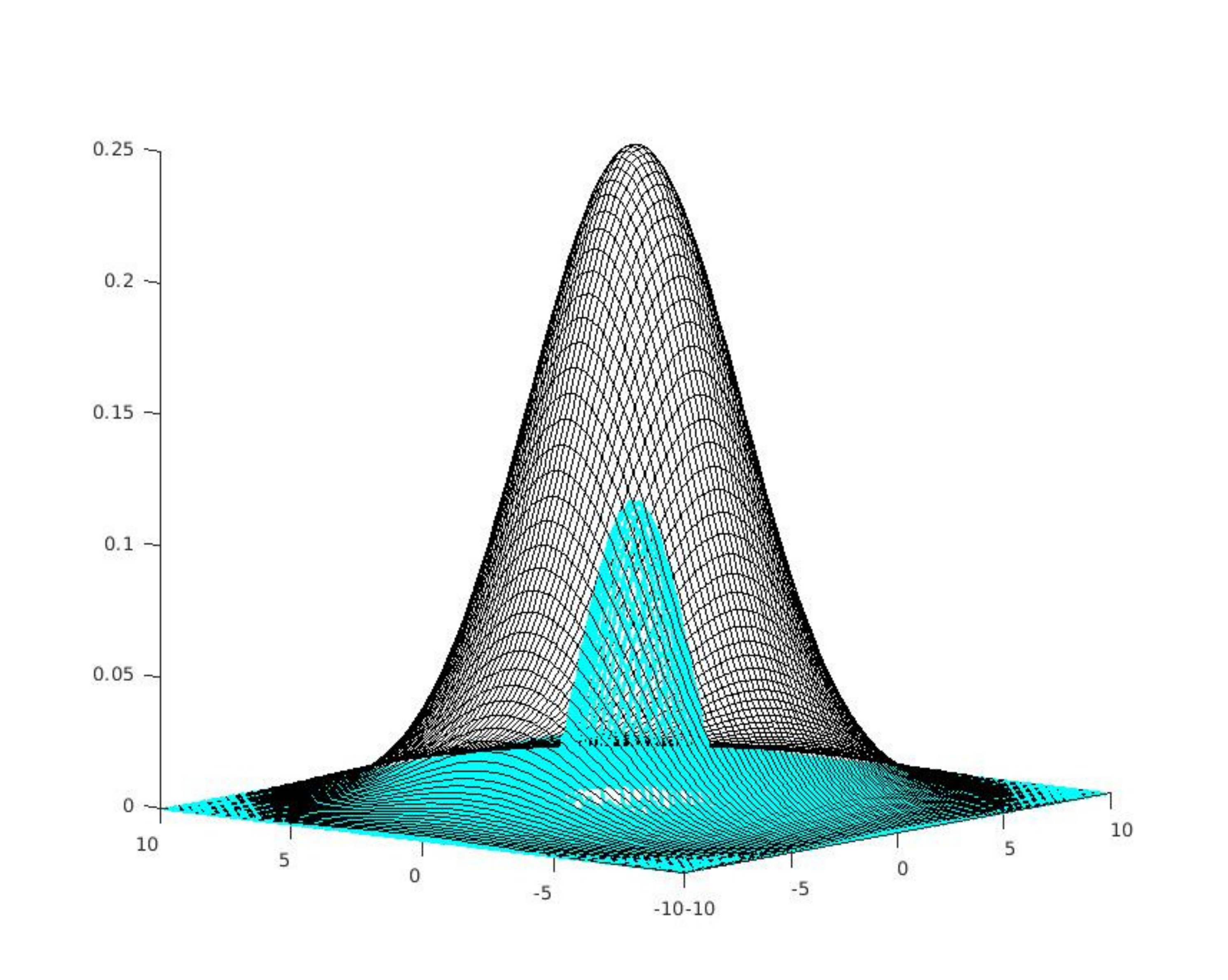} \label{fig:eta33D} } 
	\subfigure[$u(\textbf{x},\infty)$, $U(\textbf{x})$, $\eta = 5$]{\includegraphics[width=.27\linewidth]{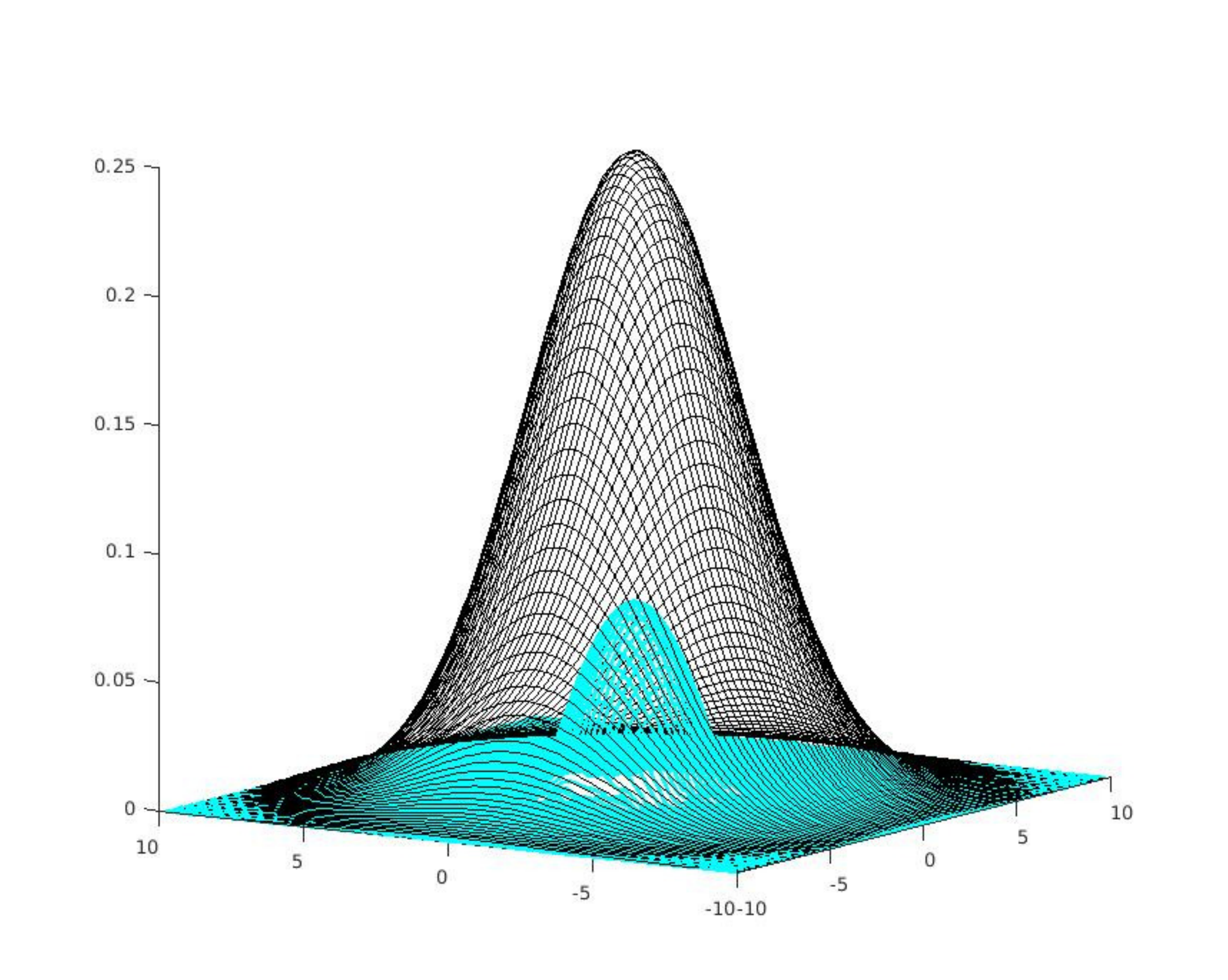}\label{fig:eta53D} } 
	\subfigure[$u(\textbf{x},\infty)$, $U(\textbf{x})$, $\eta = 7$]{\includegraphics[width=.27\linewidth]{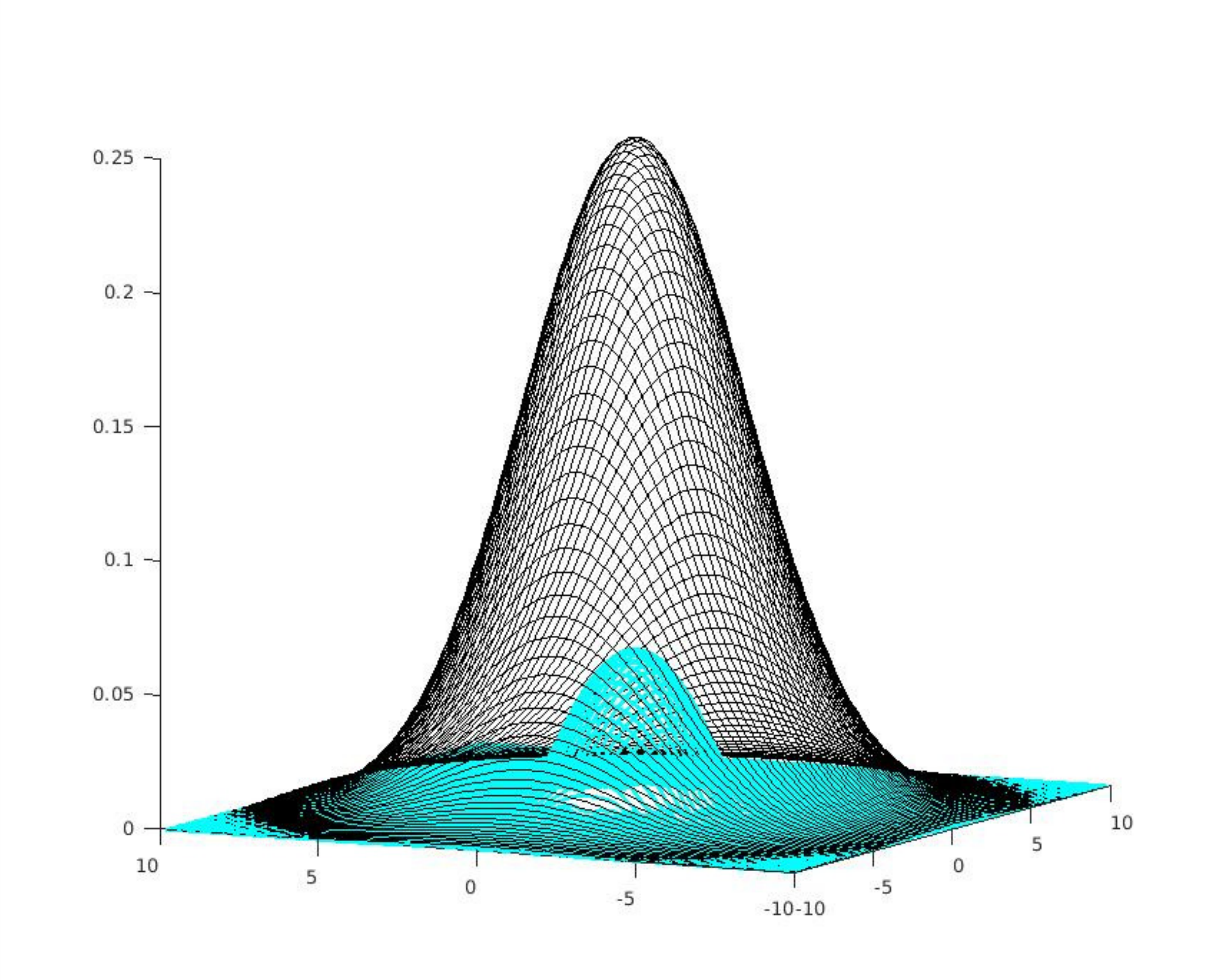}\label{fig:eta73D}}
	\caption{Equilibrium solutions to \eqref{eq:System} for $a = .25$, $b=.25$, $K$ Laplace, and varied $\eta$, shown as a contour plot (top), and a three-dimensional plot with $U(\textbf{x})$ in black (bottom)}
	\label{fig:etachange}
\end{figure}

We begin with a single group, $N=1$, in order to isolate the effect of changing potentials and parameters on a group before including inter-group interactions.
%, which allows us to investigate how the behavior of the solutions changes as our potentials and parameters change. 
As discussed, the parameter $\eta$  affects the strength of diffusivity within the group, which competes with the force of the nonlocal term, $K*u$, which aggregates the group. It is the interplay between these two terms that determines the density and size of a group's territory. 
Figure \ref{fig:etachange} illustrates an example of how the equilibrium solution changes when the diffusion strength, $\eta$, is varied. Both the environment strength and the interaction potential strength remain constant.
% Each row of the figure illustrates results from the same simulation but from different perspectives; the top row shows a contour plot of the density $u$ with the varied $\eta$ values, and the bottom row shows the three-dimensional plot of $u$ and the environment $U$ with the varied $\eta$ values. 
%The difference in population as $\eta$ increases is apparent in Figure \ref{fig:eta3}, \ref{fig:eta5}, and \ref{fig:eta7}. 
As we would expect, the territory size increases with $\eta$; the territory in Figure \ref{fig:eta7}, when $\eta = 7$, is larger than in Figure \ref{fig:eta3} and \ref{fig:eta5}, when $\eta = 3$ and $\eta=5$, respectively. The population density decreases with $\eta$; the population density is larger in \ref{fig:eta33D}, when $\eta= 3$, than in Figure \ref{fig:eta53D} and Figure \ref{fig:eta73D}, when $\eta = 5$ and $\eta = 7$. 
%The population density is more apparent in Figure \ref{fig:eta33D}, \ref{fig:eta53D}, and \ref{fig:eta73D}, as the population density is shown relative to constant $U$. As $\eta$ increases, the population spreads out to territory with a less attractive environment because the strength of diffusion is higher. 

%% SIMULATIONS WITH ETA

Next, the effect of the interaction potential is studied by changing both the strength of the interaction potential, $b$, and the shape of the interaction potential. We are looking to see how the aggregation within a group changes as we change $b$ and $K$. It is important to note that the repulsion between groups would also be affected when $N>1$, which we investigate in Section \ref{sec:multgroups}.

\begin{figure}[h!] 
	\centering
	\subfigure[$u(\textbf{x},\infty)$, $b = .25$]{\includegraphics[width=.27\linewidth]{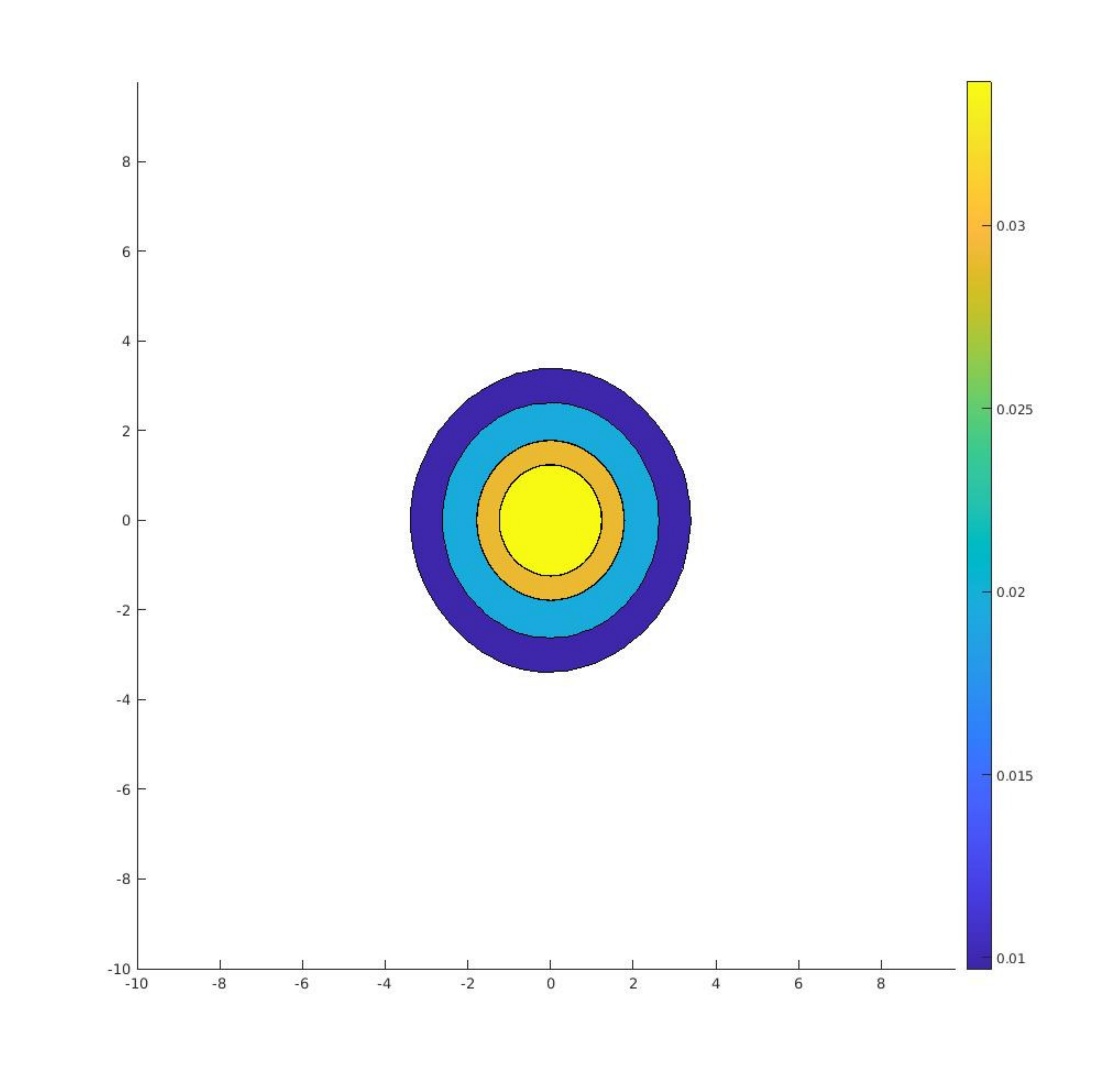}\label{fig:K25} } 
	\subfigure[$u(\textbf{x},\infty)$, $b = .5$]{\includegraphics[width=.27\linewidth]{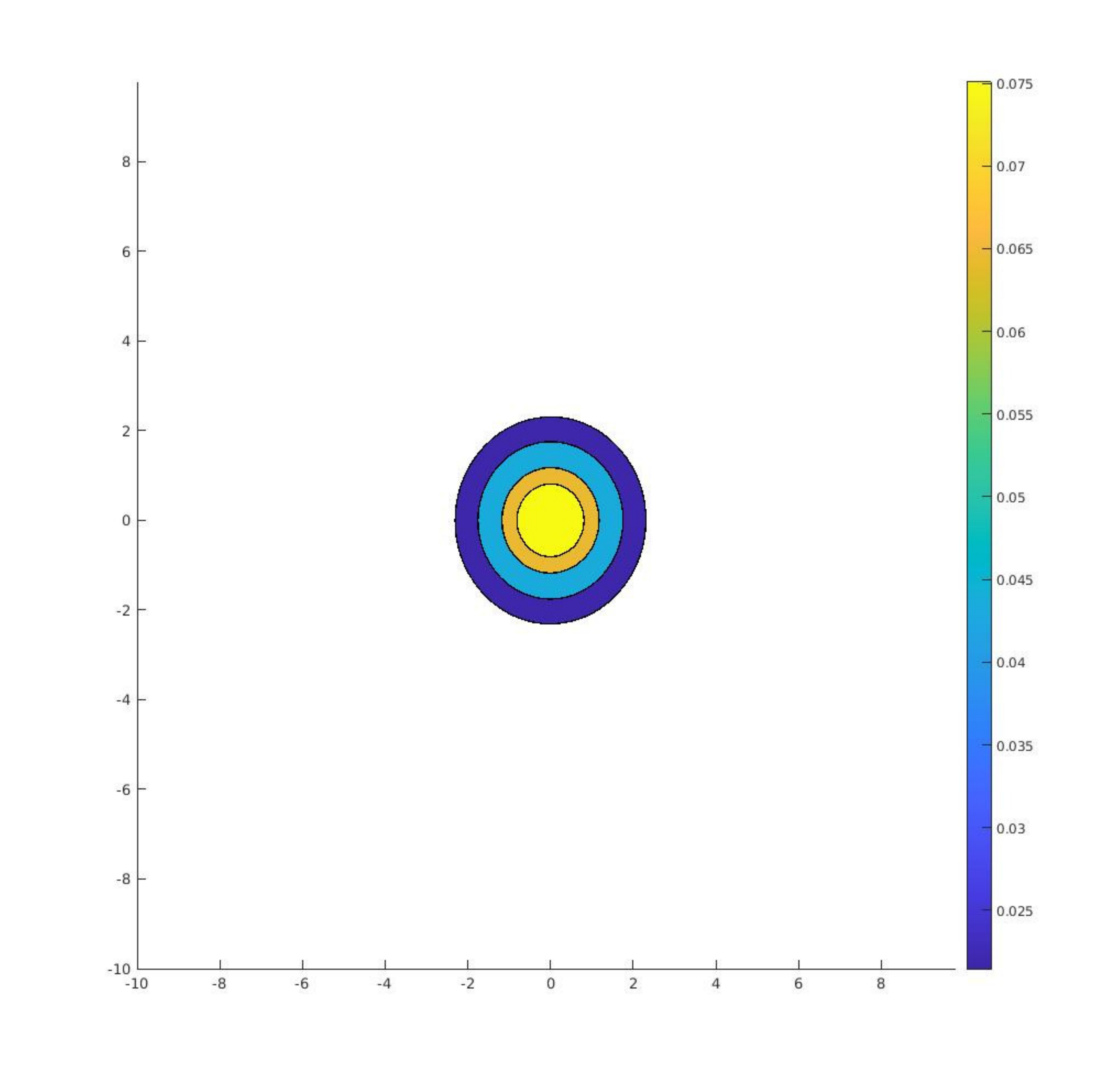}\label{fig:K5}}
	\subfigure[$u(\textbf{x},\infty)$, $b = 1$]{\includegraphics[width=.27\linewidth]{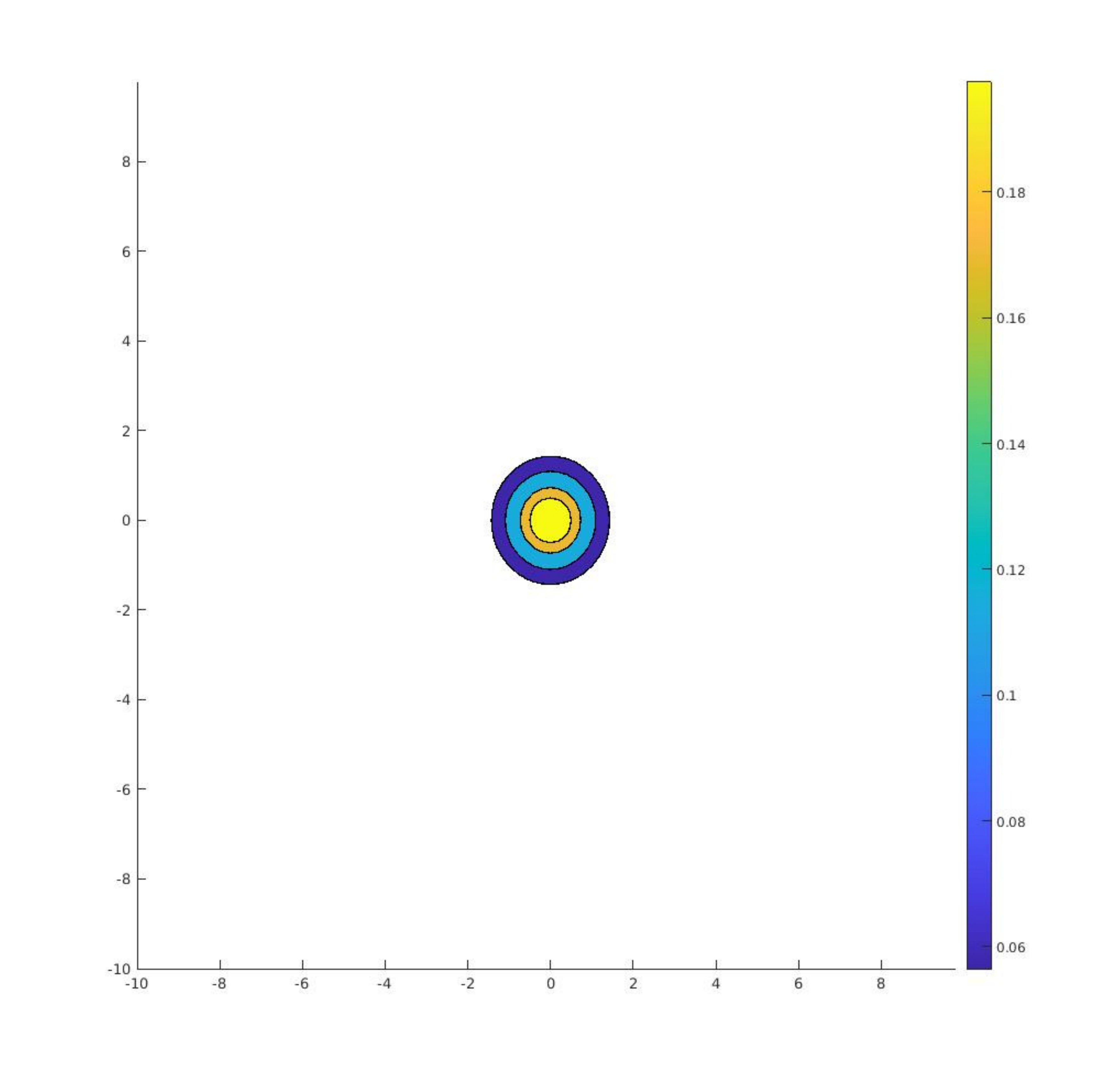}\label{fig:K1}}\\
	\subfigure[$u(\textbf{x},\infty)$, $U(\textbf{x})$, $b = .25$]{\includegraphics[width=.27\linewidth]{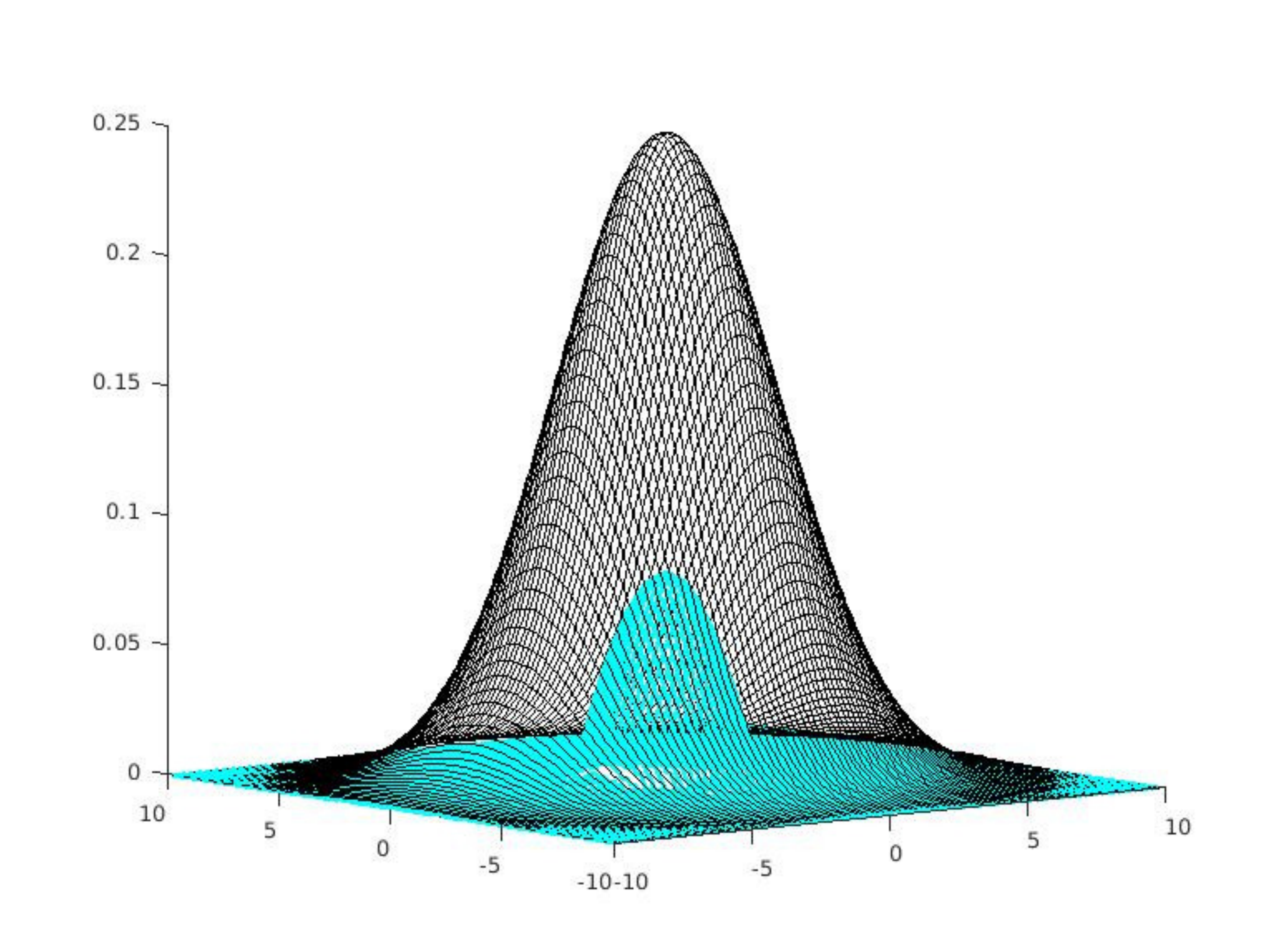}\label{fig:K253D} } 
	\subfigure[$u(\textbf{x},\infty)$, $U(\textbf{x})$, $b = .5$]{\includegraphics[width=.27\linewidth]{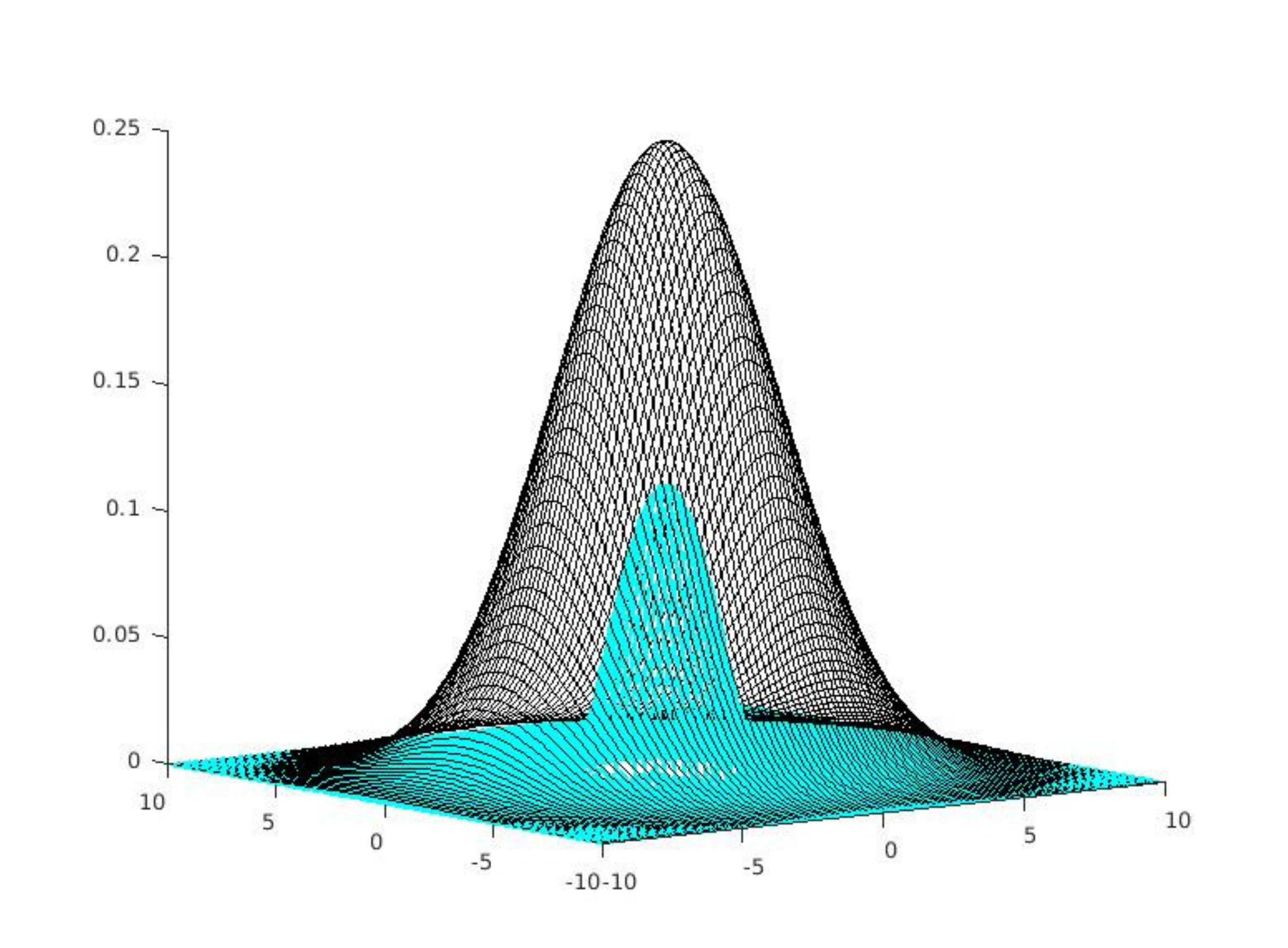}\label{fig:K53D}}
	\subfigure[$u(\textbf{x},\infty)$, $U(\textbf{x})$, $b = 1$]{\includegraphics[width=.27\linewidth]{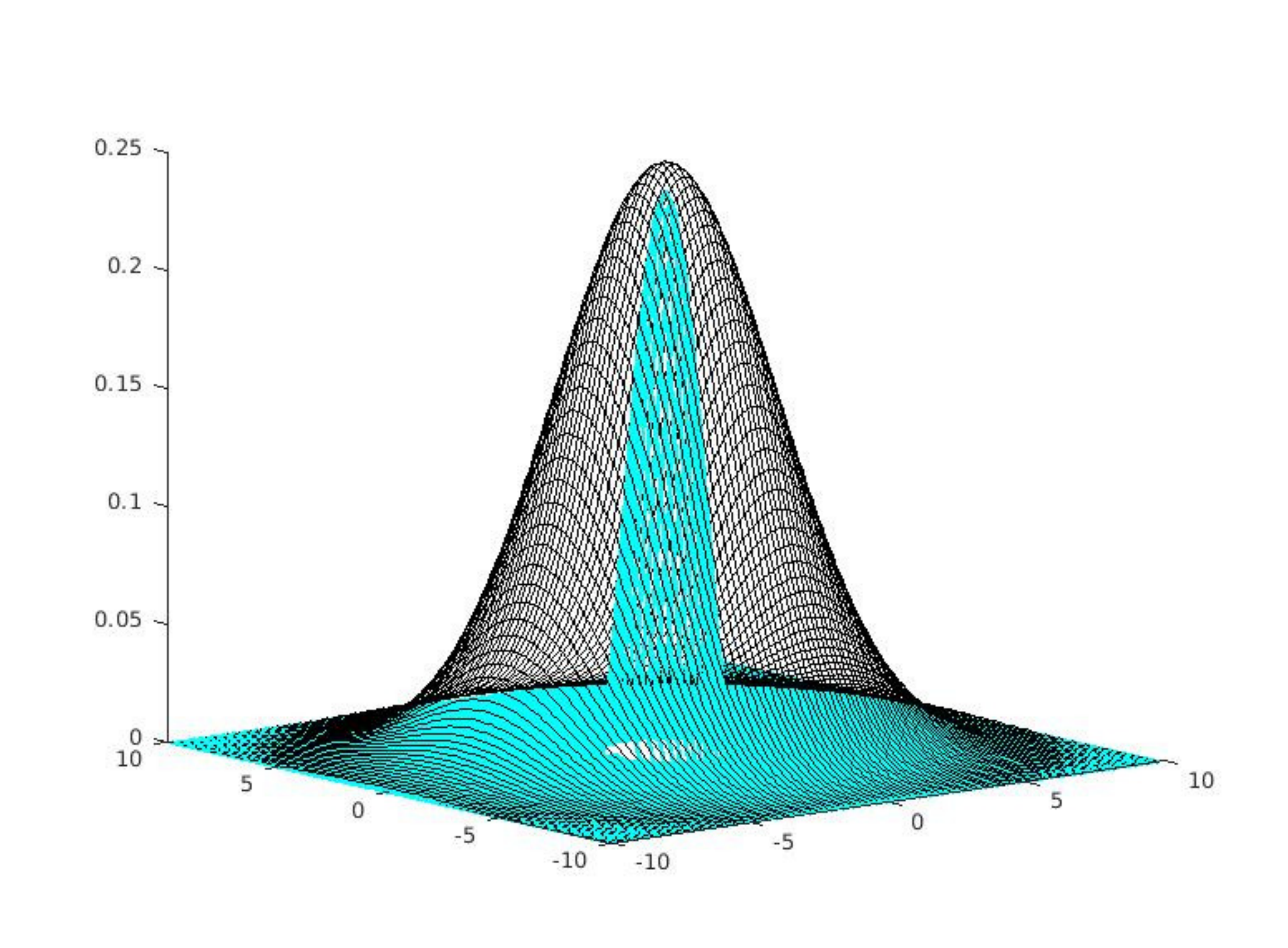}\label{fig:K13D}}	
	\caption{Equilibrium solutions to \eqref{eq:System} with $a=.25$, $\eta = 3$, $K$ Laplace, and varied $b$, shown as a contour plot (top), and a three-dimensional plot with $U(\textbf{x})$ in black (bottom)}
	\label{fig:Kchange}
\end{figure}

Figure \ref{fig:Kchange} illustrates how the strength of the interaction potential changes the behavior of the solution. In this simulation, $K$ is the Laplace potential, described above. 
% As with the previous case, the same simulation is shown with two different viewpoints: the top row is a contour plot of $u$ with the varied $b$ values while the bottom row is a three-dimensional plot of $u$ and $U$ with the varied $b$ values. 
As we would expect, when the strength of the interaction potential, $b$, increases, the aggregation strength within $u$ increases. Thus, size of the territory of $u$ decreases. The territory in Figure \ref{fig:K1}, when $b=1$, is smaller than in Figure \ref{fig:K5} and \ref{fig:K25}, when $b = .5$ and $b=.25$, respectively. Additionally, the population density of a group increases as $b$ increases; the population density in Figure \ref{fig:K13D}, when $b=1$, is larger than in Figure \ref{fig:K5} and \ref{fig:K25}, when $b = .5$ and $b=.25$, respectively.
%This behavior is demonstrated in Figure \ref{fig:K25}, \ref{fig:K5}, and \ref{fig:K1}; the territory in Figure \ref{fig:K1} is much smaller than in Figure \ref{fig:K25}. The increase in population density as $K$ increases is more apparent in Figure \ref{fig:K253D}, \ref{fig:K53D}, and \ref{fig:K13D}. When $b=1$, as in \ref{fig:K13D}, the population density is much higher than in Figure \ref{fig:K253D} when $b=.25$.
 The inverse relationship between $b$ and $\eta$ proves important in our simulations and in computation time, and it is evident when comparing Figure \ref{fig:etachange} and Figure \ref{fig:Kchange}. We see similar responses in the territories when we increase $b$ as when we decrease $\eta$, thus the relationship between these two parameters is important in determining population density and territory size.

%%SIMULATIONS WITH K

\begin{figure}[h!]
	\centering
	\subfigure[$u(\textbf{x},\infty)$]{\includegraphics[width=.27\linewidth]{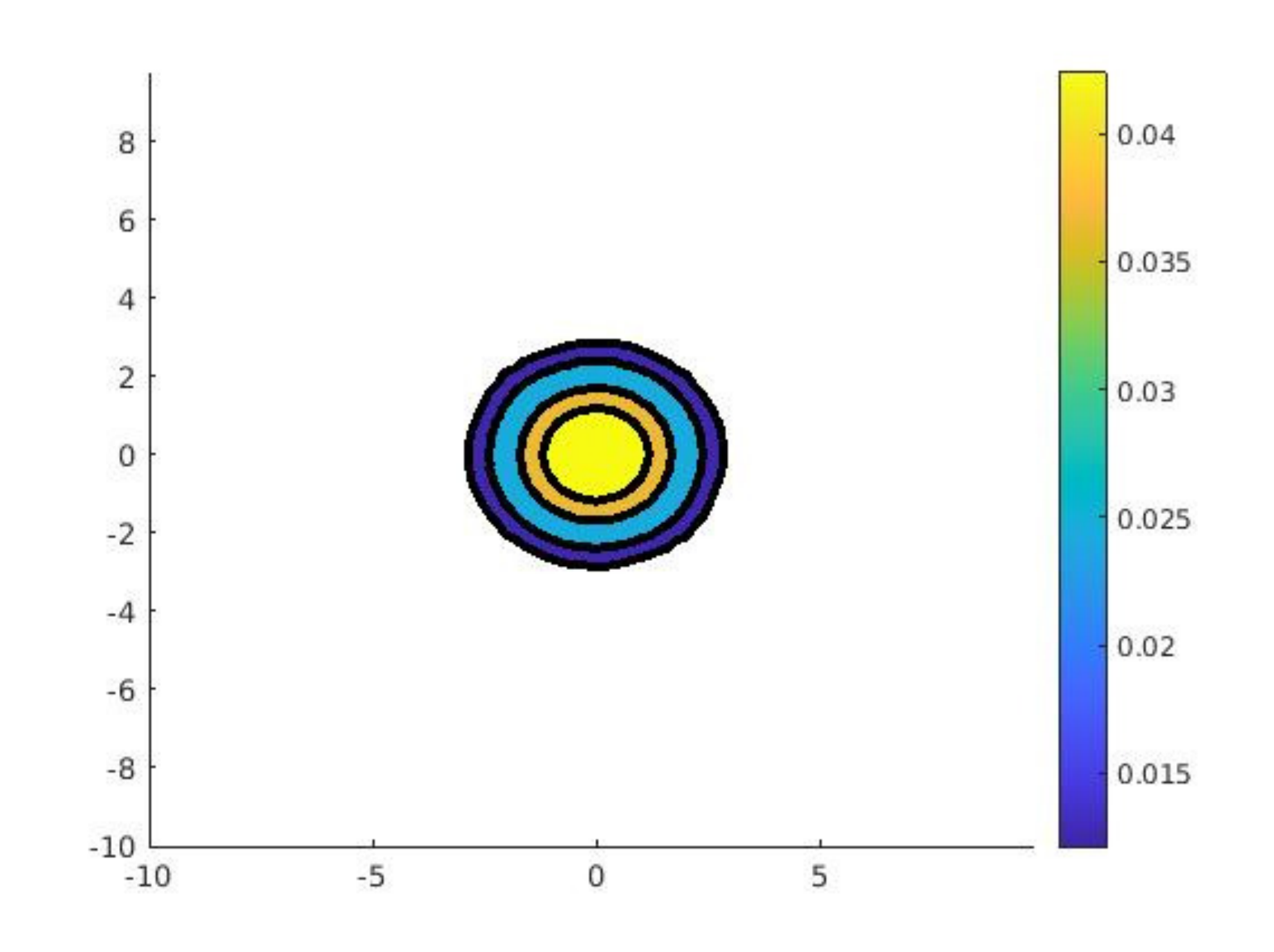}\label{fig:KG1v2}}
	\subfigure[$u(\textbf{x},\infty)$]{\includegraphics[width=.27\linewidth]{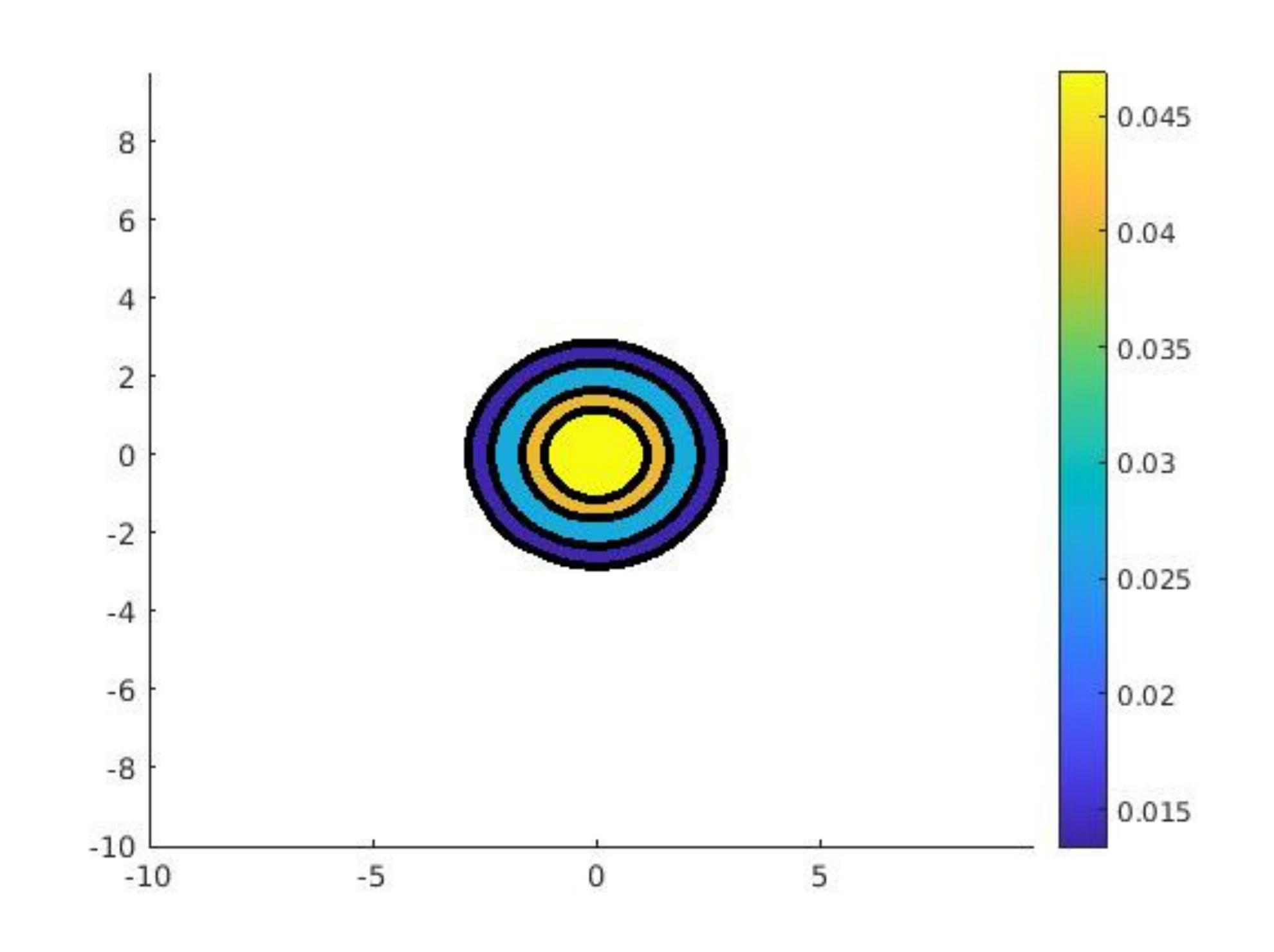}\label{fig:KG1v5}}
	\subfigure[$u(\textbf{x},\infty)$]{\includegraphics[width=.27\linewidth]{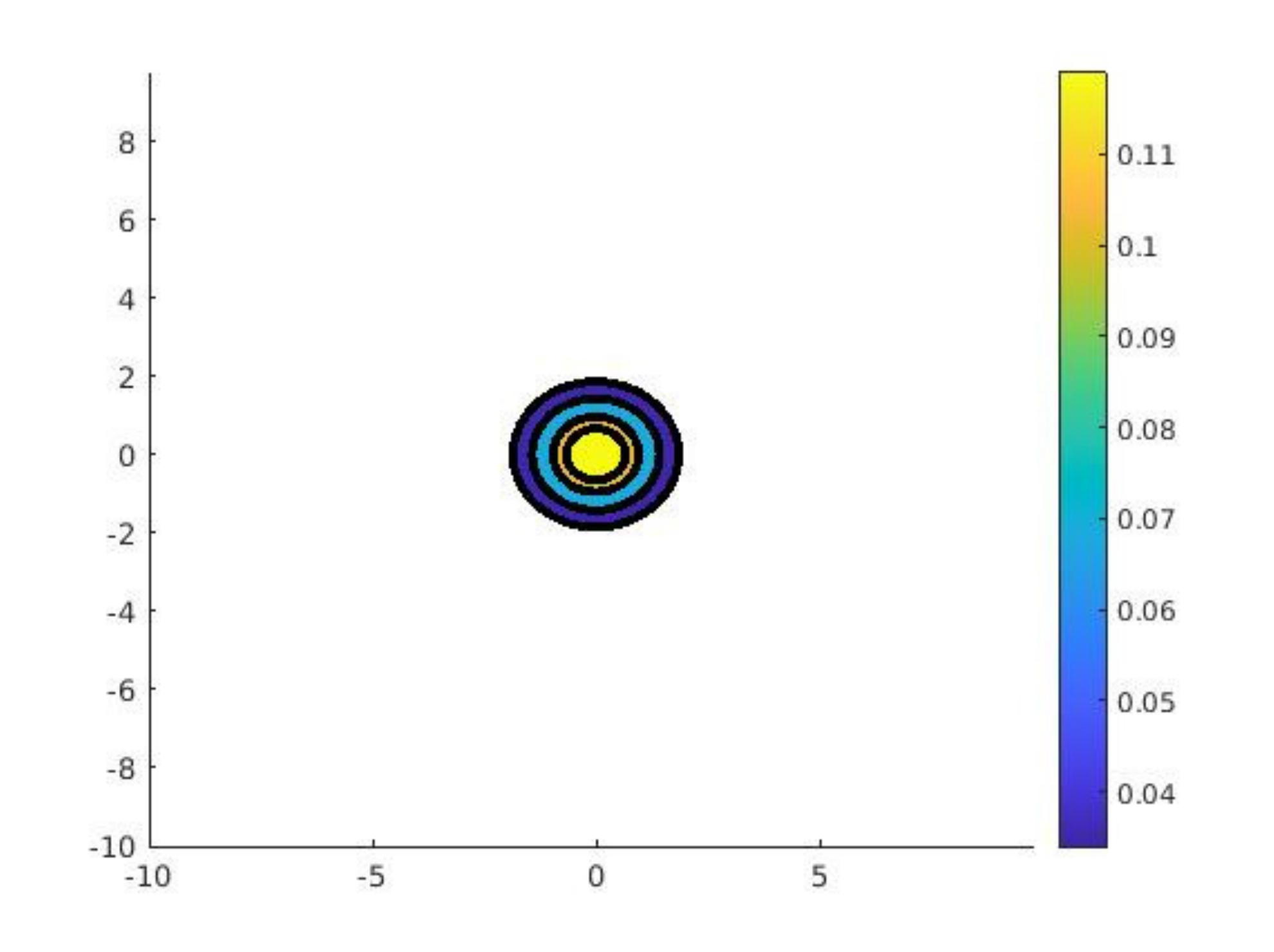}\label{fig:KL1}}\\
	\subfigure[$u(\textbf{x},\infty)$ and $U(\textbf{x})$]{\includegraphics[width=.27\linewidth]{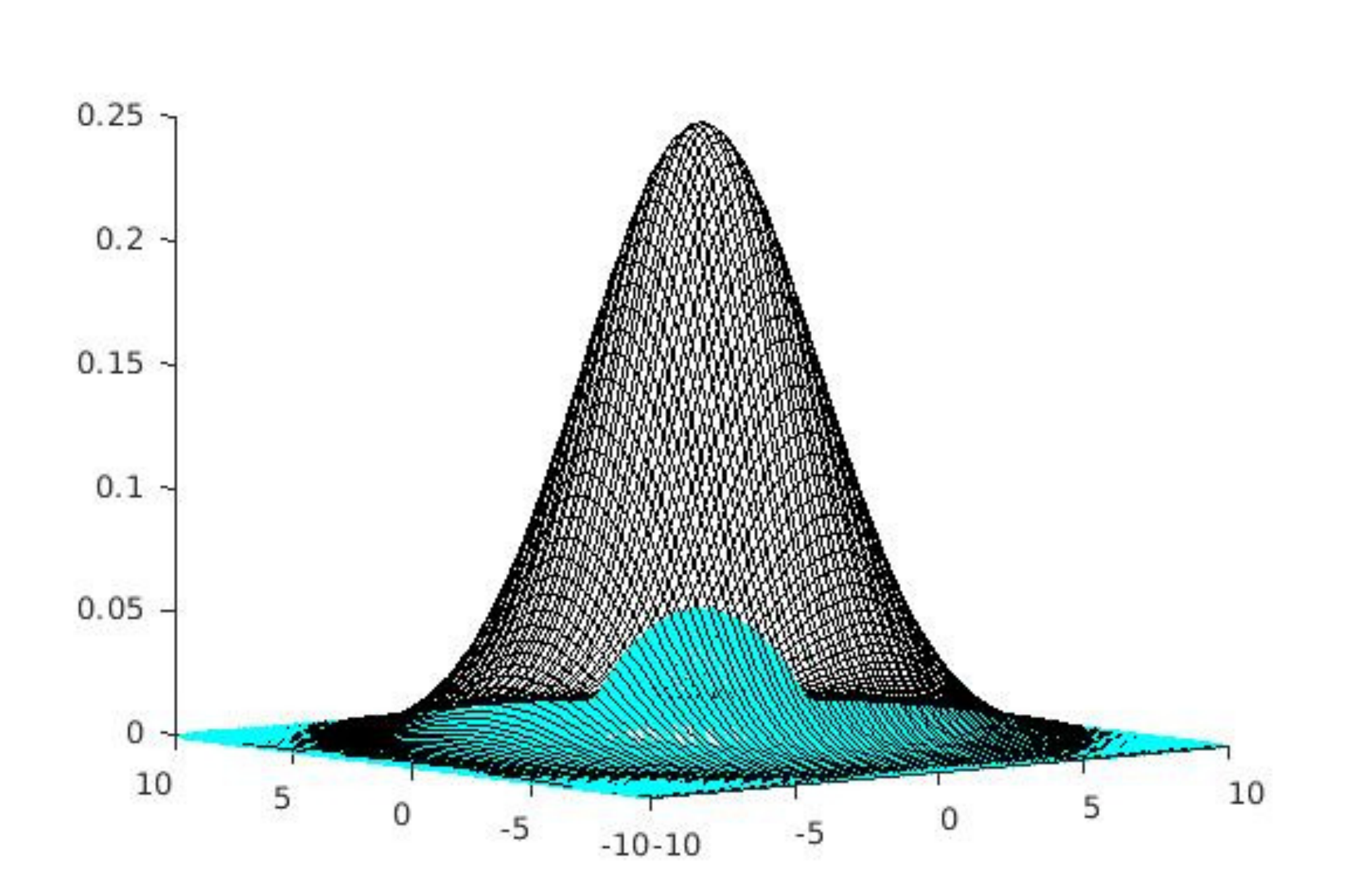}\label{fig:KG1v23D}}
	\subfigure[$u(\textbf{x},\infty)$ and $U(\textbf{x})$]{\includegraphics[width=.27\linewidth]{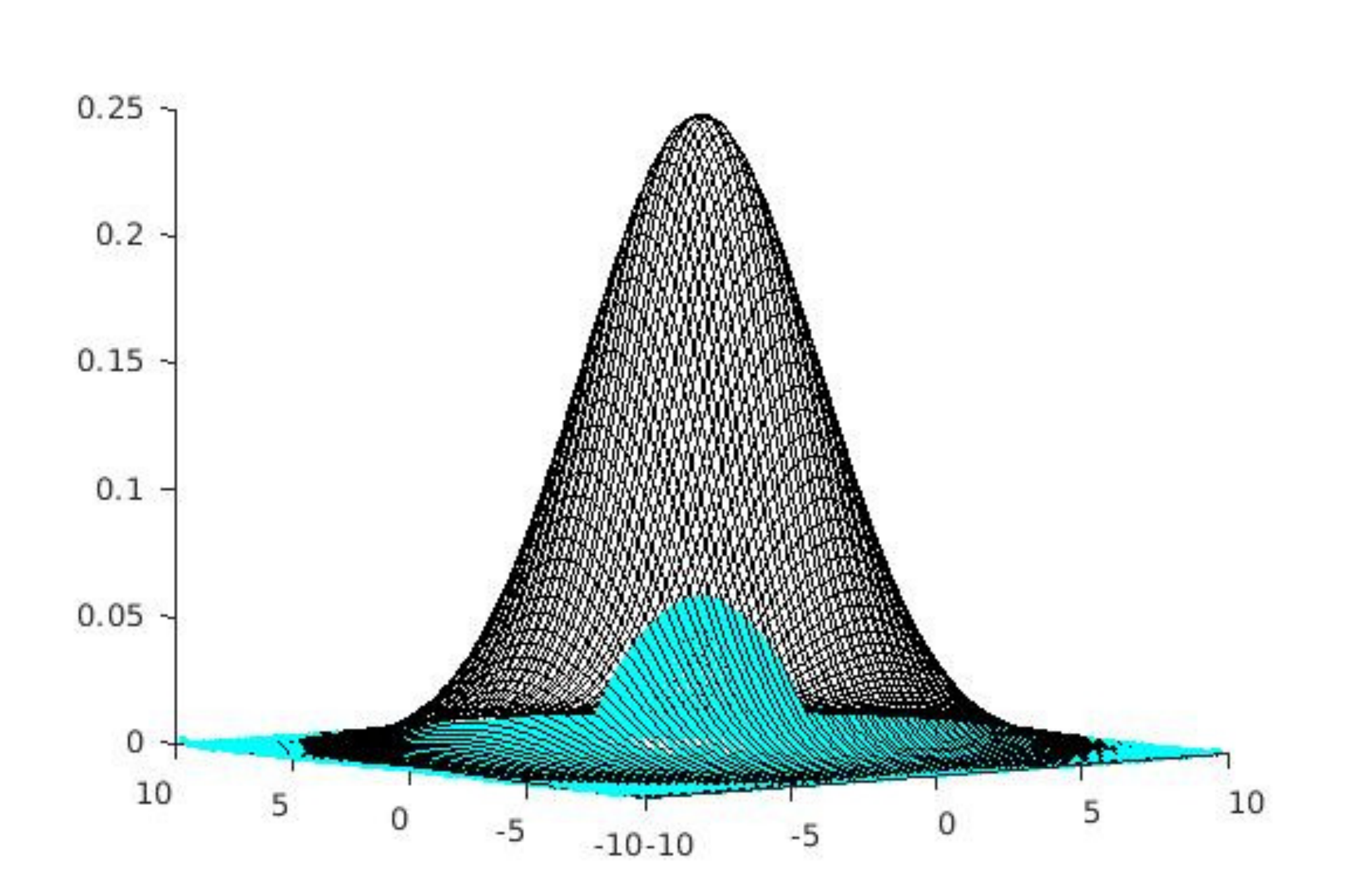}\label{fig:KG1v53D}}
	\subfigure[$u(\textbf{x},\infty)$ and $U(\textbf{x})$]{\includegraphics[width=.27\linewidth]{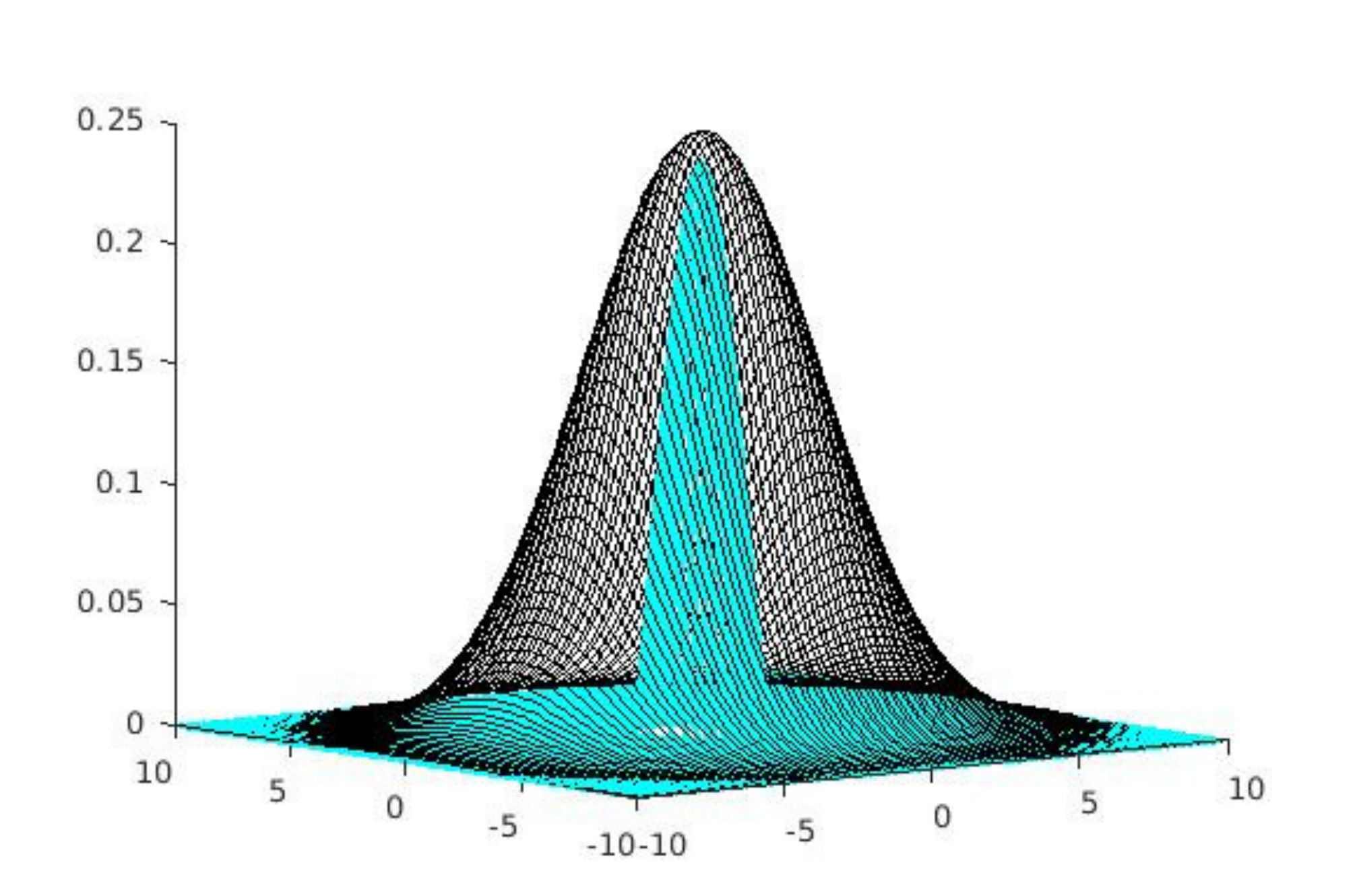}\label{fig:KL13D}}\\
	\subfigure[$\sigma =2$ ]{\includegraphics[width=.27\linewidth]{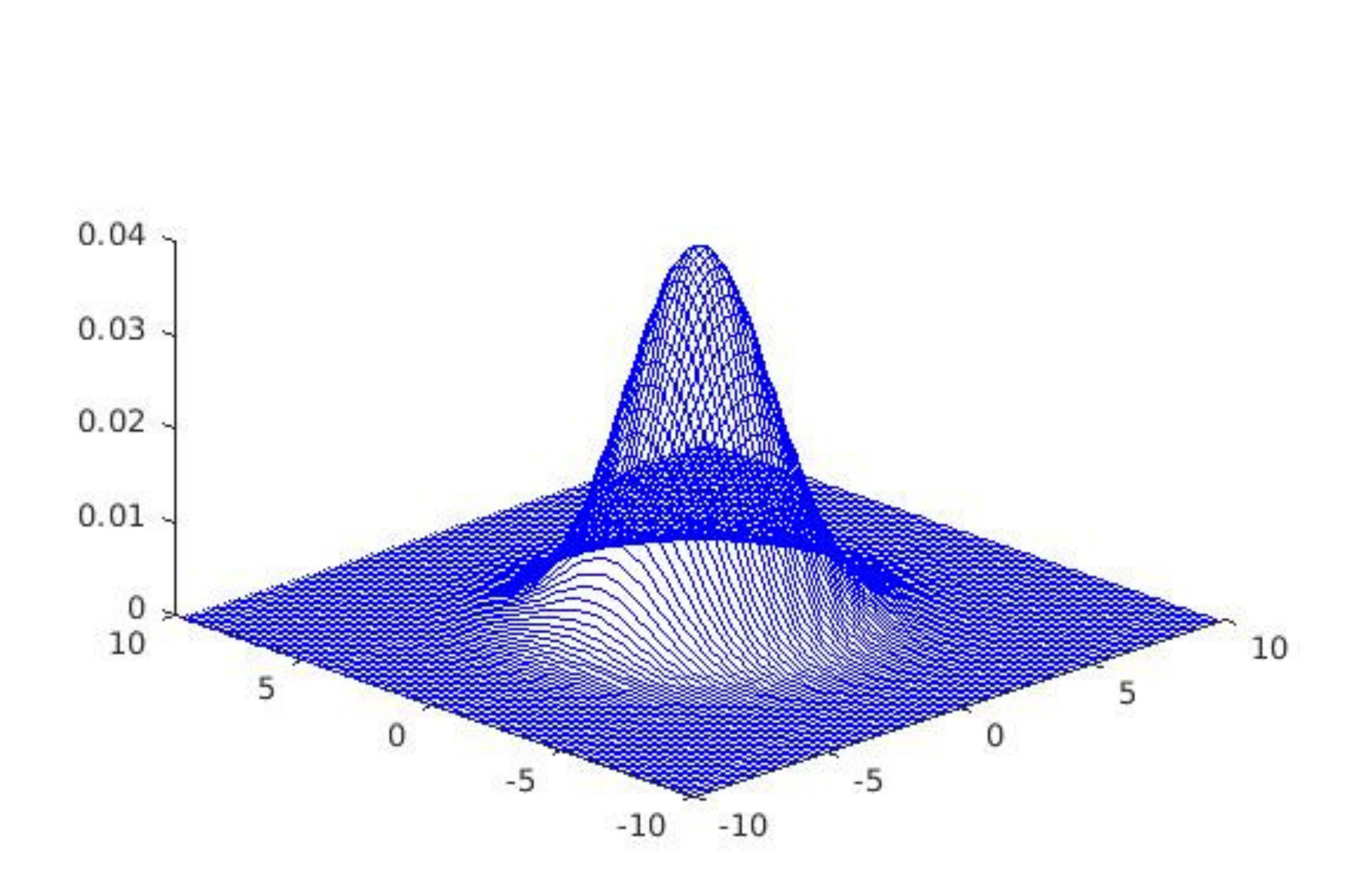}\label{fig:KGv2}}
	\subfigure[$\sigma = .5$]{\includegraphics[width=.27\linewidth]{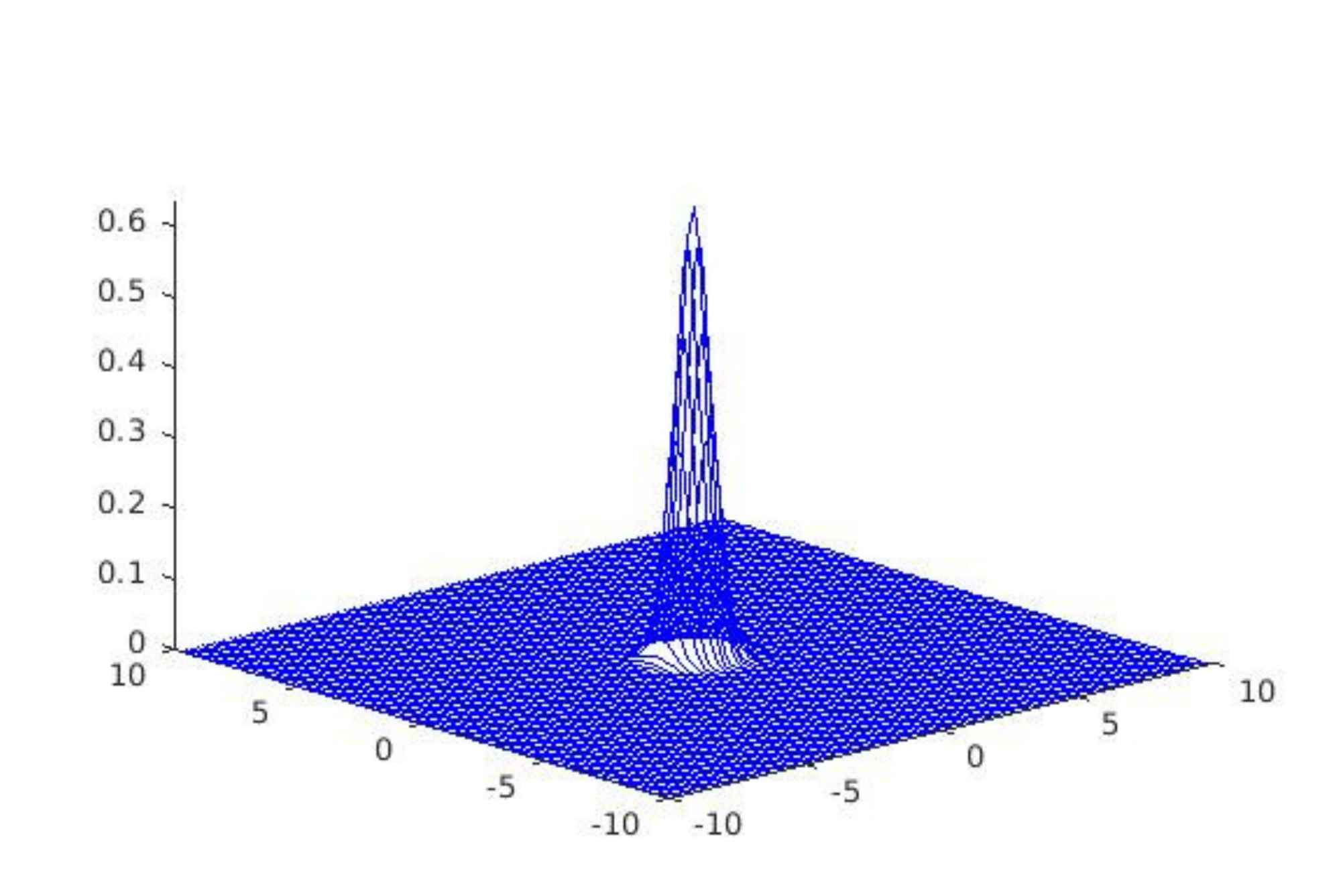}\label{fig:KGv5}}
	\subfigure[$K$ Laplace]{\includegraphics[width=.27\linewidth]{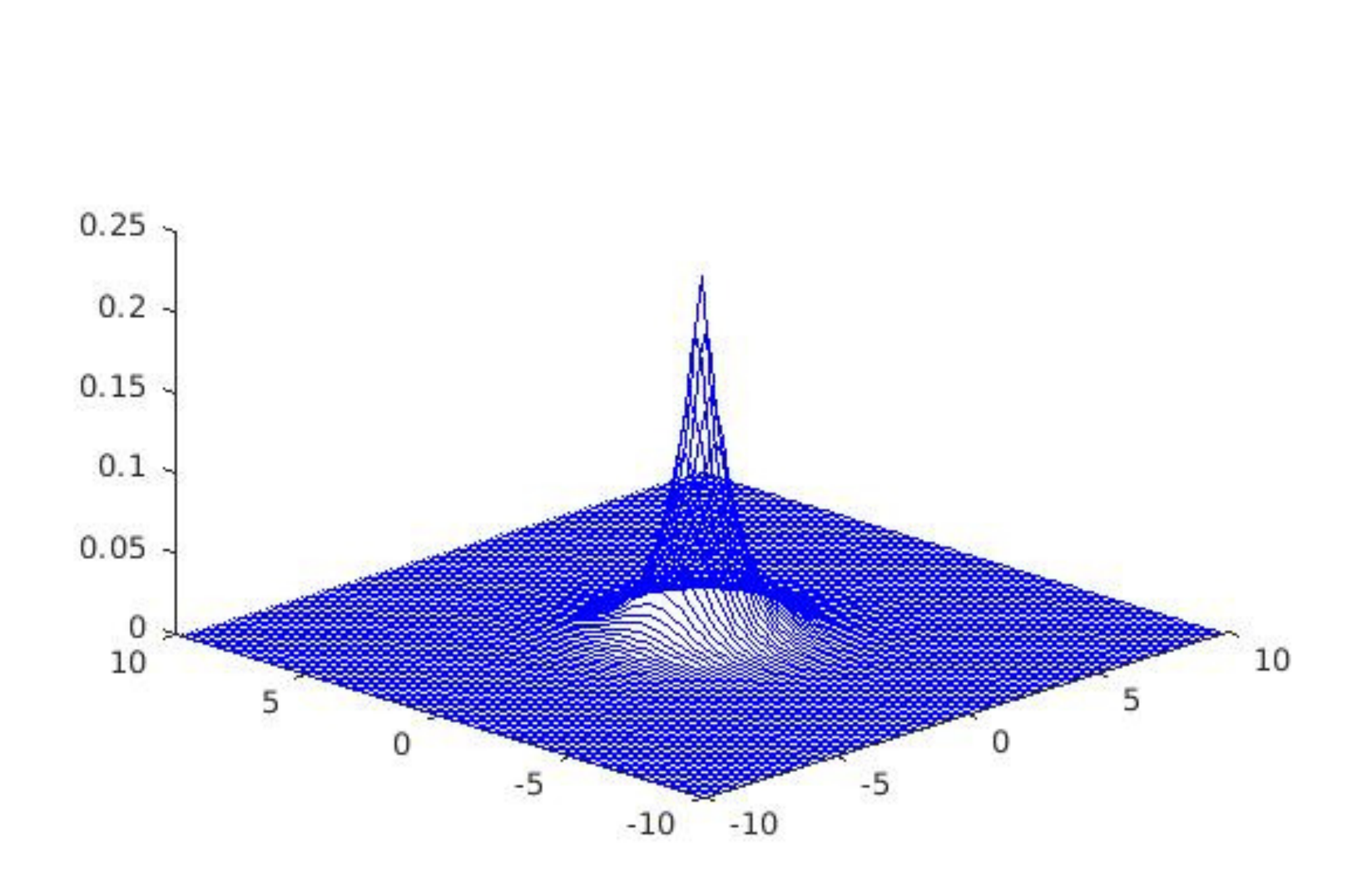}\label{fig:KL}}
	
	\caption{Equilibrium solutions to \eqref{eq:System} with $a = .25$, $\eta = 3$, $b = 1$, and varied $K$, shown as a contour plot (top), a three-dimensional plot with $U(\textbf{x})$ in black (middle), and the respective potentials $K(x,y)$ (bottom)}
	\label{fig:KShape}
\end{figure}

It is also important to consider how the shape of the interaction potential changes the behavior of the solution. We run simulations with $K$ equal to the Laplace potential and compare the behavior of the equilibrium solutions to equilibrium solutions with the interaction potential equal to a Gaussian.
Figure \ref{fig:KShape} compares the equilibrium solutions from the different interaction potentials. Each column represents a single simulation
 with the bottom row depicting a three-dimensional plot of the interaction potential used. 
Figure \ref{fig:KGv2} and Figure \ref{fig:KGv5} are Gaussian potentials with $\sigma = 2$ and $\sigma = .5$, respectively. Figure \ref{fig:KL} is the Laplace potential. 
%The first row of figures, Figure \ref{fig:KG1v2}, \ref{fig:KG1v5}, and \ref{fig:KL1}, are the contour plots of the equilibrium solutions with with the respective $K$. The second row of figures, Figure \ref{fig:KG1v23D}, \ref{fig:KG1v53D}, and \ref{fig:KL13D}, are the three-dimensional plots of the equilibrium solutions and $U$ from the same simulation.
 When comparing the two simulations with $K$ Gaussian, it is interesting to note that although $K$ is qualitatively quite different in Figure \ref{fig:KGv2} and \ref{fig:KGv5}, the predicted territory and population density are  quite similar, with the population density in Figure \ref{fig:KGv5} being slightly larger. Although $K$ in Figure \ref{fig:KGv5} and \ref{fig:KL} qualitatively look most similar, the difference in equilibrium solutions is most drastic when we shift $K$ from the Gaussian to the Laplace Potential. In Figure \ref{fig:KL1} and \ref{fig:KL13D}, the population density is much higher and the territory is smaller than in the simulations with $K$ Gaussian. The steep slope of the potential in these simulations has a bigger effect on the equilibrium solutions than the variance of $K$.

\begin{figure}[h!] 
	\centering
	\subfigure[ $u(\textbf{x},\infty)$, $a = .25$]{\includegraphics[width=.45\linewidth]{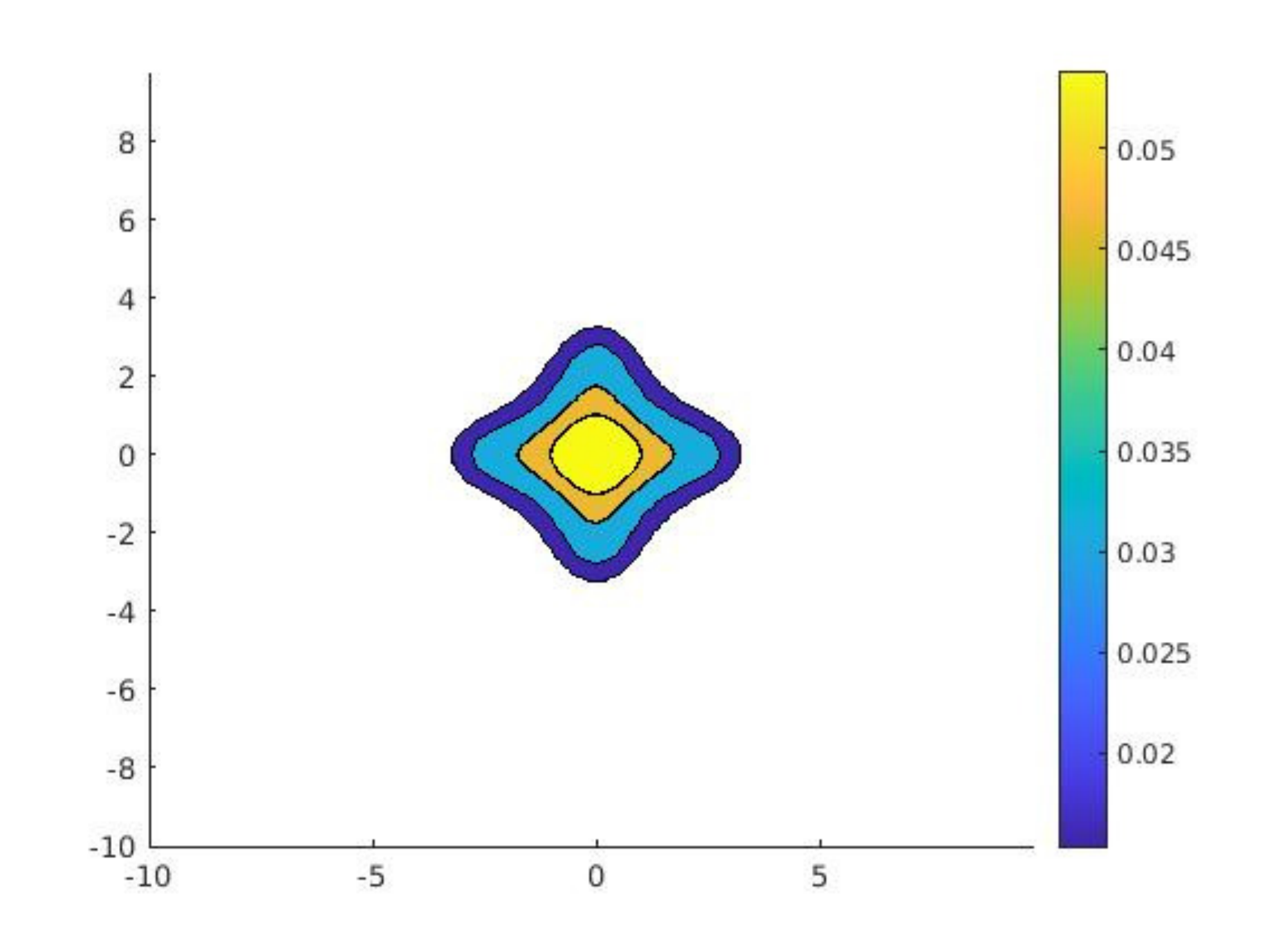}\label{fig:e2part2}}
	\subfigure[$u(\textbf{x},\infty)$, $a = .5$]{\includegraphics[width=.45\linewidth]{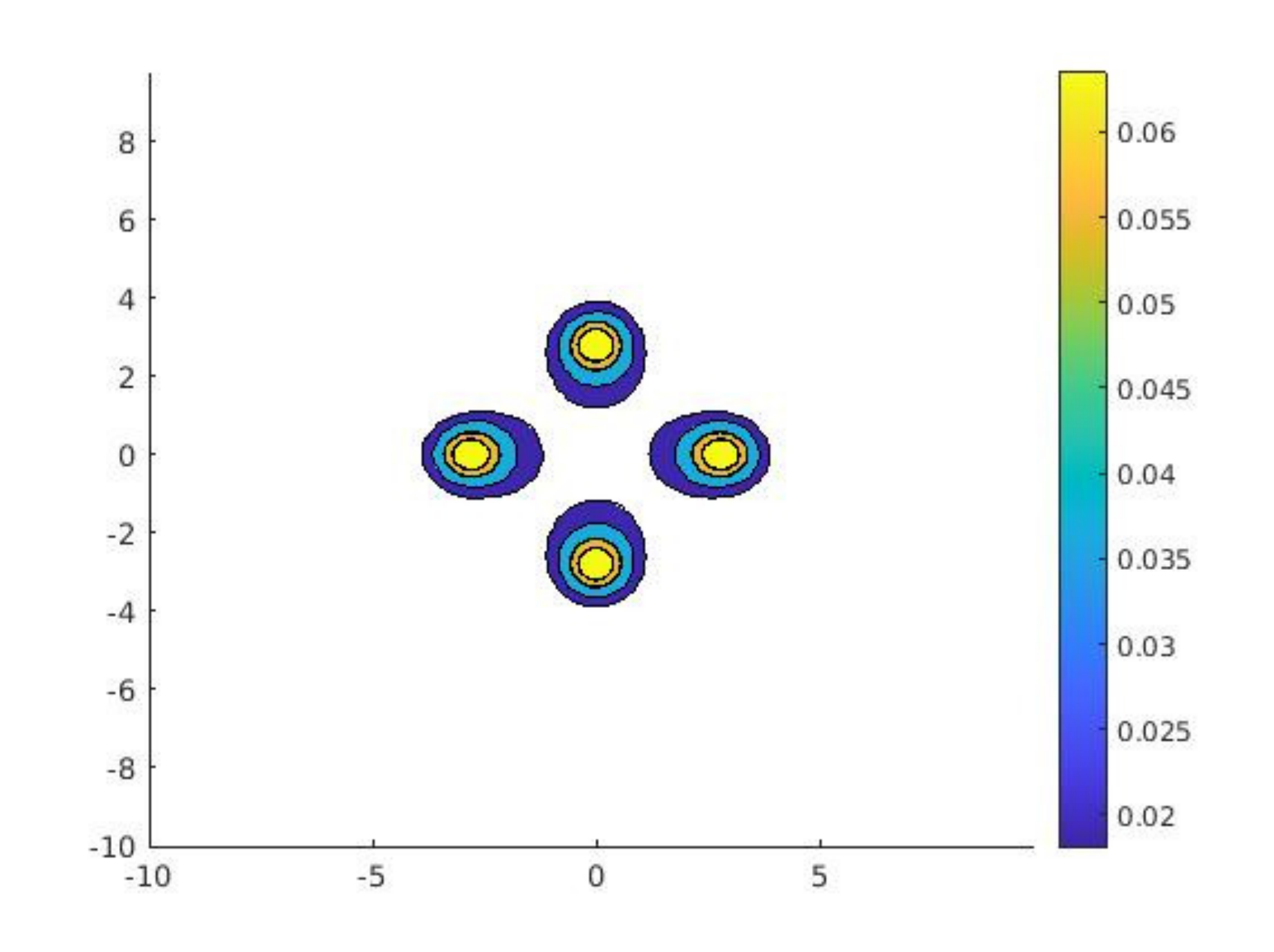}\label{fig:e2}}
	\\
	\subfigure[$u(\textbf{x},\infty)$, $U(\textbf{x})$, $a = .25$]{\includegraphics[width=.45\linewidth]{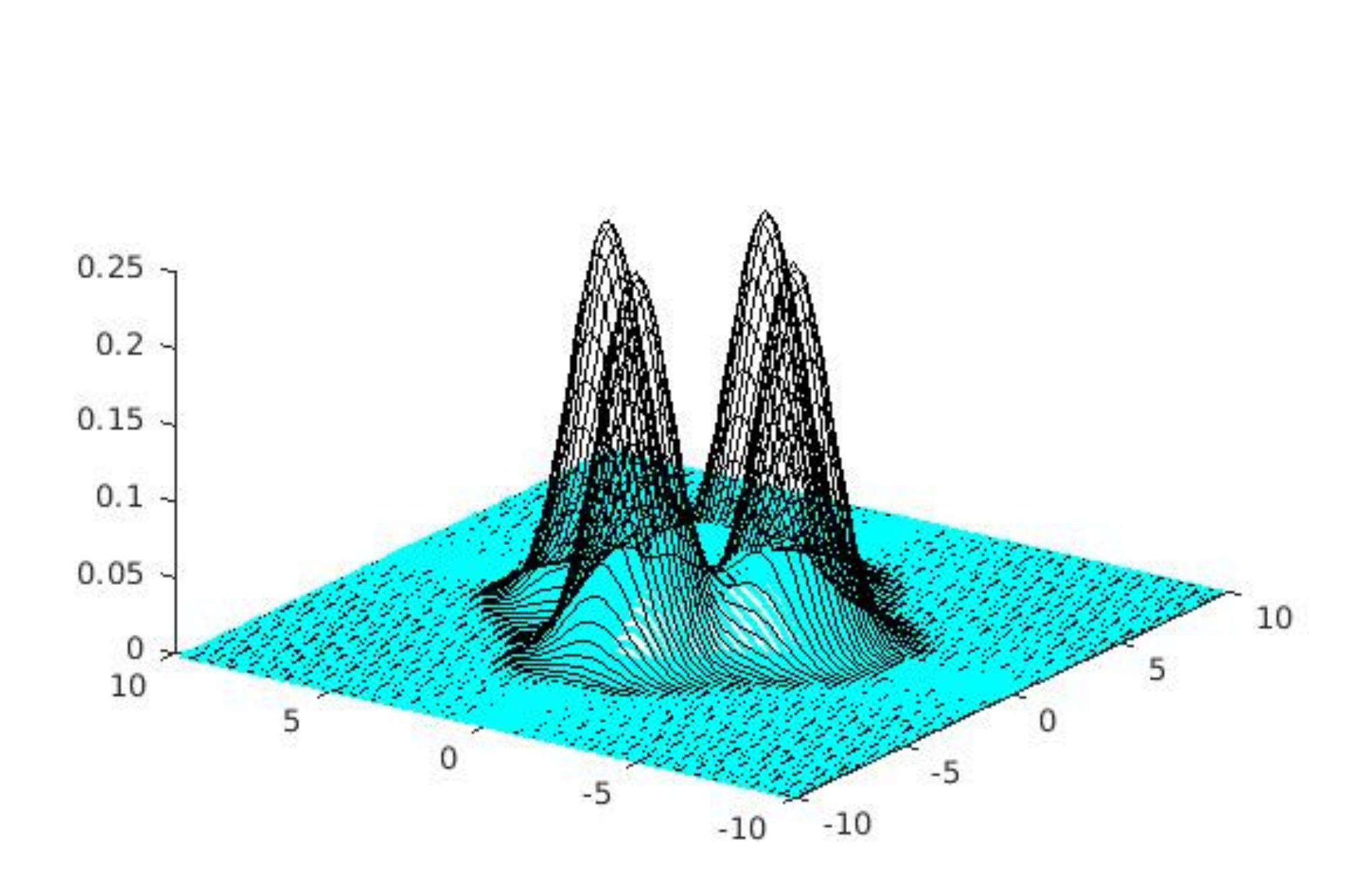}\label{fig:e2part23D}}
	\subfigure[$u(\textbf{x},\infty)$, $U(\textbf{x})$,  $a = .5$]{\includegraphics[width=.45\linewidth]{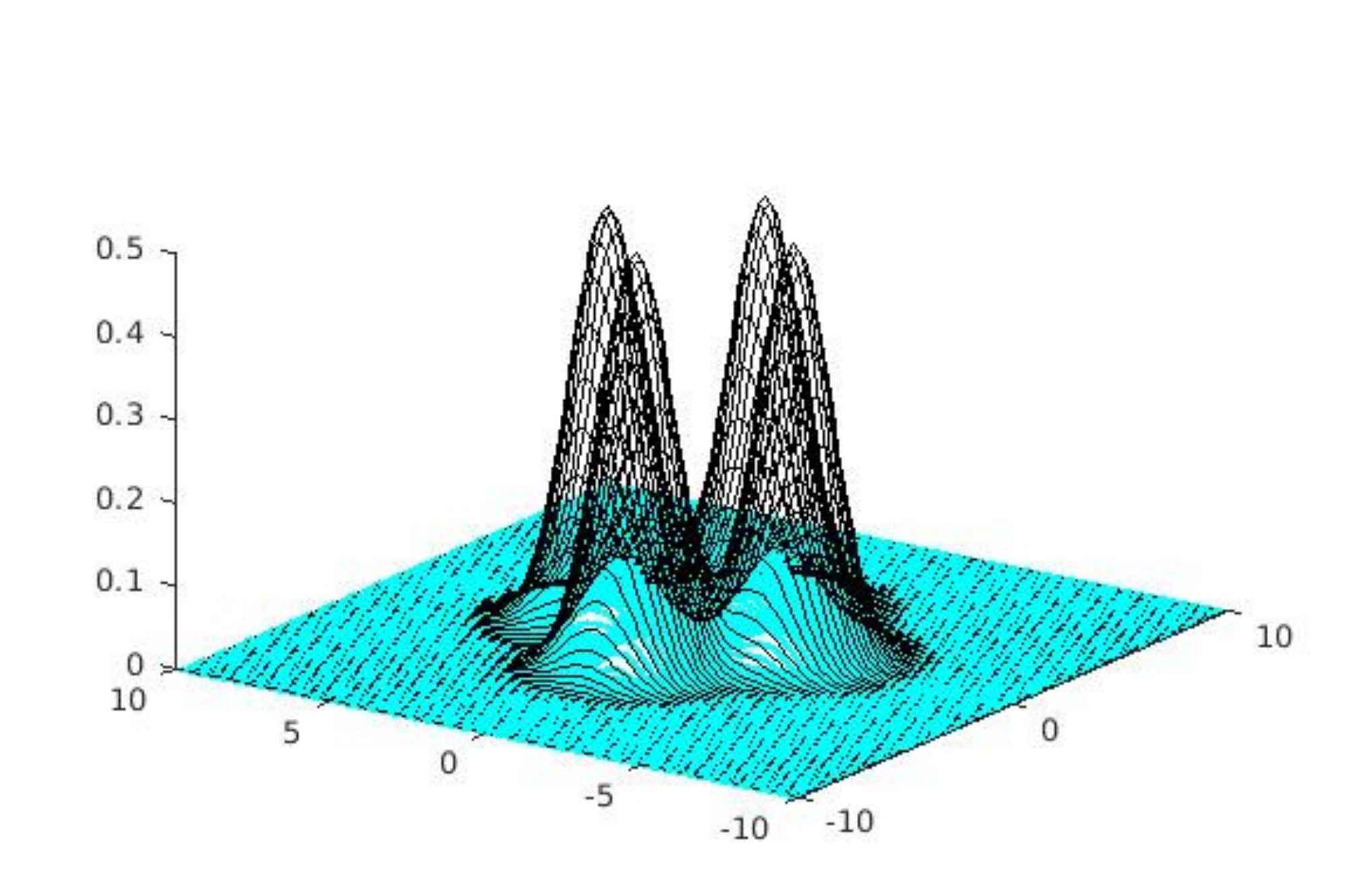}\label{fig:e23D}}

	\caption{Equilibrium solutions to \eqref{eq:System} with varied $a$, $\eta = 3$, $b=.5$, and $K$ Laplace, shown as a contour plot (top) and a three-dimensional plot with $U(\textbf{x})$ in black (bottom)}
	\label{fig:echange}
\end{figure}

We also investigate how the environment potential affects solutions. One interesting simulation shows how an environment can compete with the aggregation strength. This is demonstrated in Figure \ref{fig:echange}, which shows two simulations with attractive environment in four different regions of the domain, $\eta$ and $b$ constant, and varying values of $a$. In Figure \ref{fig:e2part2}, the territory of $u$ is connected, although attracted to the environment in four different directions. In Figure \ref{fig:e2}, $a$ is stronger, and thus overpowers the aggregation strength within $u$, and the territory splits into the four distinct regions of environment. Note that the four regions of $u$ are skewed towards the center due to the aggregation force. 
% We further study how the strength and shape of the environment changes the solutions in Section \ref{sec:Appendix}, investigating how the environment can compete with other forces in the system such as diffusion.
%strength of the environment potential, $a$, and the shape of the environment potential changes the behavior of equilibrium solutions. 

\subsubsection{Multi-group Dynamics}\label{sec:multgroups}

\begin{figure}[h!]
	\centering
	\subfigure[$u_i(\textbf{x},\infty)$, $b = .5$]{\includegraphics[width=.3\linewidth]{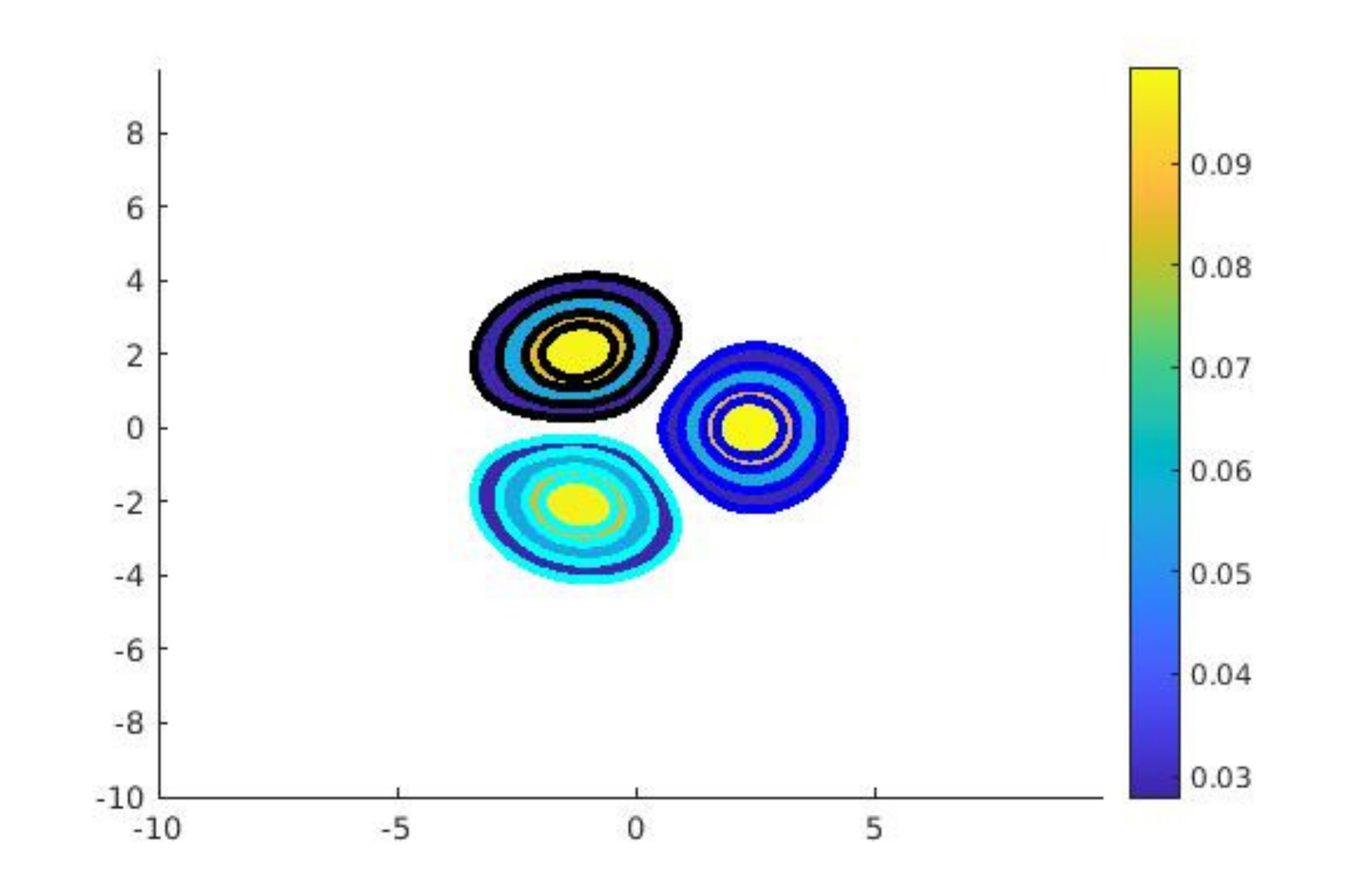}\label{fig:3K5} } 
	\subfigure[$u_i(\textbf{x},\infty)$, $b = .75$]{\includegraphics[width=.3\linewidth]{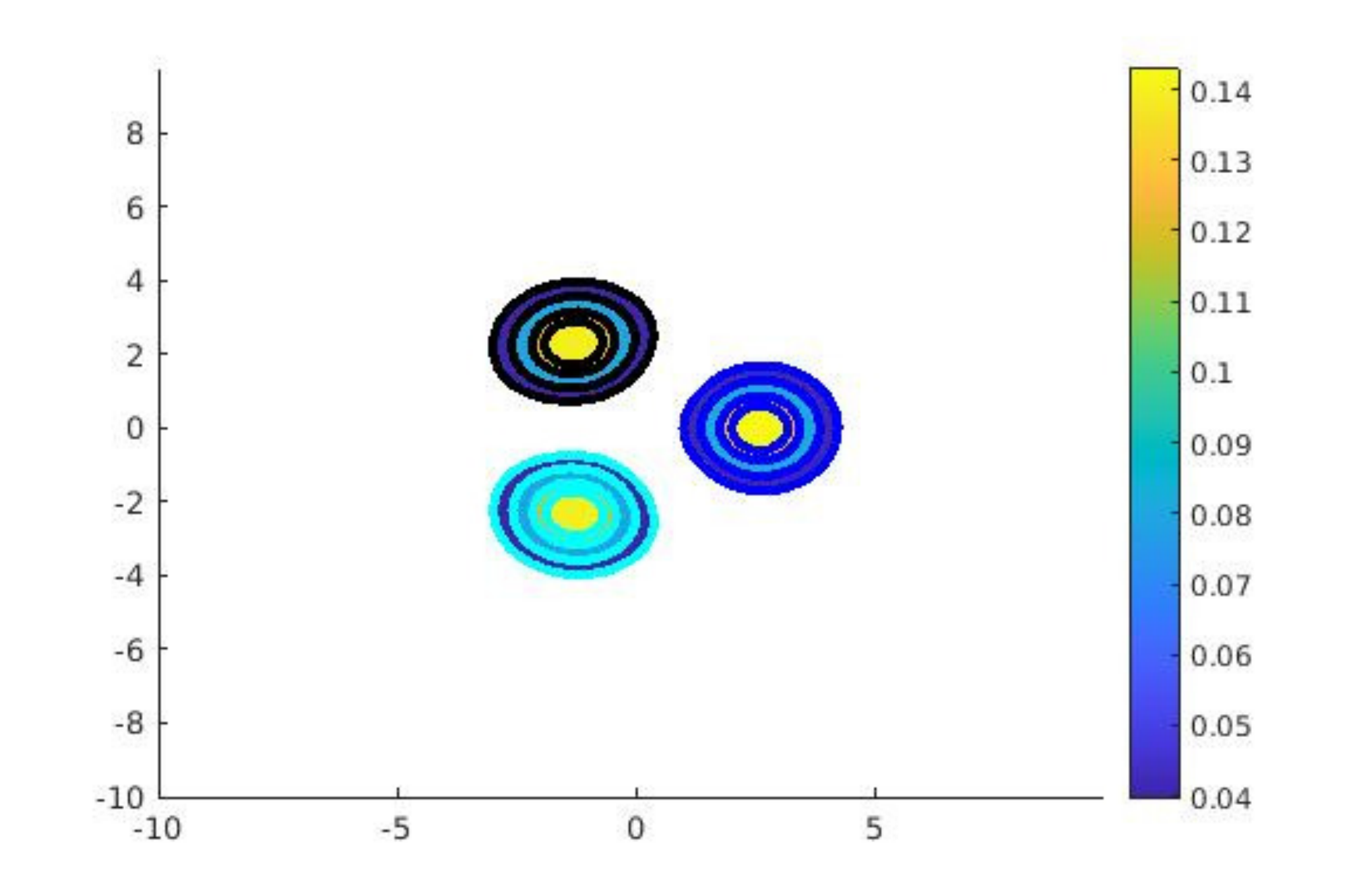}\label{fig:3K75}}
	\subfigure[$u_i(\textbf{x},\infty)$, $b = 1$]{\includegraphics[width=.3\linewidth]{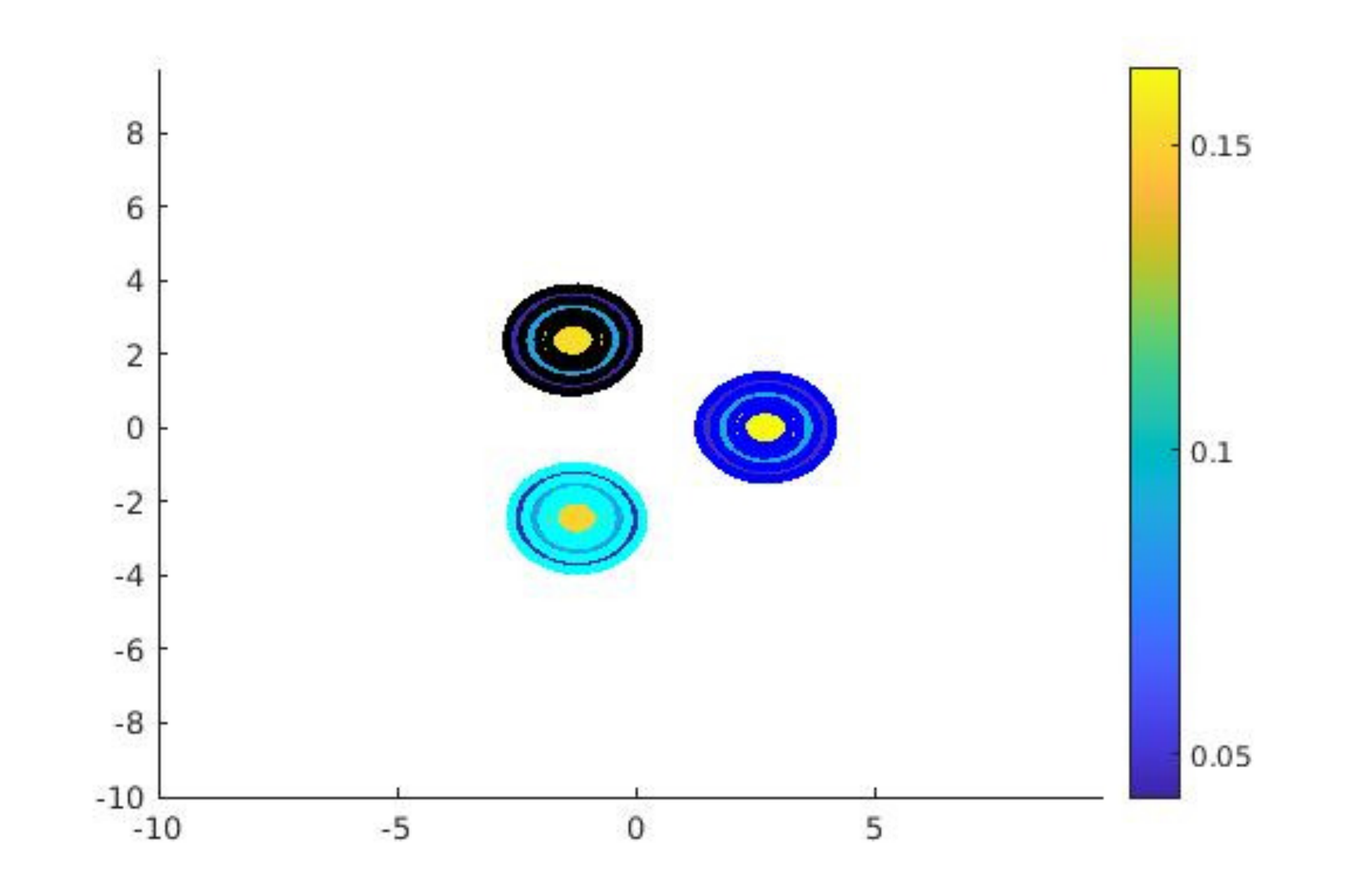}\label{fig:3K1}}\\
	\subfigure[$u_i(\textbf{x},\infty)$, $U(\textbf{x})$, $b = .5$]{\includegraphics[width=.3\linewidth]{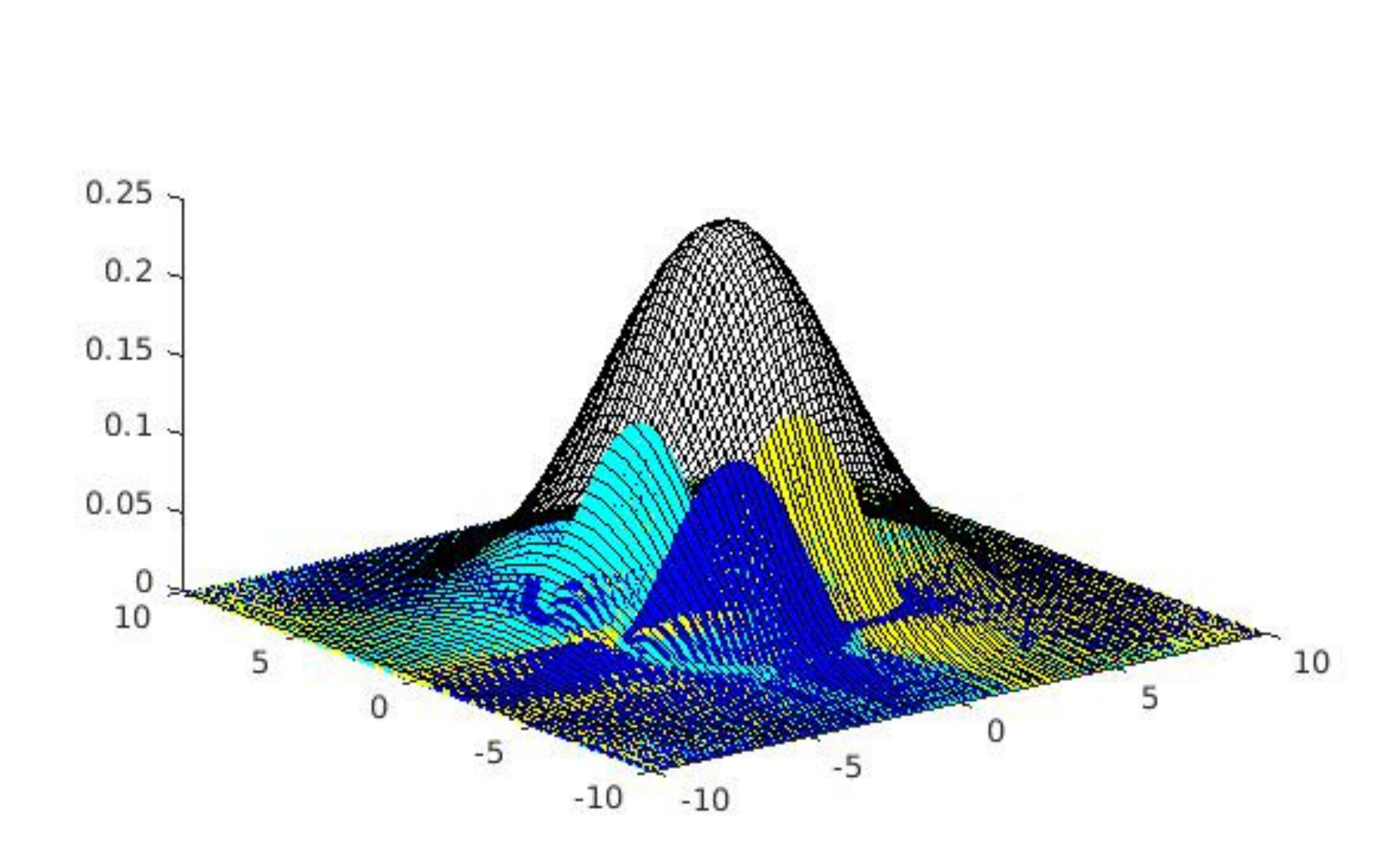}\label{fig:3K5-3D} } 
	\subfigure[$u_i(\textbf{x},\infty)$, $U(\textbf{x})$, $b = .75$]{\includegraphics[width=.3\linewidth]{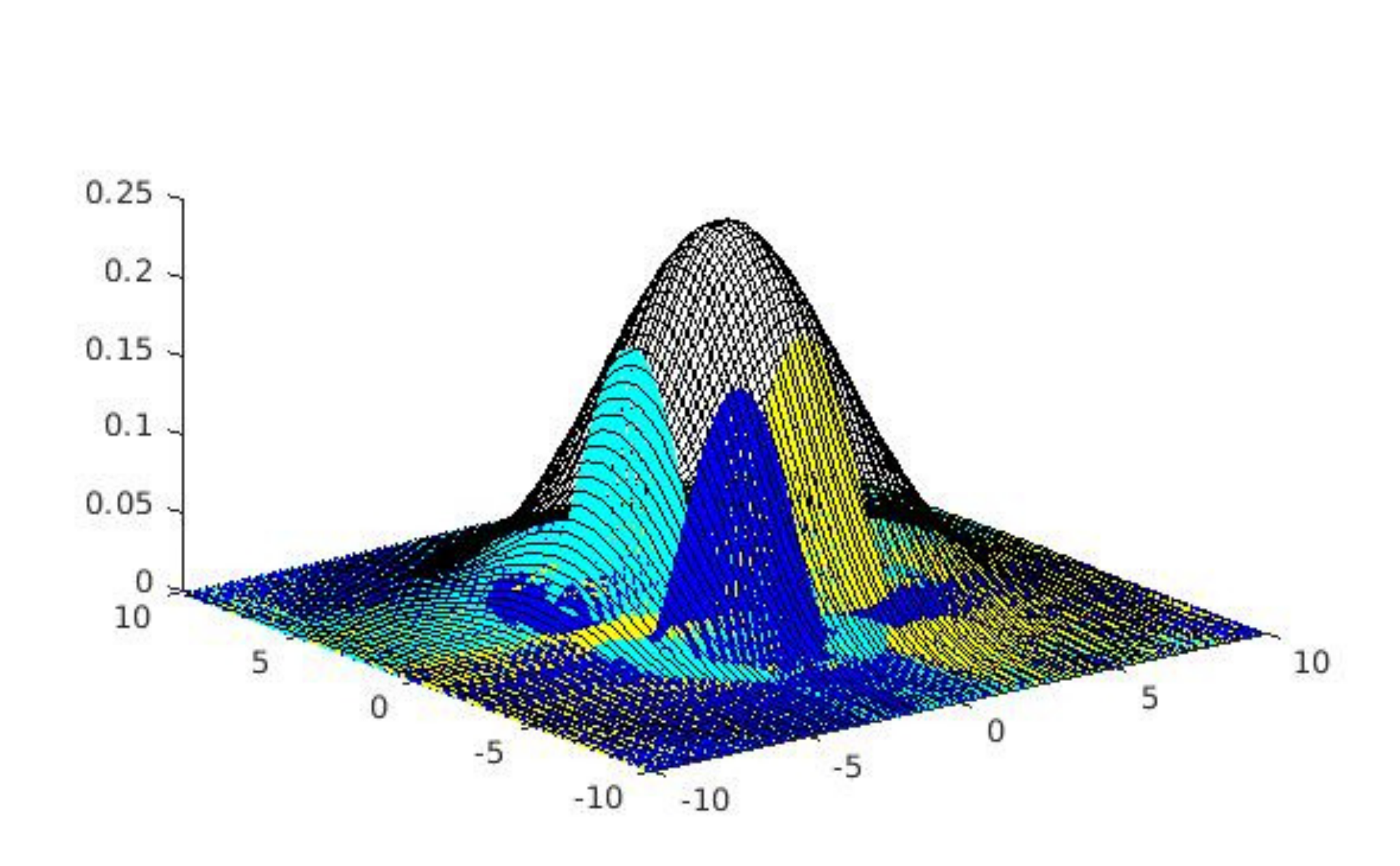}\label{fig:3K75-3D}}
	\subfigure[$u_i(\textbf{x},\infty)$, $U(\textbf{x})$, $b = 1$]{\includegraphics[width=.3\linewidth]{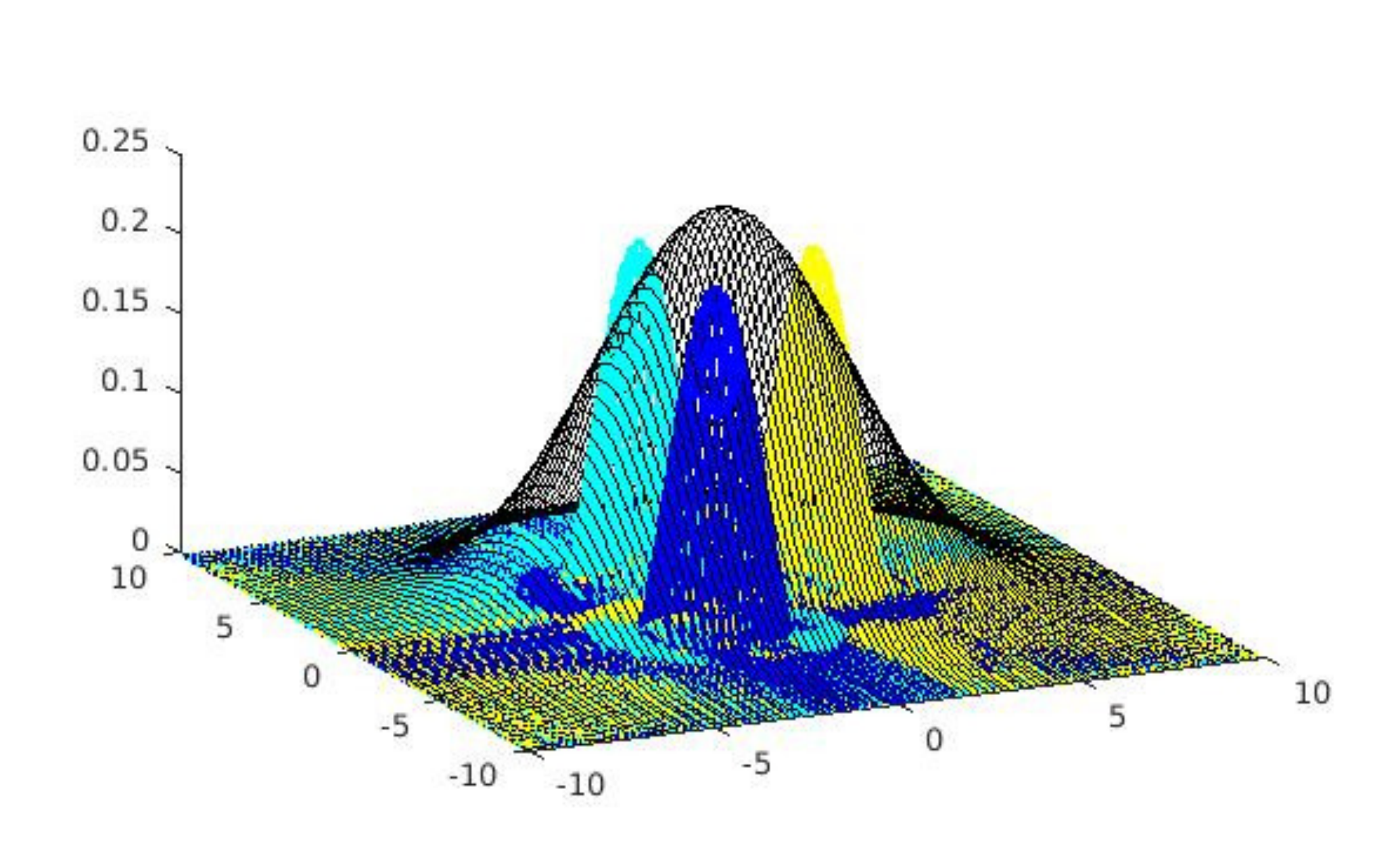}\label{fig:3K1-3D}}
	\caption{Equilibrium solution to \eqref{eq:System} for $i=1,2,3$ with $\eta = 3$, $a = 1$, $K$ Laplace, and varied $b$, shown with a contour plot (top) and a three dimensional plot with $U(\textbf{x})$ in black (bottom)}
	\label{fig:multiKchange}
\end{figure}

\begin{figure}[h!]
	\centering
	\subfigure[Initial Condition]{\includegraphics[width=.3\linewidth]{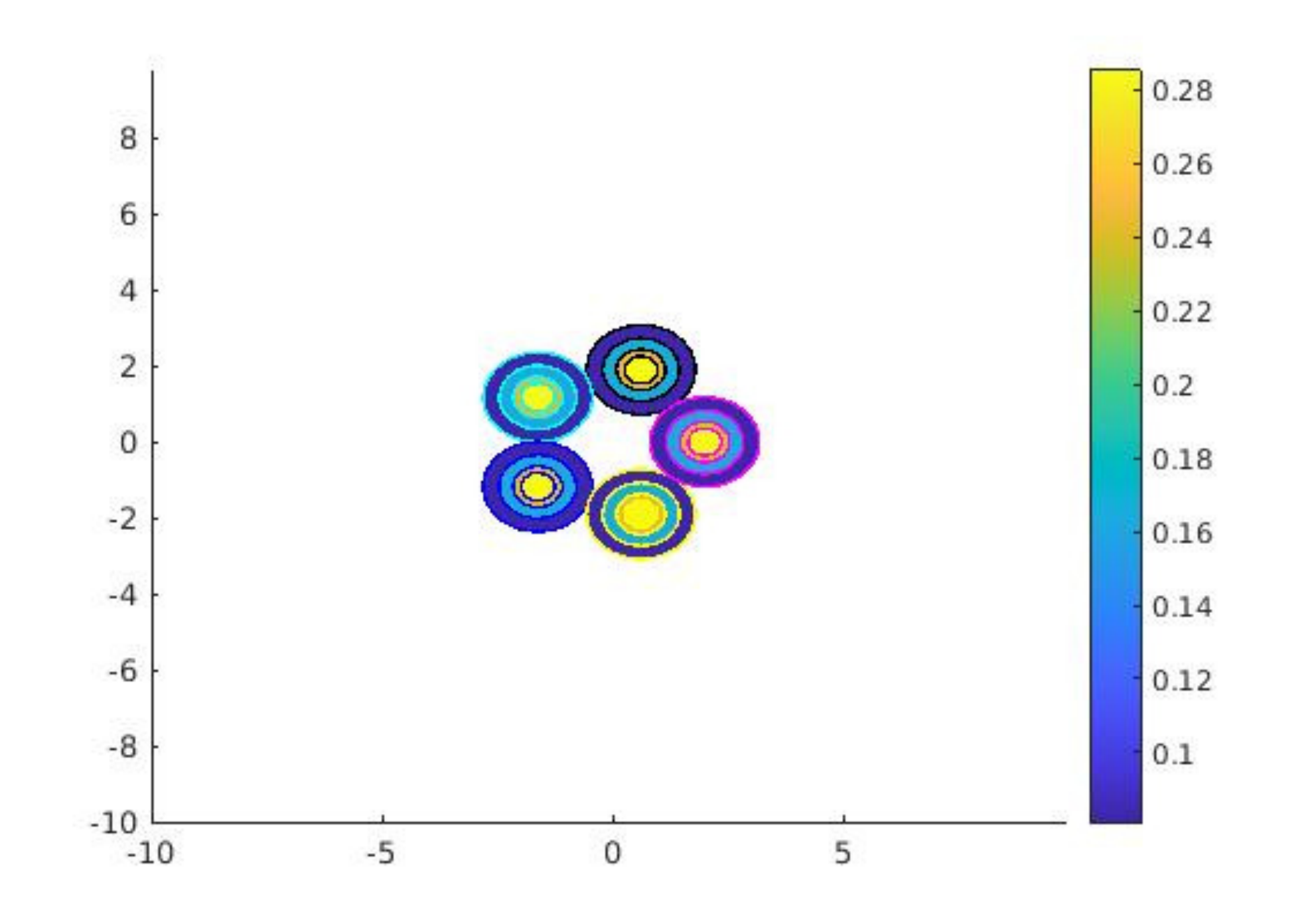}}
	\subfigure[$u_i(\textbf{x},.08)$]{\includegraphics[width=.3\linewidth]{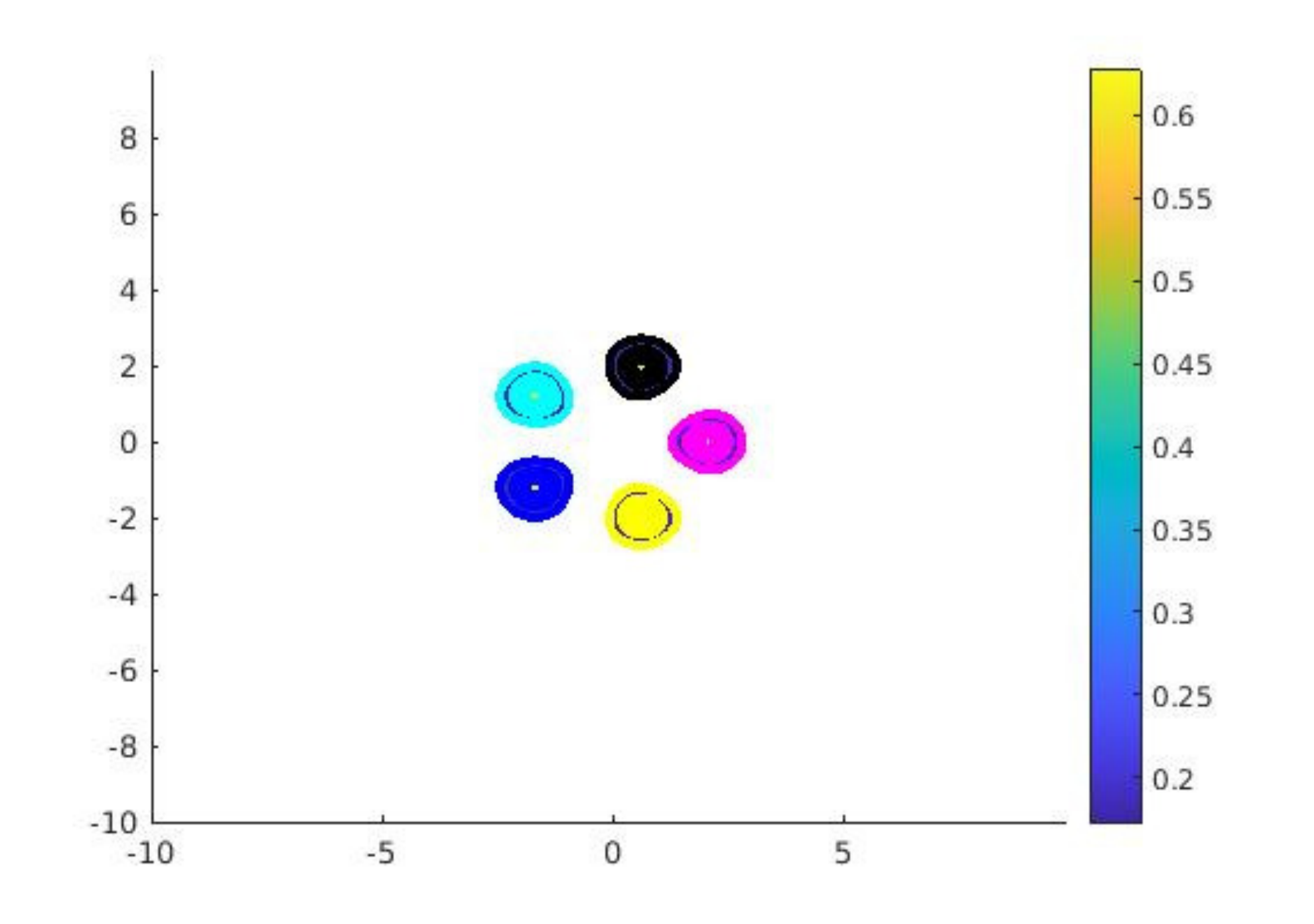}}
	\subfigure[$u_i(\textbf{x},.15)$]{\includegraphics[width=.3\linewidth]{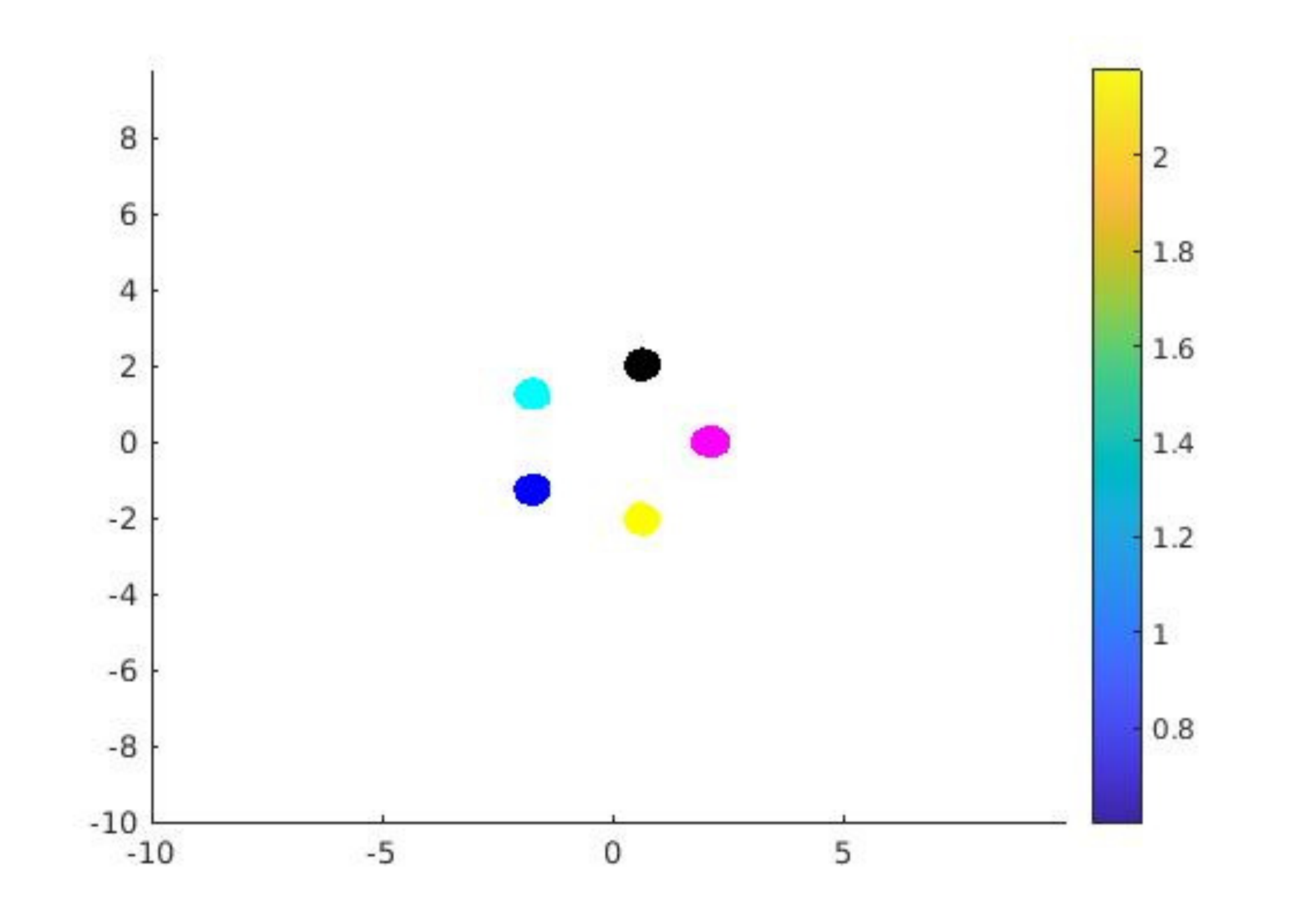}\label{fig:noetafinal}}\\
	\subfigure[Initial Condition]{\includegraphics[width=.3\linewidth]{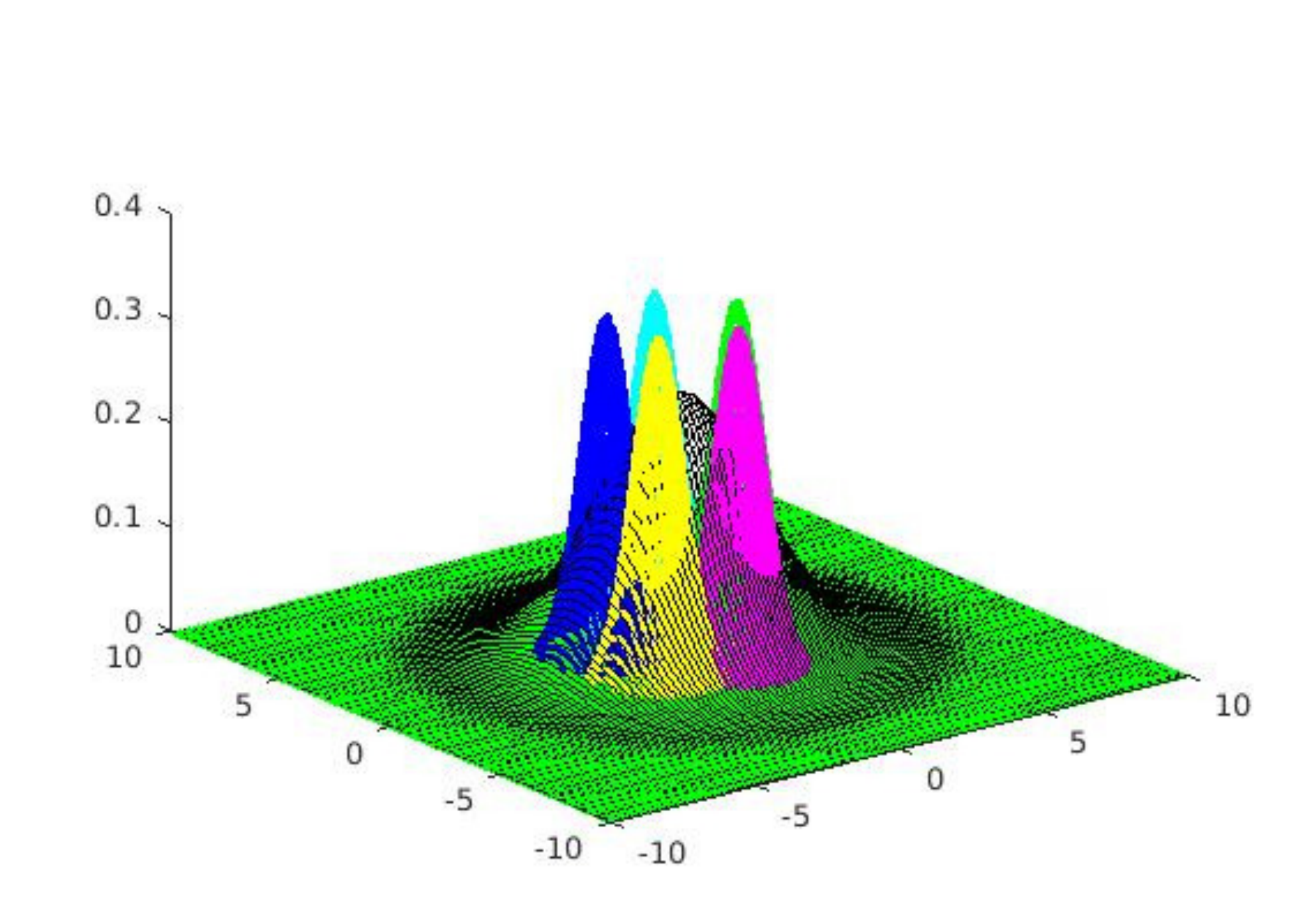}}
	\subfigure[$u_i(\textbf{x},.08)$]{\includegraphics[width=.3\linewidth]{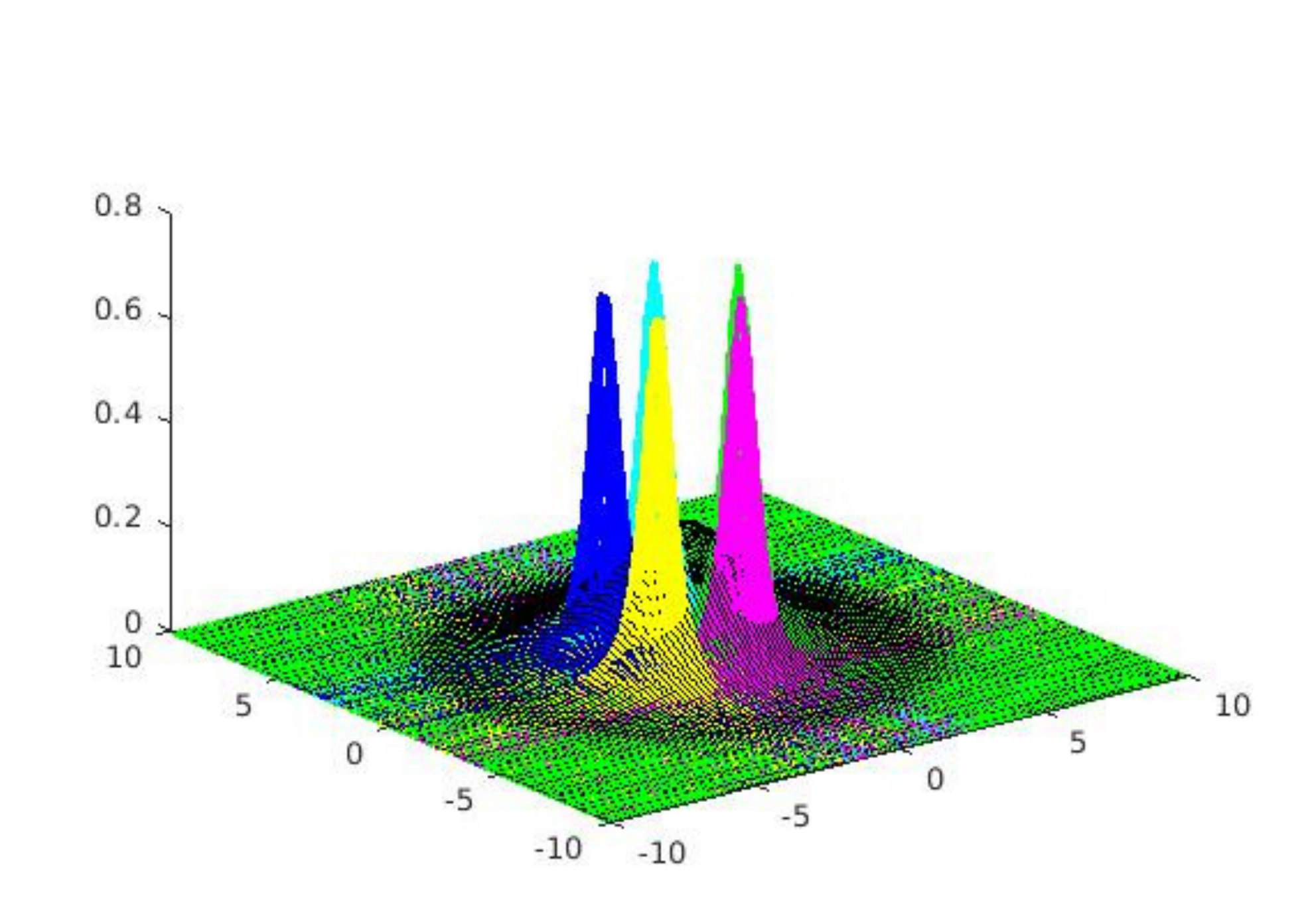}}
	\subfigure[$u_i(\textbf{x},.15)$]{\includegraphics[width=.3\linewidth]{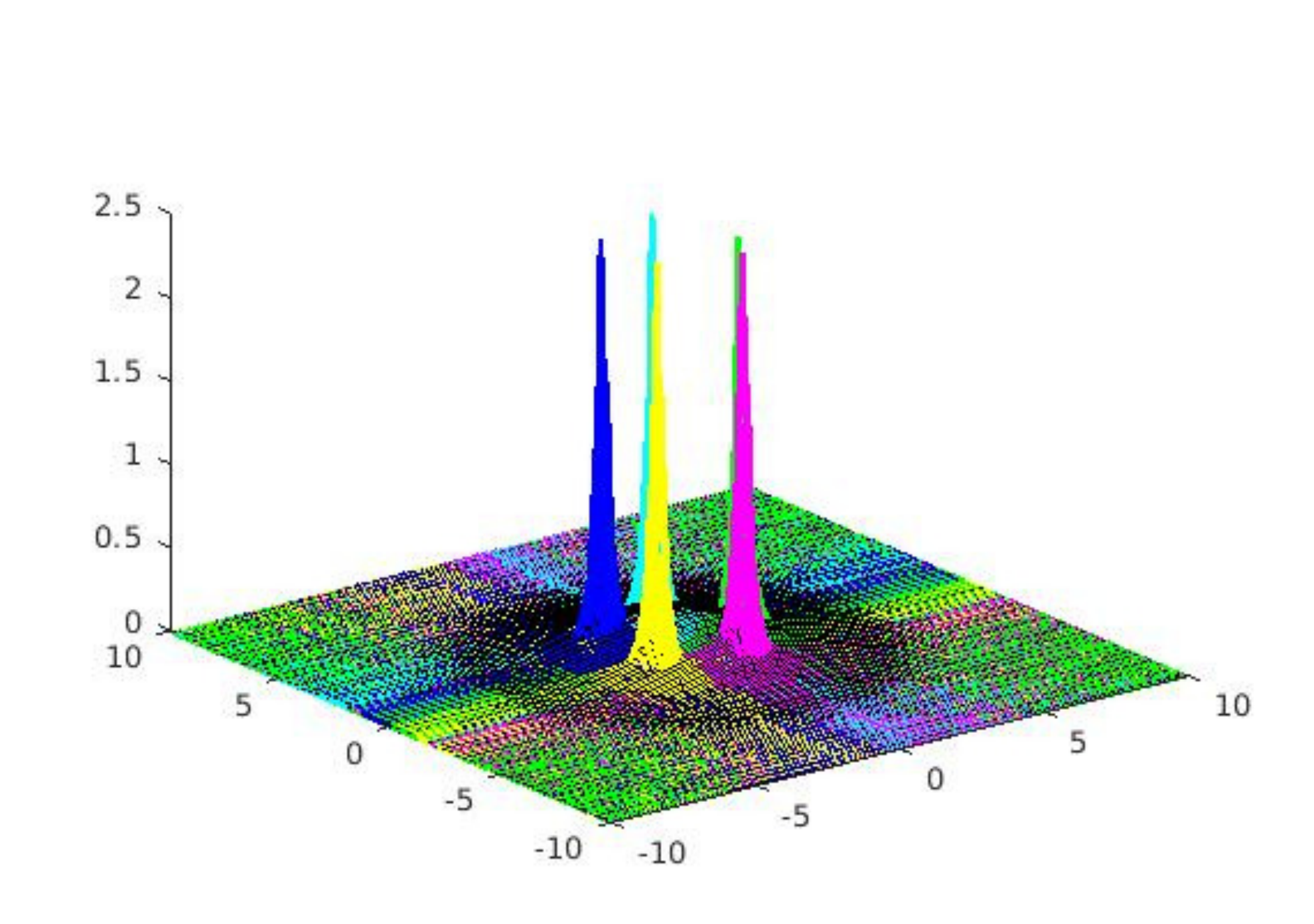}\label{fig:noetafinal3D}}
	\caption{Dynamics of the solution to \eqref{eq:System} for $i=1,...,5$ with $\eta = 0$, $b=1$, $a = .25$, and $K$ Laplace, shown with a contour plot (top) and three-dimensional plot with $U(\textbf{x})$ in black (bottom)}
	\label{fig:noeta}
\end{figure}

\begin{figure}[h!]
	\centering
	\subfigure[Initial Coniditon]{\includegraphics[width=.3\linewidth]{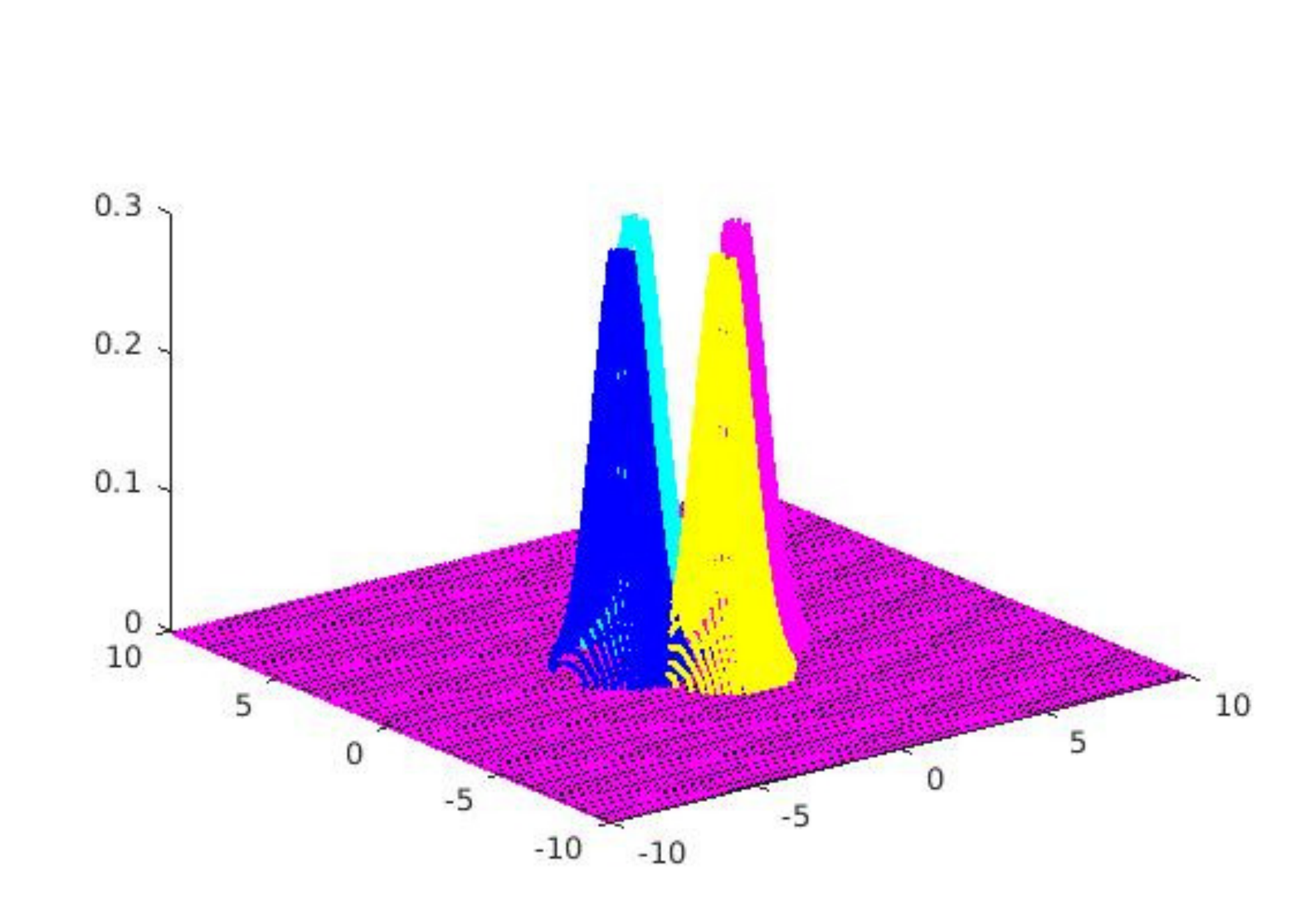}\label{fig:noAggIC}}
	\subfigure[$u_i(\textbf{x},1.65)$]{\includegraphics[width=.3\linewidth]{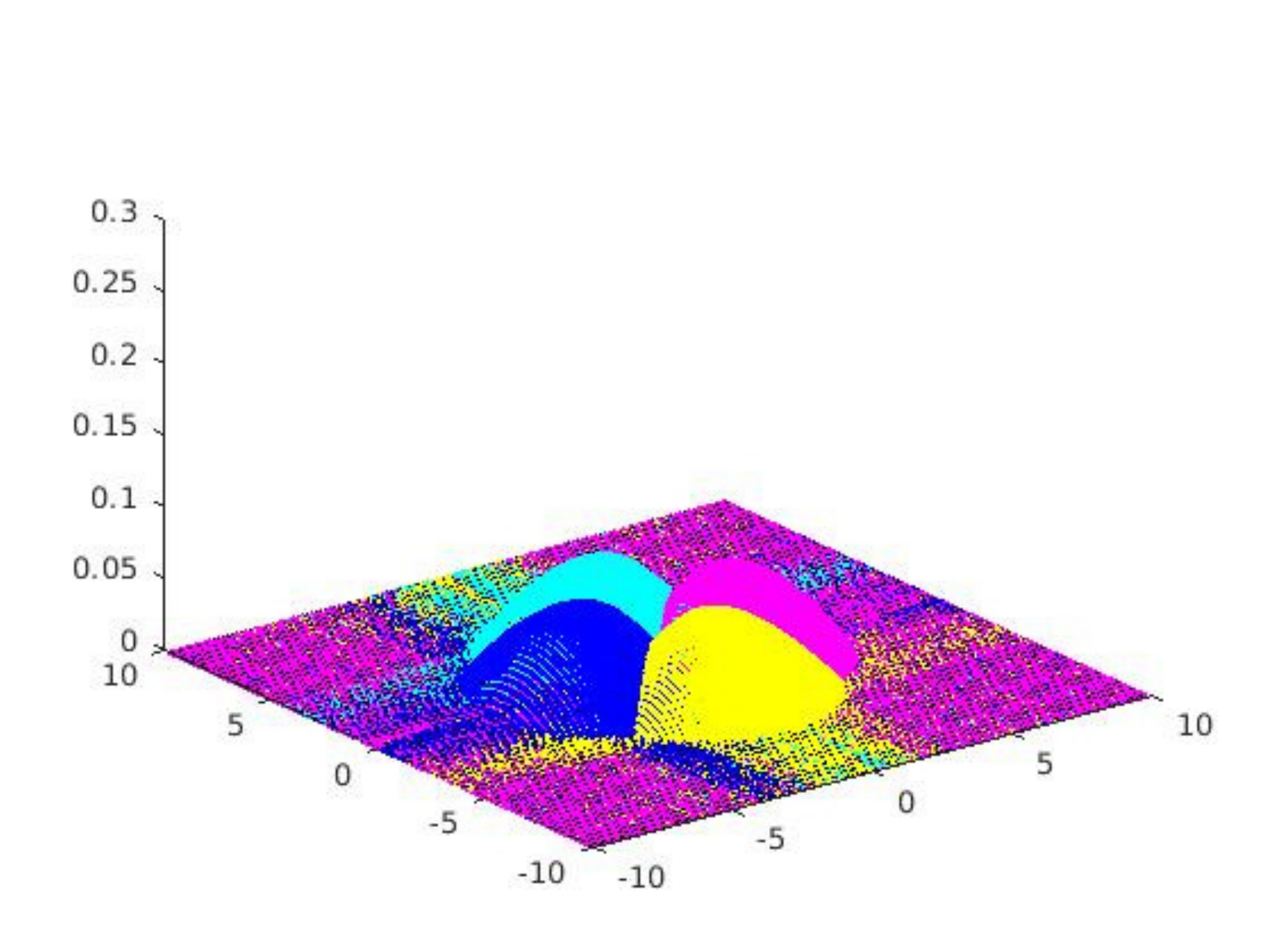}\label{fig:noAgginter}}
	\subfigure[$u_i(\textbf{x},60.4)$]{\includegraphics[width=.3\linewidth]{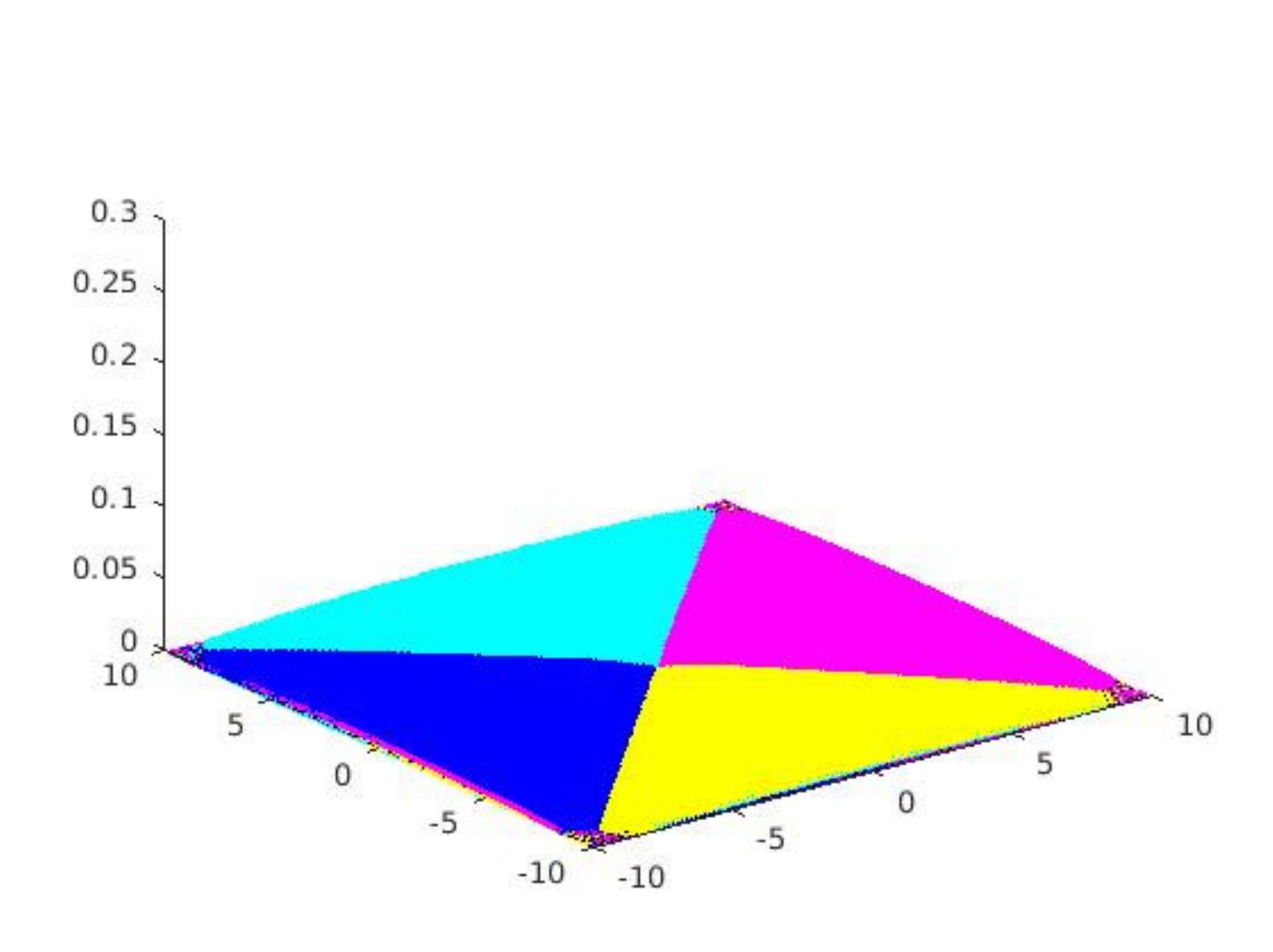}\label{fig:noAggfinal}}
	\caption{Dynamics of the solution to \eqref{eq:System} for $i=1,2,3,4$ with $\eta = 5$, $b=1$, $a = .25$, $K_2$ Laplace, $U(\textbf{x})=0$, and $K_1=0$}
	\label{fig:noAgg}
\end{figure}
\begin{figure}[h!]
	\centering
	\subfigure[Initial Condition]{\includegraphics[width=.3\linewidth]{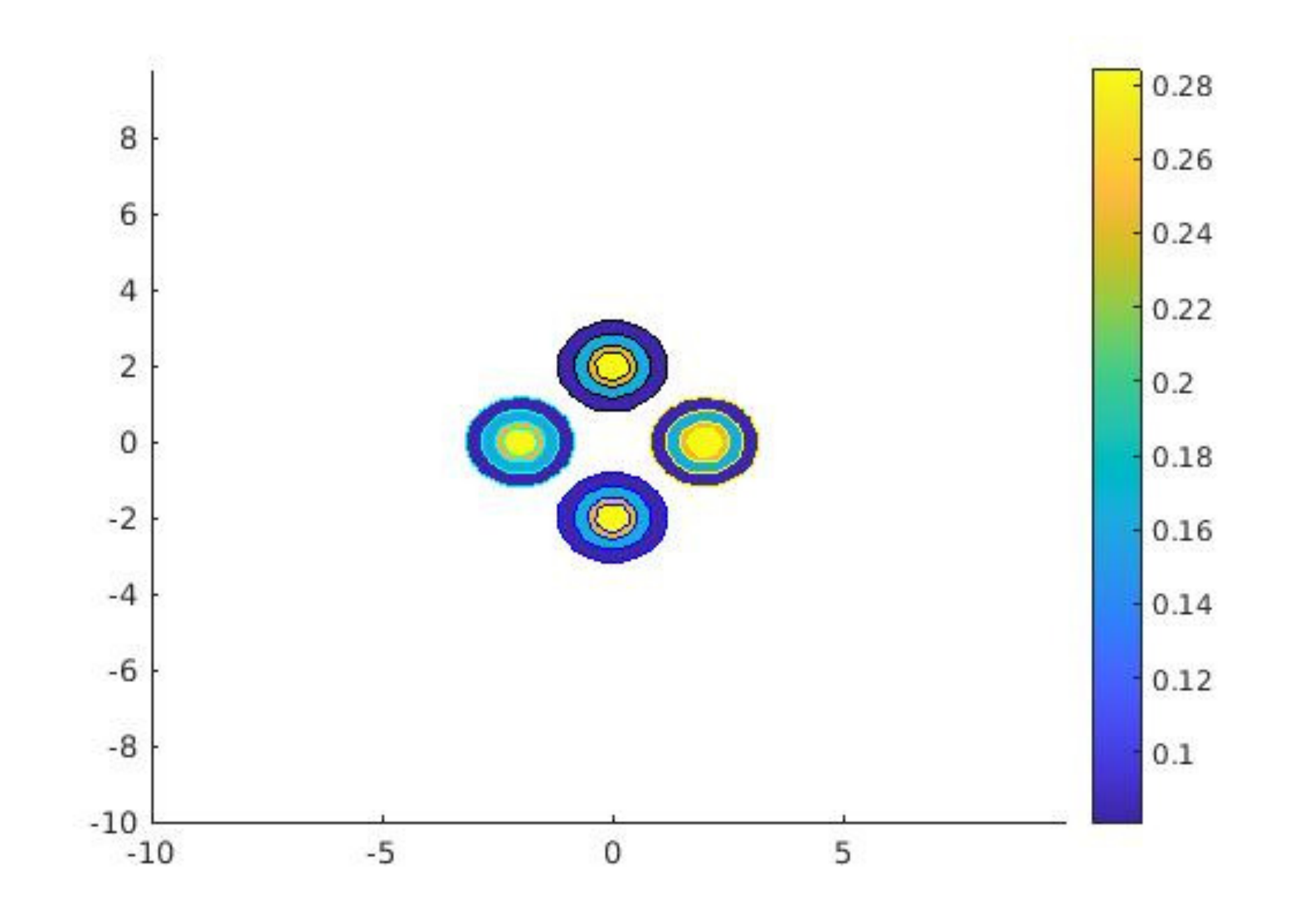}}
	\subfigure[$u_i(\textbf{x},9.982)$]{\includegraphics[width=.3\linewidth]{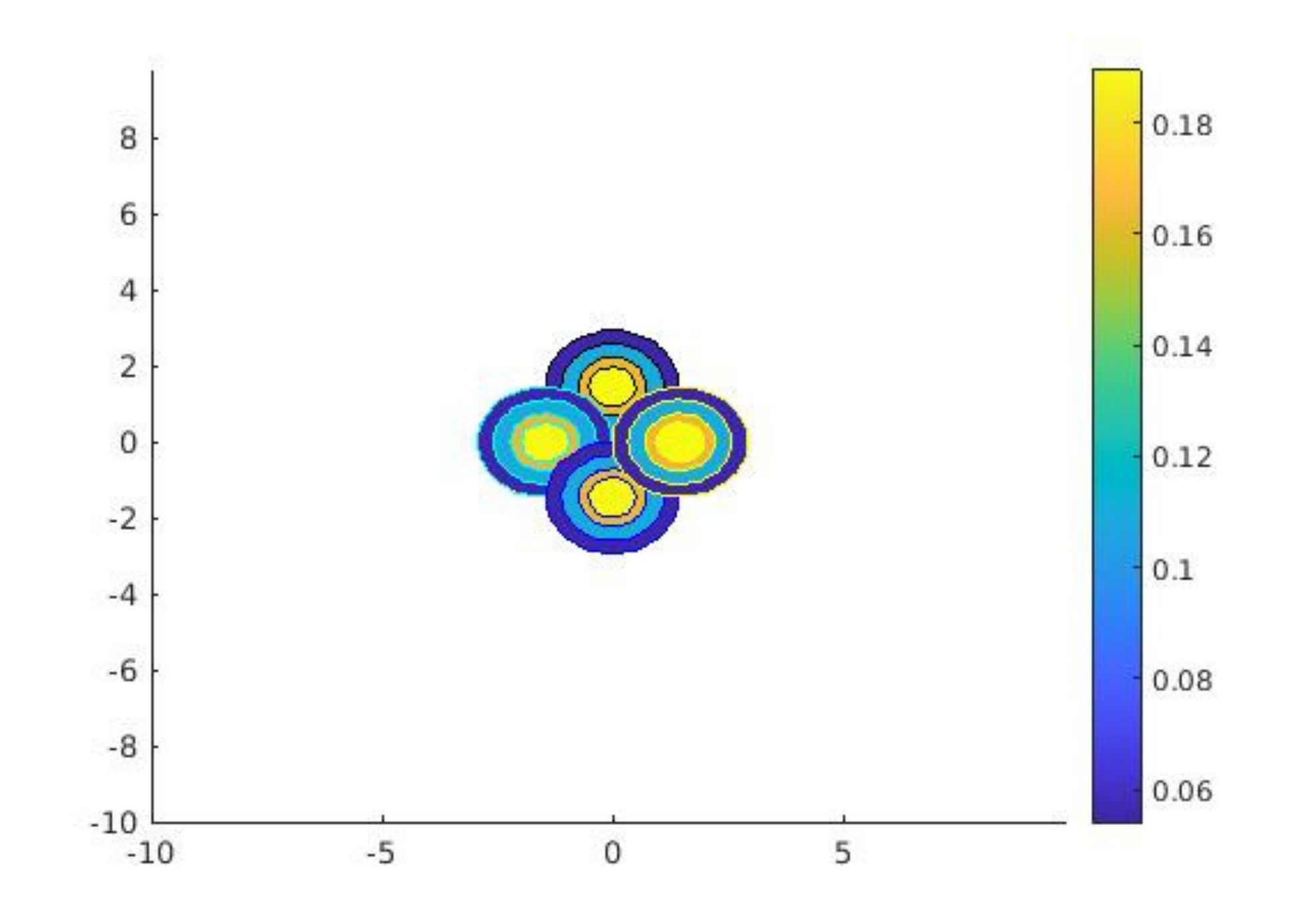}}
	\subfigure[$u_i(\textbf{x},49.33)$]{\includegraphics[width=.3\linewidth]{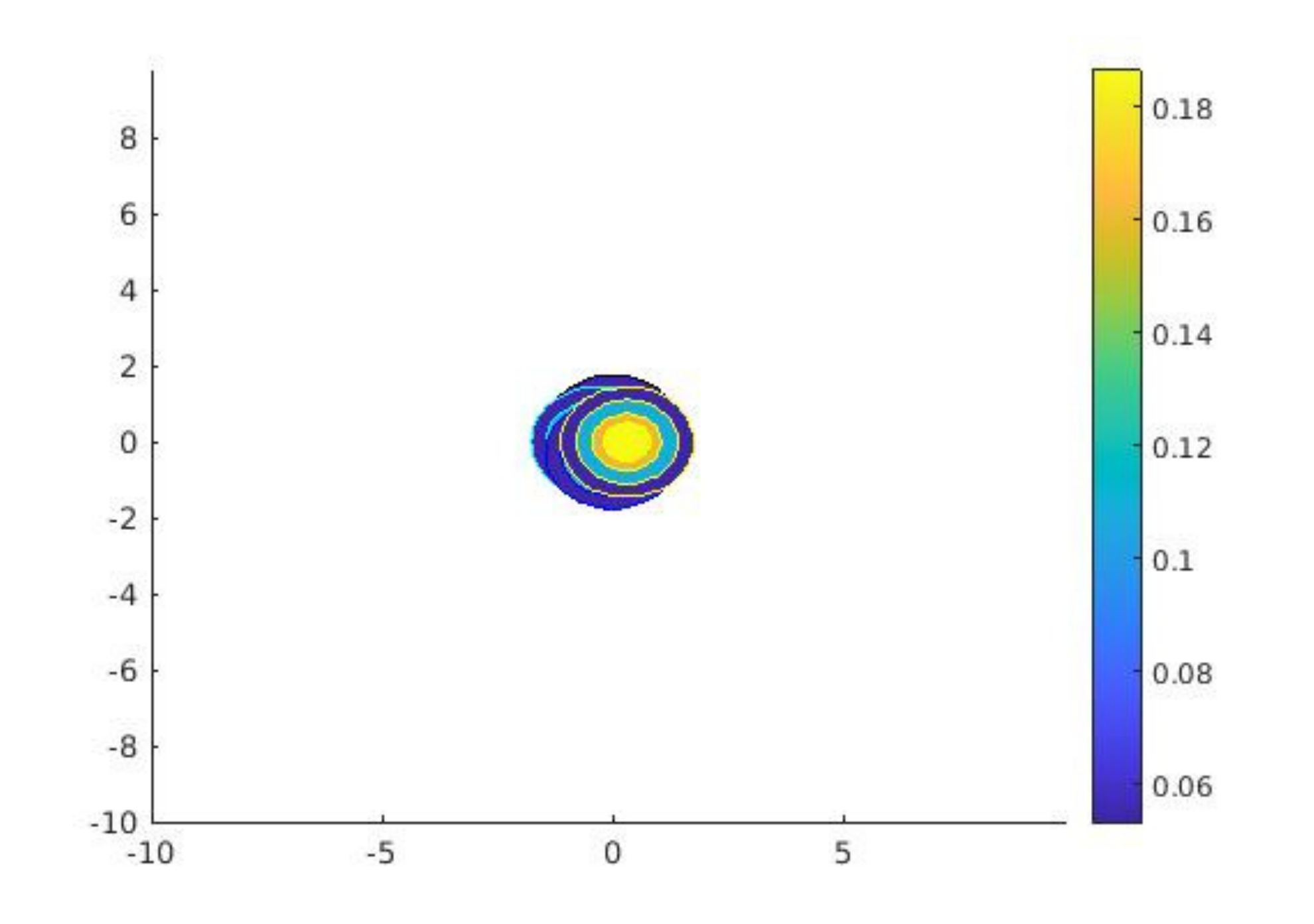}\label{fig:noSegc}}
	\caption{Dynamics of the solution to \eqref{eq:System} for $i=1,2,3,4$ with $\eta = 5$, $b=1$, $a = .25$, $K_1$ Laplace, $K_2=0$, and $U(\textbf{x})$ from Figure \ref{fig:noeta}}
	\label{fig:noSeg}
\end{figure}

We increase the number of groups and investigate the effect of the interaction potential $K$. With multiple groups, the potential $K$ acts as an aggregation potential within groups and a segregation potential between groups. We also consider the importance of each of the terms in system \eqref{eq:System} to the behavior of solutions.
%We run simulations with $a$ and $\eta$ constant, varying $b$. As we change the strength of the potential $K$, the segregation strength between groups will change as well as the aggregation within a group. 

Figure \ref{fig:multiKchange} demonstrates what occurs as the interaction potential strength changes for multiple groups.
% In this figure, each column is a simulation with a different value of $b$.The rows are two different perspectives: a contour plot and a three-dimensional plot of $u$ and $U$. 
In these simulations, $\eta$ and the environment potential, which is a Gaussian, are kept constant. 
%The environment potential is the same as Figure \ref{fig:eG1}. 
When comparing Figure \ref{fig:3K5-3D} and \ref{fig:3K1-3D}, as $b$ increases, there is higher population density within groups and more segregation between groups. It is clear in Figure \ref{fig:3K1}, the groups have a smaller territory, but also have more space separating the territories between groups than when $b$ is smaller, shown in Figure \ref{fig:3K5}.

We also consider the importance of each of the terms in the system by exploring the dynamics of the solution when we remove them. Figure \ref{fig:noeta} considers the dynamics of the solution when $\eta= 0$, thus no diffusion within each group. In this case, the aggregation within a group has nothing to balance it out, and each group begins to aggregate to a single point, seen in Figure \ref{fig:noetafinal} and Figure \ref{fig:noetafinal3D}. Figure \ref{fig:noAgg} demonstrates the importance of nonlocal aggregation within a group. When there is no aggregation within a group, $K_1 = 0$, the groups do not maintain coherence. Figure \ref{fig:noAggIC} shows the initial condition, and Figure \ref{fig:noAggfinal} shows the groups diffusing to a constant in the area not occupied by opposing groups. With only one group, we would see the solution go towards a constant over the whole domain, and with $U(\textbf{x})>0$, the group would be attracted to larger values of $U$ without regard to whether it split into disjoint territories. Finally, Figure \ref{fig:noSeg} demonstrates the importance of the nonlocal segregation term by removing segregation between groups, $K_2 = 0$. The groups are attracted to the environment, but not repelled from each other, and territories in Figure \ref{fig:noSegc} overlap. Thus, in order for territories of groups to be distinct, the  nonlocal segregation is necessary.

\subsection{Different Inter-group versus Intra-group potentials}\label{sec:diffPotentials}
 As noted in Section \ref{sec:NLMechanistic}, we can use the same potential to control aggregation within a group and repulsion between groups in order to have an energy structure in the system. However, if having this structure is not necessary, we can use two different potentials to govern these interactions. In this section, we consider different potentials for aggregation within groups and repulsion between groups. We use a Morse-type potential for the aggregation potential, $K_1(x,y) = Be^{-c(x^2+y^2-r)} - e^{-2c(x^2+y^2- r)}$ with $c = .5$ and $r=2$. As with the previous cases, $B$ is found so $K_1$ integrates to one over the domain. This potential allows us to consider a larger range of $b$ and $\eta$ values, because it repels a population at close distances. %Therefore, a higher $b$ value does not need a higher $\eta$ value to balance the strong aggregation force.
 
  In  case where $N=1$, when $b$ was large and $\eta$ small, solving the system took longer to find an equilibrium solution with the requested tolerance, if the simulation terminated at all. However, when we choose $K_1$ to be the Morse-type potential, there is repulsion within a group at shorter distances and aggregation at longer distances. Therefore, the balance between diffusion and aggregation is encapsulated in $K_1$. We completed simulations with $N=1$, $a$ constant, and varying $b$, for both $K_1$ equal to the Laplace potential and $K_1$ equal to the Morse-type potential and determined a stable range of $\eta$ for the numerical simulations. For each simulation where $K_1$ was the Laplace potential, there was a lower bound for a stable $\eta$ value in order for it to balance out the aggregation force. However, when $K_1$ was the Morse-type potential, we can set $\eta = 0$ and incorporate both the diffusivity and aggregation in $K_1$.

\begin{figure}[h!]
	\centering
	\subfigure[Equilibrium solution]{\includegraphics[width=.3\linewidth]{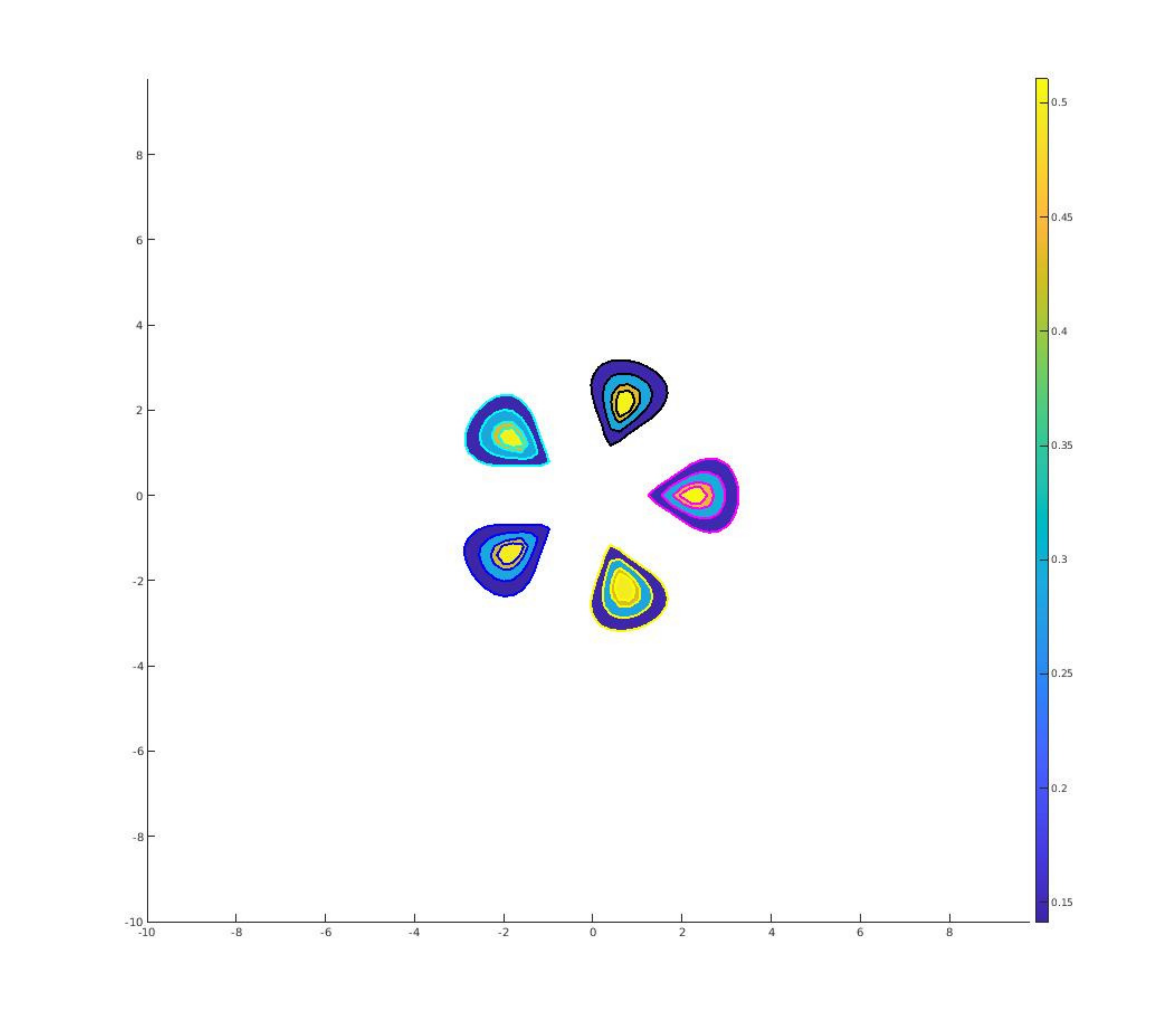}\label{fig:Morsec}}
	\subfigure[Equilibrium solution]{\includegraphics[width=.3\linewidth]{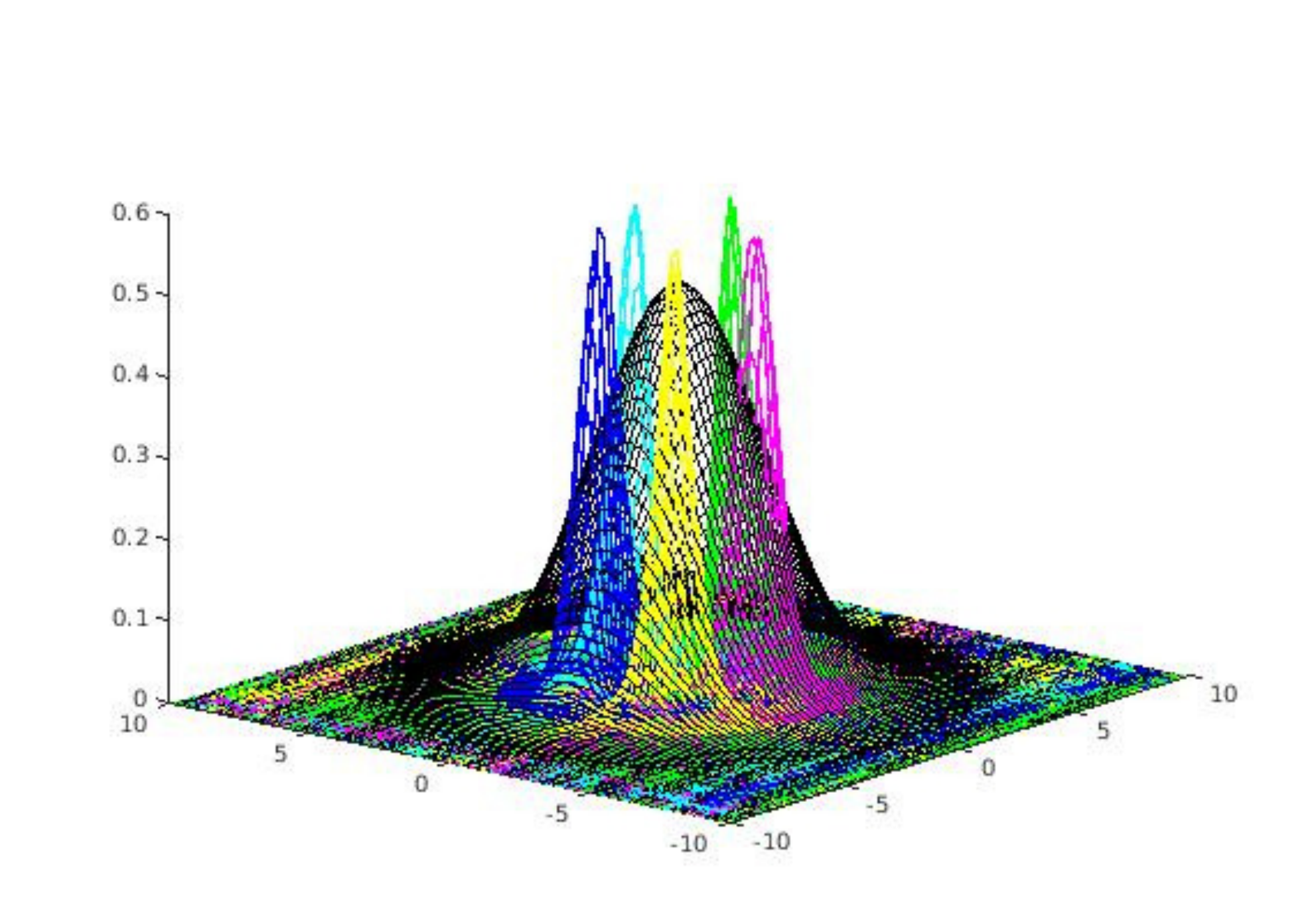}\label{fig:Morse3D}}
	\subfigure[Aggregation Potential]{\includegraphics[width=.3\linewidth]{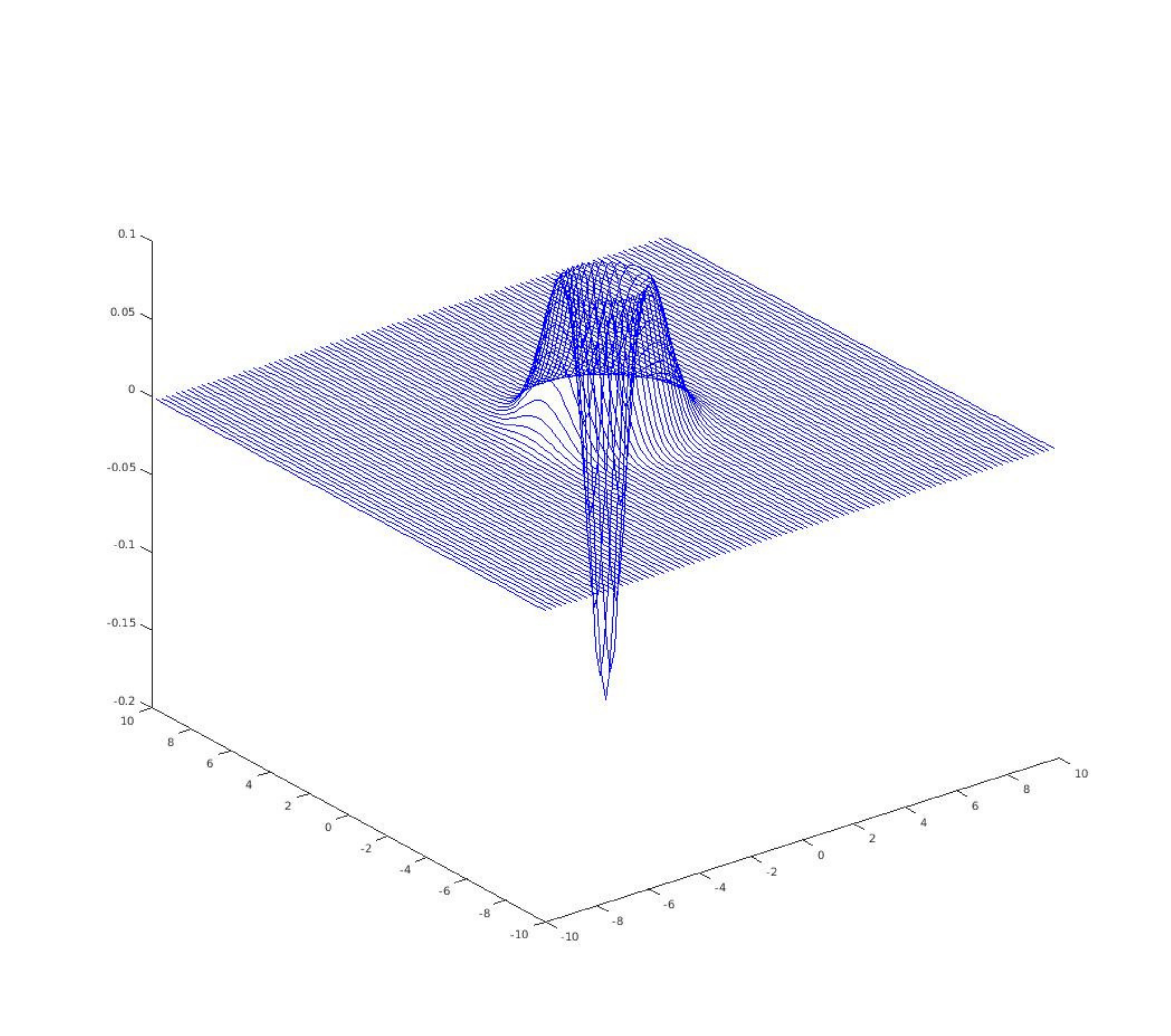}\label{fig:Morsep}}
	\caption{Equilibrium solution  to \eqref{eq:System} for $i=1,...,5$ with $\eta = 0$, $a = .5$, $b=.5$ shown with (a) a contour plot, (b) a three-dimensional plot with $U(\textbf{x})$ in black, and (c) Morse-type potential $K_1$ }
	\label{fig:Morse}
\end{figure}

In the case where $N>1$, when $K$ was equal to the Laplace potential and $\eta = 0$, each group aggregated to a single point. However, Figure \ref{fig:Morse} shows a simulation done with $\eta = 0$, and $K_1$ is equal to the Morse-type potential, shown in Figure \ref{fig:Morsep}. The interaction potential, $K_2$, is the Laplace potential. The repulsion at close range in the Morse-type potential allows for $\eta=0$ while preventing groups from aggregating to a single point. It is worth noting that changing the shape of the aggregation potential has a large effect on the shape of the territories when comparing it to simulations done with $K_1$ equal to Laplace potential.

 \subsection{Environment and Location Data}

Finally, we find equilibrium solutions informed by the data we intend to incorporate into the system. This allows us to understand challenges that come with using more complicated environments and computation times associated with more detailed environments and large $N$. We use the location data, Figure \ref{fig:data}, take the kernel density estimation of this data, Figure \ref{fig:KDE}, and use this as an initial condition. Then, for a fixed set of parameters, we find the equilibrium solutions using the two environment types, mollified with extended domains, $U=\text{EDGE}$, shown in Figure \ref{fig:EDGE}, and $U=\text{SAND}$, shown in Figure \ref{fig:SAND}.

 \begin{figure}[h]
 	\centering 
 	\subfigure[$U =$ EDGE]{\includegraphics[width = .45\linewidth]{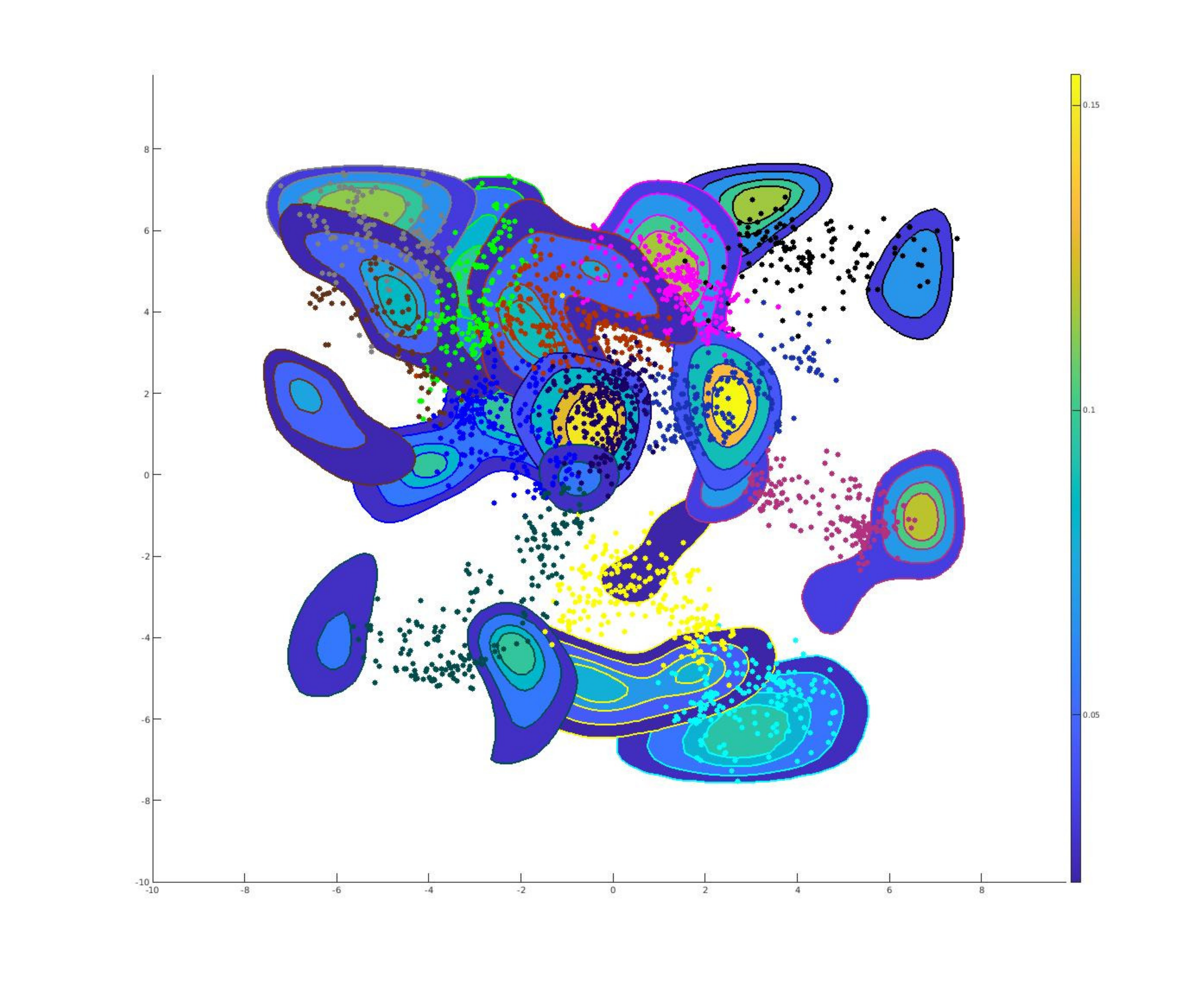} \label{fig:EquilEDGE}}
 	\subfigure[$U=$ SAND]{\includegraphics[width = .45\linewidth]{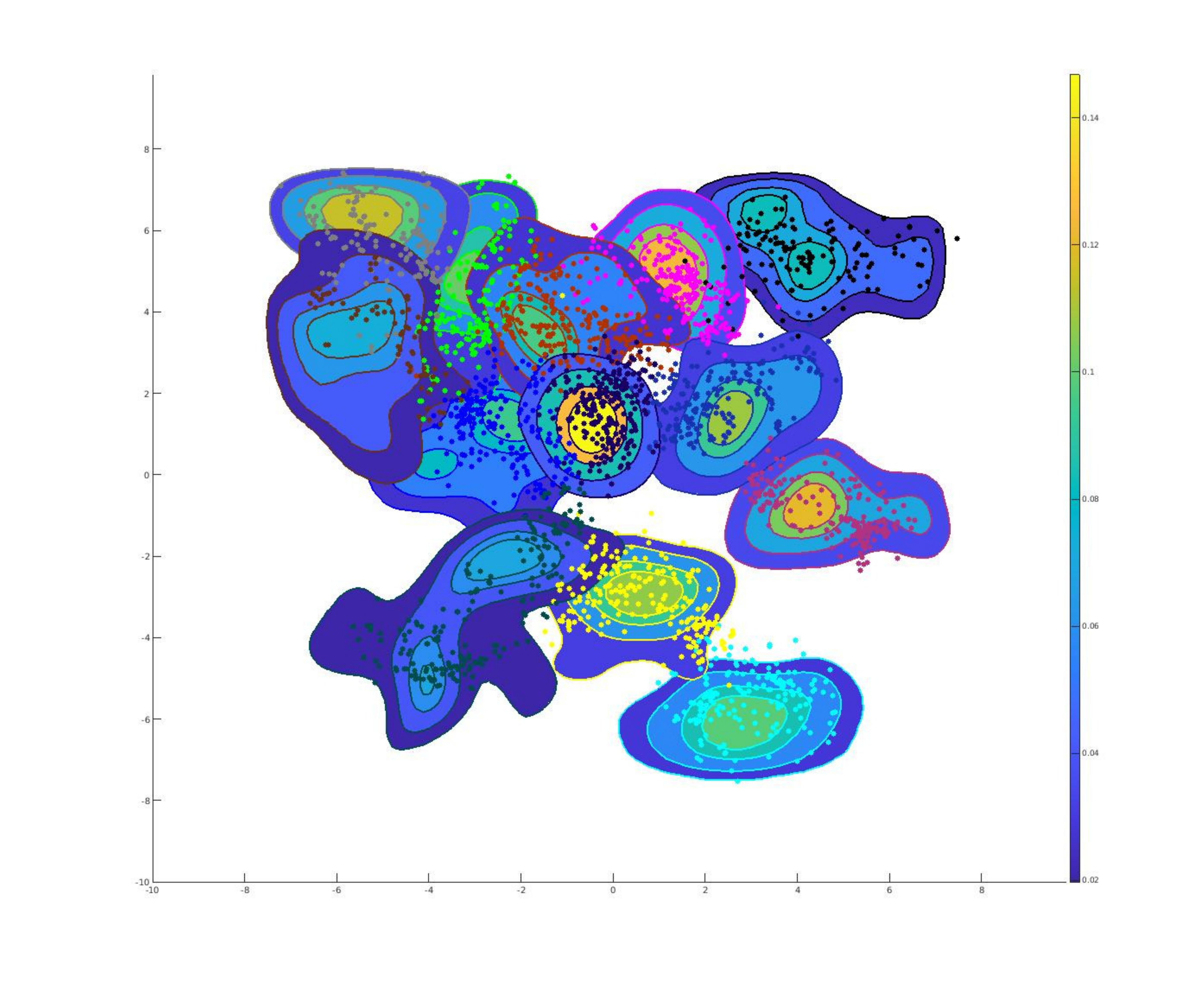}\label{fig:EquilSAND}}
 	\caption{Equilibrium solutions to \eqref{eq:System} with $b=.25$, $a=1$, $\eta=3$, $u_0(x)$ from Figure \ref{fig:KDE}, two different mollified environment types from Figure \ref{fig:Environment}, and the meerkat location data}
 	\label{fig:DataEquil}
 \end{figure}

 Figure \ref{fig:DataEquil} shows equilibrium solutions with two different environments, as well as the location data. The figures were both generated with the same parameter values, but Figure \ref{fig:EquilEDGE} was generated with the environment seen in Figure \ref{fig:EDGE} and Figure \ref{fig:EquilSAND} was generated with the environment seen in Figure \ref{fig:SAND}.  These different environments lead to markedly different territories. From visual inspection it is clear that for these parameter values, the EDGE data is a better predictor of the meerkat territories than the SAND data. It is worth noting that these equilibrium solutions found with starting points from location data were reached more quickly than those with a similar number of groups and a Gaussian environment. It is likely the detailed environment and initial data in already popular locations decreased the computation time.

\section{Data Incorporation}\label{sec:MLE}
\begin{figure}[h!]
	\centering
	\subfigure[Equilibrium solution to \eqref{eq:System}]{\includegraphics[width = .45\linewidth]{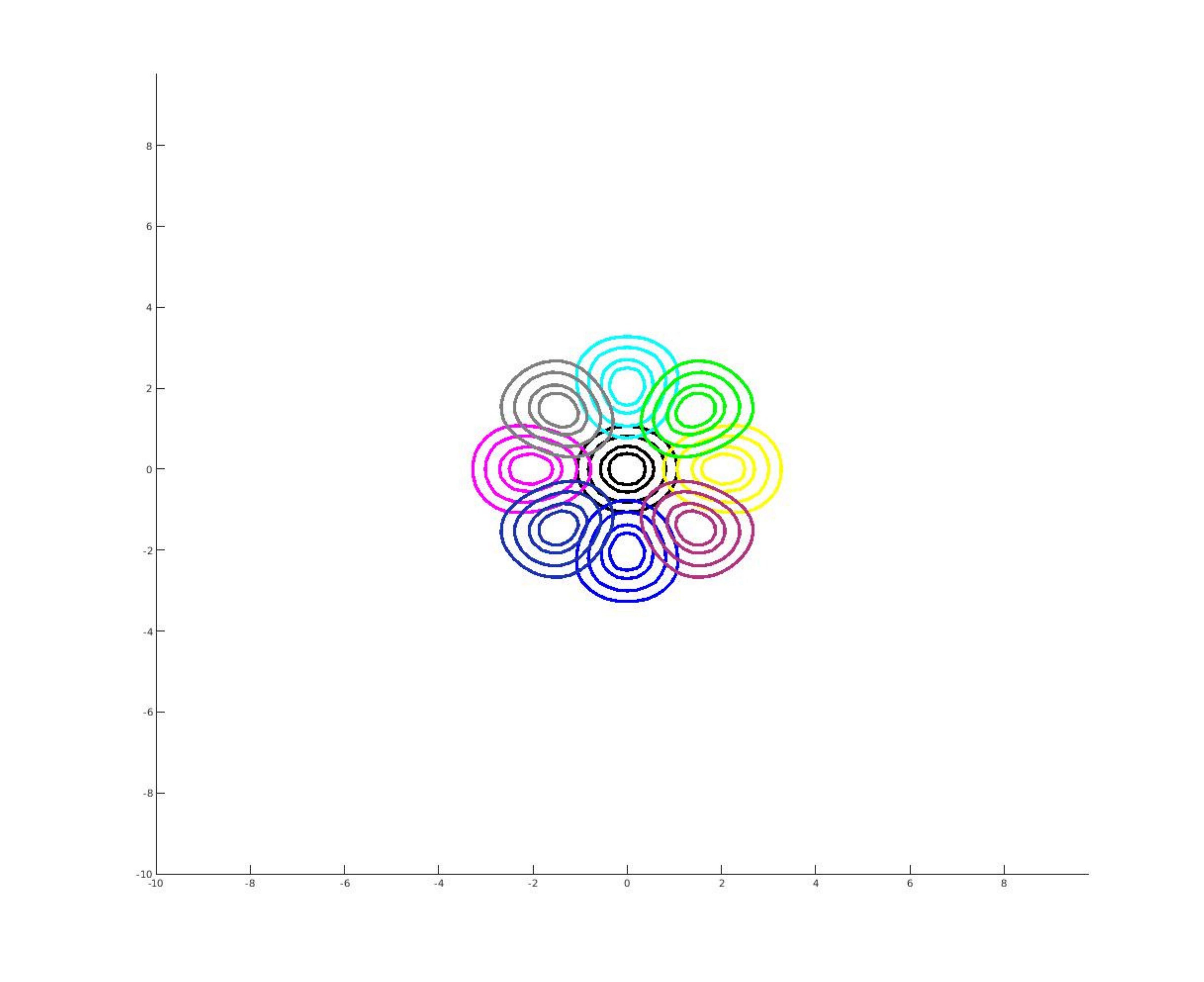} \label{fig:dataa}}
	\subfigure[Synthetic data generated from (a)]{\includegraphics[width = .45\linewidth]{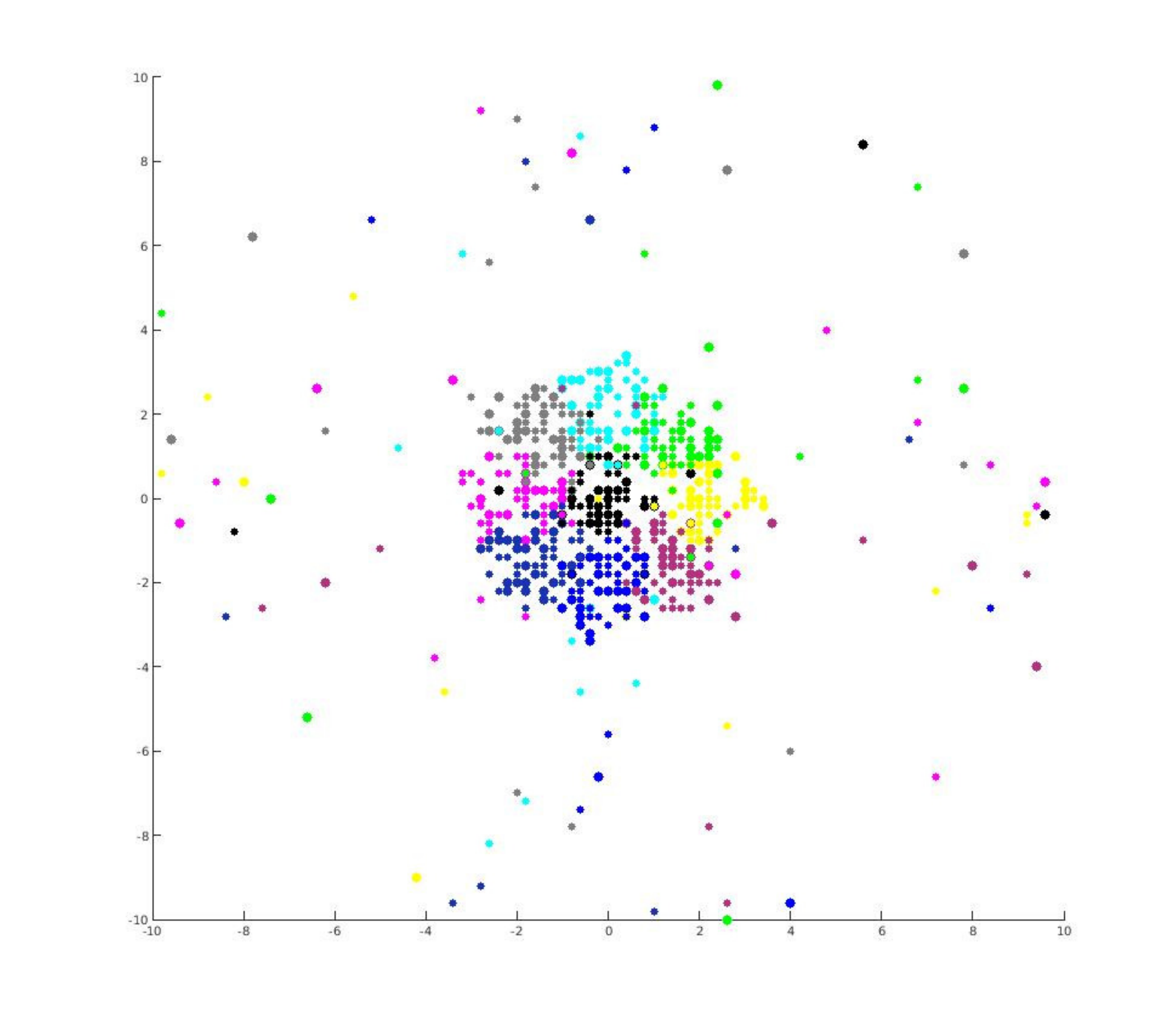} \label{fig:datab}}
	\caption{Equilibrium solution to system \eqref{eq:System} for $N=9$ and corresponding synthetic data generated from the equilibrium solution}
	\label{fig:syndata}
\end{figure}

With an efficient way to solve system \eqref{eq:System} for a large number of groups, we are able to use synthetic data obtained from equilibrium solutions to the model to explore parameter inferencing via maximum likelihood estimation. Figure \ref{fig:dataa} illustrates an equilibrium solution to system \eqref{eq:System} with nine interacting groups. Figure \ref{fig:datab} illustrates the synthetic data generated using the distribution provided by the solution in Figure \ref{fig:dataa}

Given the set of synthetic data with $M$ data points for each of $N$ groups, $\{y_i^m \vert i = 1,... N, m = 1, ... M\}$, we minimize the negative log-likelihood function over the set of parameters, $\theta$, 
\begin{align*}
	-\ell(y,\theta) = -\sum_{i=1}^N\sum_{m=1}^M\log\left(f(y^m_i\vert \theta)\right).
\end{align*}
The probability density functions, $f(\textbf{x} \vert \theta)$, are found using the methods described in Section \ref{sec:numerics_method} with initial condition determined by the kernel density estimation of the set of synthetic data. 
We use stochastic gradient descent (SGD) to minimize the negative log-likelihood function. SGD uses one or a few data points to update the parameter at each iteration. This leads to more variance in the update, thus allowing a possibility to move away from a local minimizer. We update the parameter as follows: 
\begin{align}
\theta &= \theta - \alpha \nabla_{\theta}\left(-\sum_{i=1}^N\log(f(y^m_i\vert \theta))\right) \label{eq:SGD},
\end{align}
where $m$ is a randomly chosen from the set $\{1, ... M\}$, and $\alpha$ is the learning parameter, which typically decreases in time. 

 \subsection{Data fitting with one parameter and one group}

\begin{figure}[ht]
 	\centering
 	\subfigure[Negative log-likelihood function]{\includegraphics[width=.45\linewidth]{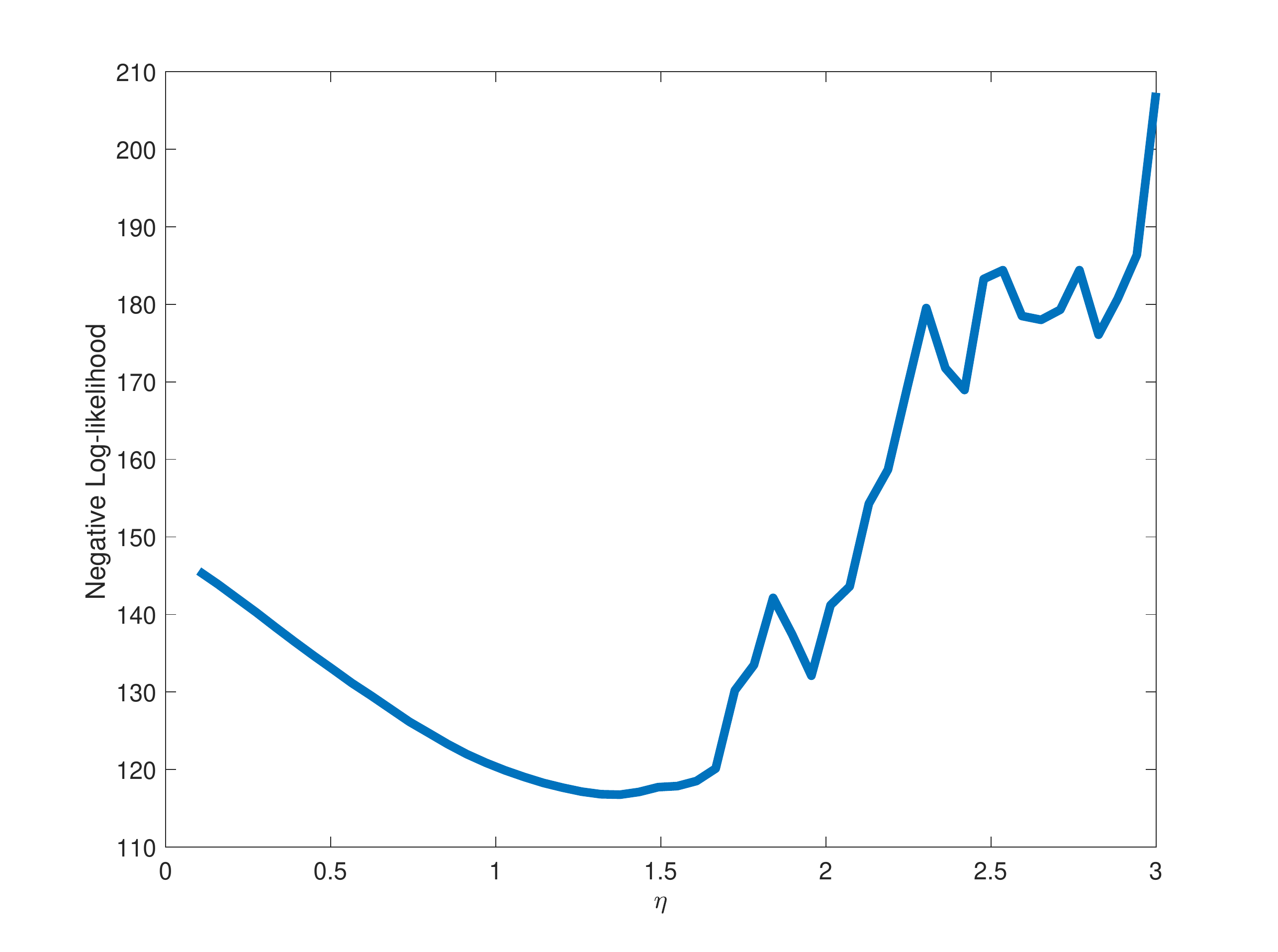}\label{fig:fitkLL}}
 	\subfigure[Iterations of SGD and corresponding estimated parameter (top) and negative log-likelihood value (bottom) ]{\includegraphics[width=.45\linewidth]{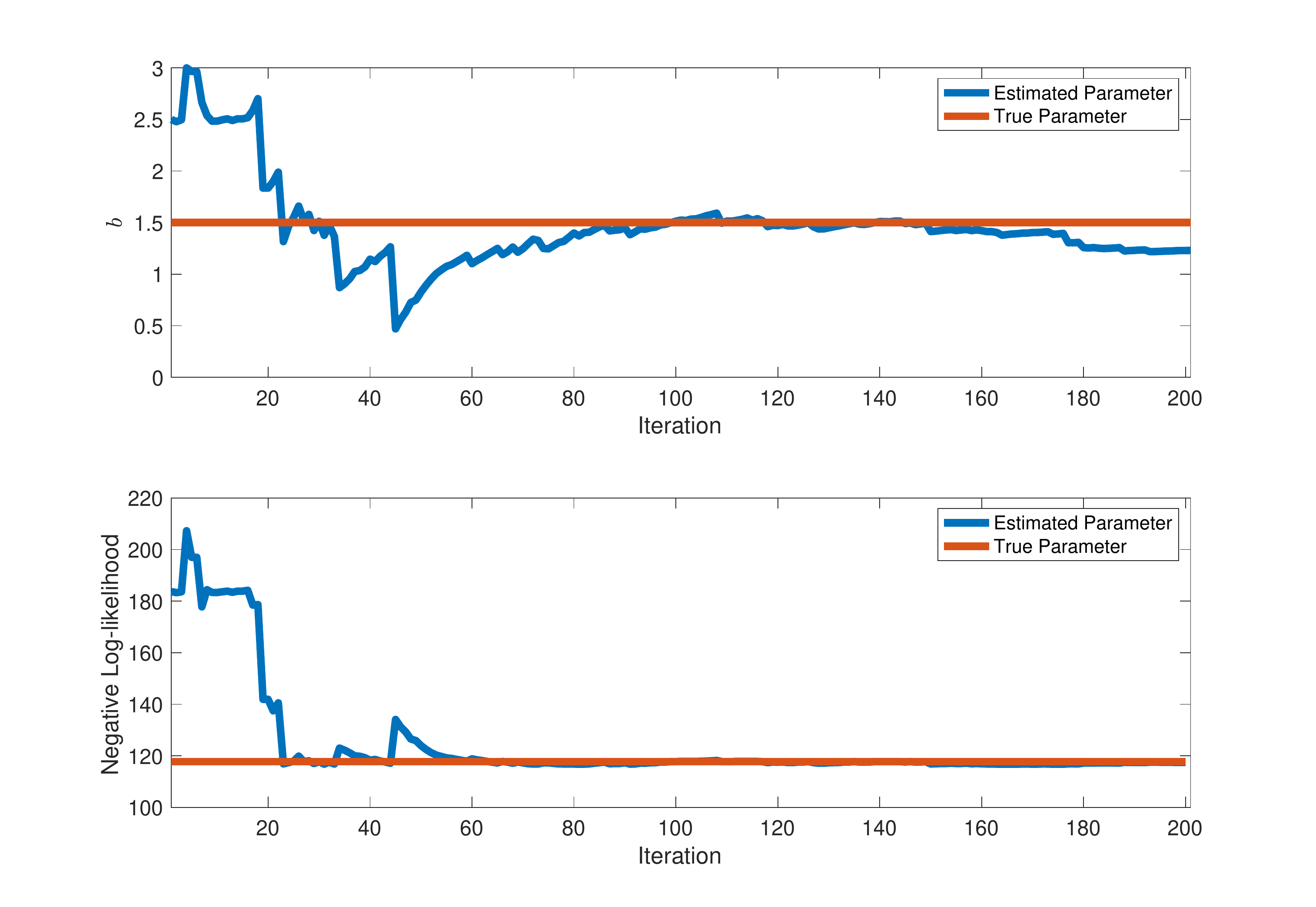}\label{fig:fitkit}}
 	\caption{ Iterations of SGD minimizing the negative log-likelihood function to estimate $b$ using data generated with $M=50$, $n=1$, $b = 1.5$, $a = .25$, $\eta =5$ }
 	\label{fig:fitk}
 \end{figure}

\begin{figure}[ht]
 	\centering
 	\subfigure[Negative log-likelihood function]{\includegraphics[width=.45\linewidth]{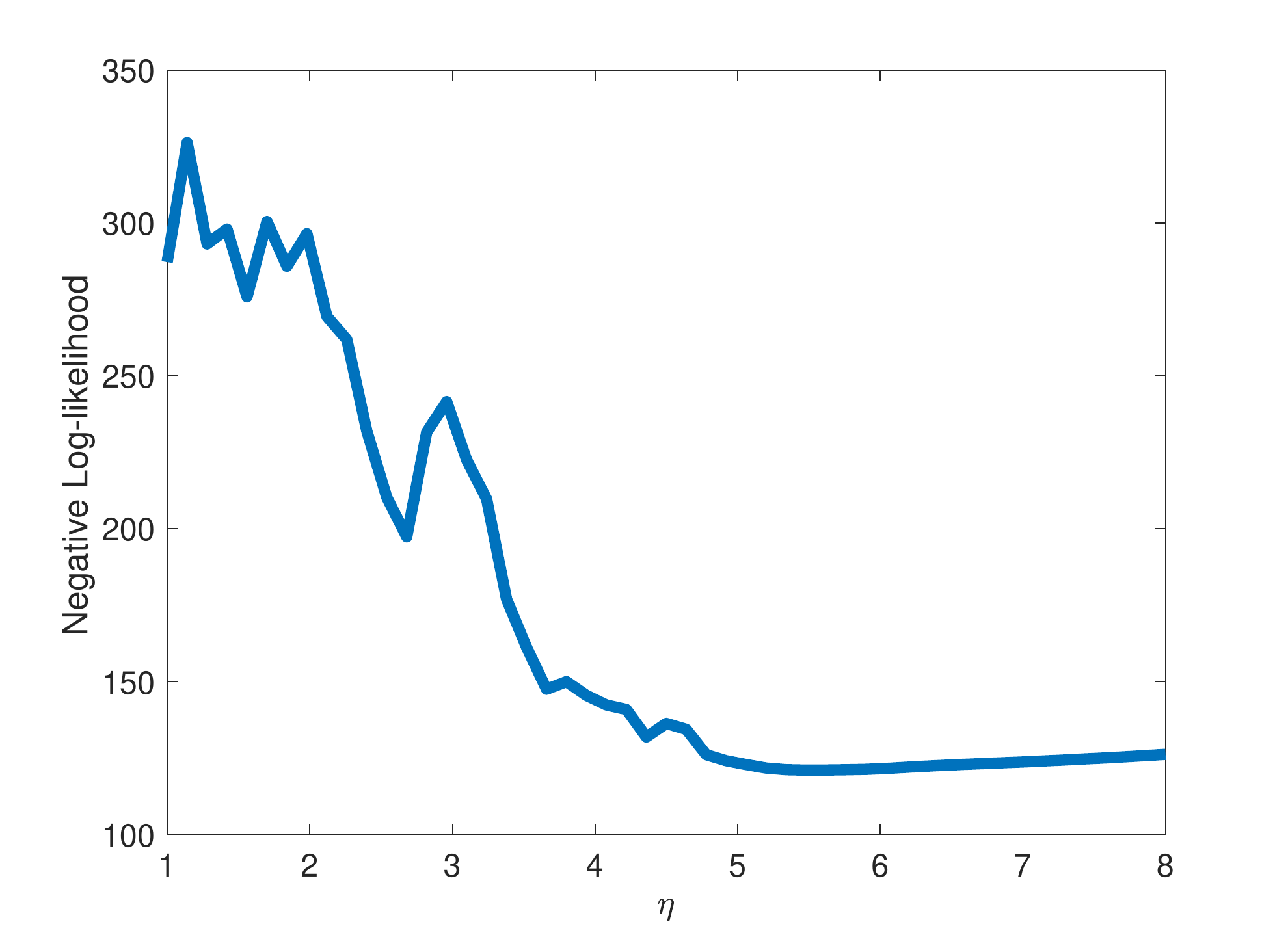}\label{fig:fitetaLL}}
 	\subfigure[Iterations of SGD and corresponding estimated parameter (top) and negative log-likelihood value (bottom) ]{\includegraphics[width=.45\linewidth]{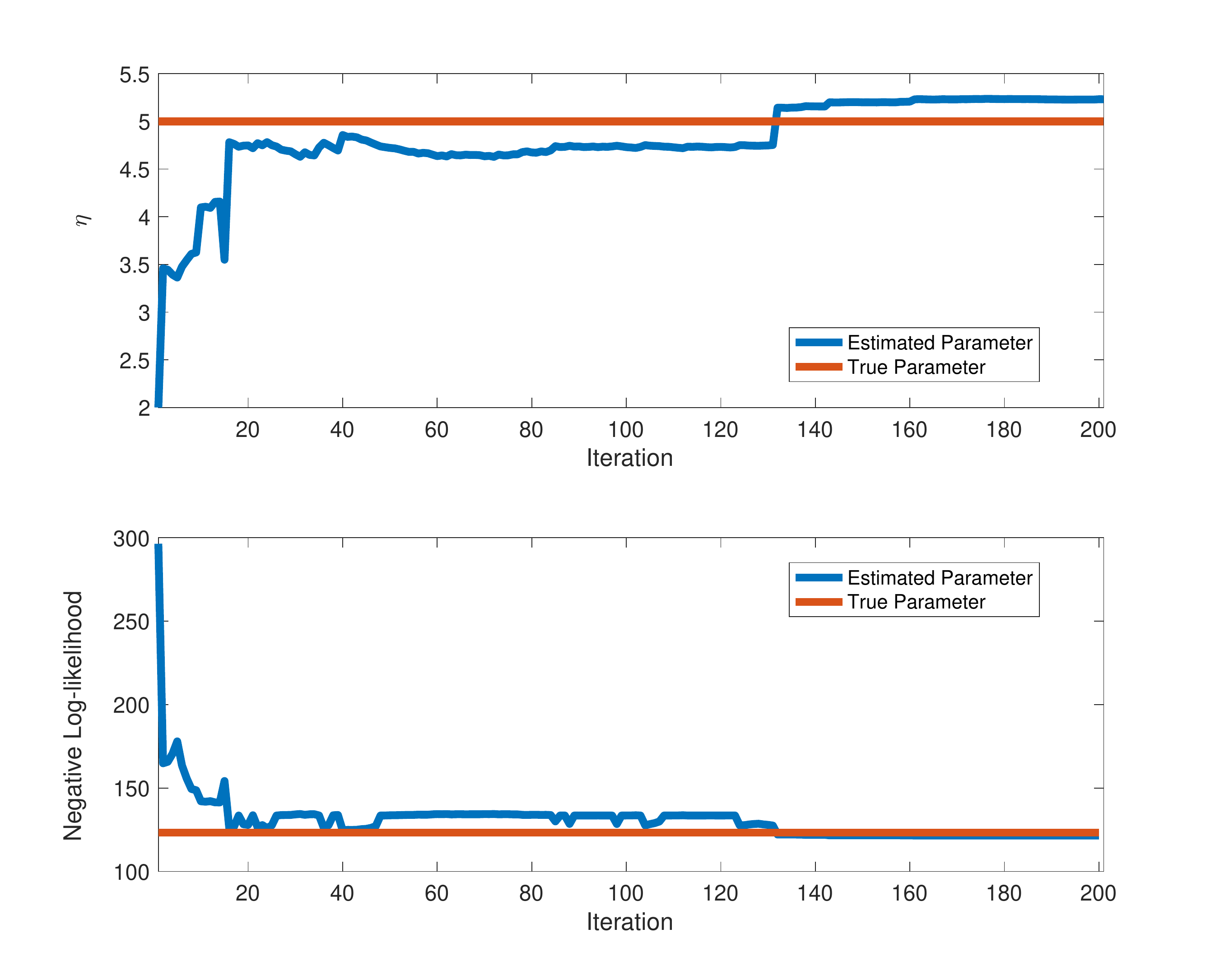}\label{fig:fitetait}}
 	\caption{Iterations of SGD minimizing the negative log-likelihood function to estimate $\eta$ using data generated with $M=50$ , $\eta = 5$, $b = 1.5$, $a = .25$}
 	\label{fig:fiteta}
 \end{figure}

  Figure \ref{fig:fitk} and Figure \ref{fig:fiteta} demonstrate two examples of fitting a single parameter to sythentic data for one group using the methods above. We fit $b$ and $\eta$, respectively. Figure \ref{fig:fitkLL} and Figure \ref{fig:fitetaLL} illustrate the negative log-likelihood functions. Note that the shallower slopes in the negative log-likelihood functions occur in the parameter regimes that lead to larger territories. There is a greater expense for data points lying outside of a territory due to the steepness of the log function near zero. This observation can serve as information to keep in mind when choosing starting estimates for parameters; it is possible choosing a starting point on a steeper gradient can lead to faster convergence. In Figure \ref{fig:fitk}, 
 %$\eta=5$, $a =.25$, and the number of data points generated was $50$. 
 %The step size for iteration $m$ was $\alpha(m) = .01(.9)^m$. 
% We see in Figure \ref{fig:fitkLL} 
 there are local minimizers in the negative log-likelihood function that the algorithm successfully avoids and converges to the minimum of the function. 
 %$b = 1.5$, $a = .25$, and the number of data points generated was $50$. The step size for iteration $m$ was $.1(.9)^m$. Thus, when we fit $\eta$, we found a larger step size to be appropriate. 
Similarly, we see local minimums in Figure \ref{fig:fitetaLL}, and the algorithm converges to the minimum of the log-likelihood function, shown in Figure \ref{fig:fitetait}. We find a larger step size is appropriate when fitting $\eta$ than when fitting $b$. In addition, as illustrated in Figure \ref{fig:fitetaLL}, the log-likelihood function for $\eta$ often had shallower slopes and a less-pronounced minimum than the log-likelihood function for $b$, possibly making it difficult to minimize. 
 
 \begin{figure}[ht]
	\centering
	\subfigure[$M=50$]{\includegraphics[width = .3\linewidth]{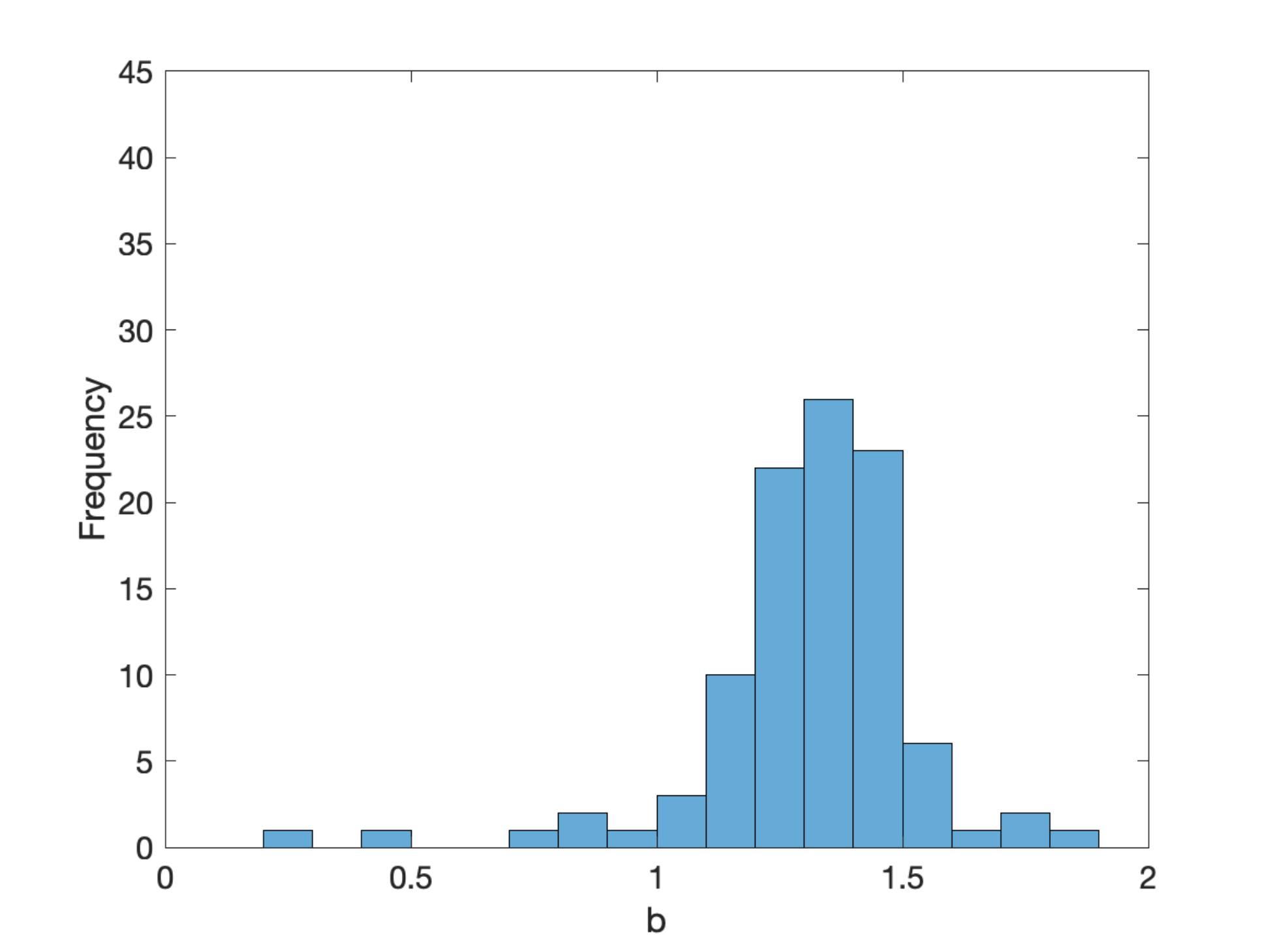}}
	\subfigure[$M=100$]{\includegraphics[width = .3\linewidth]{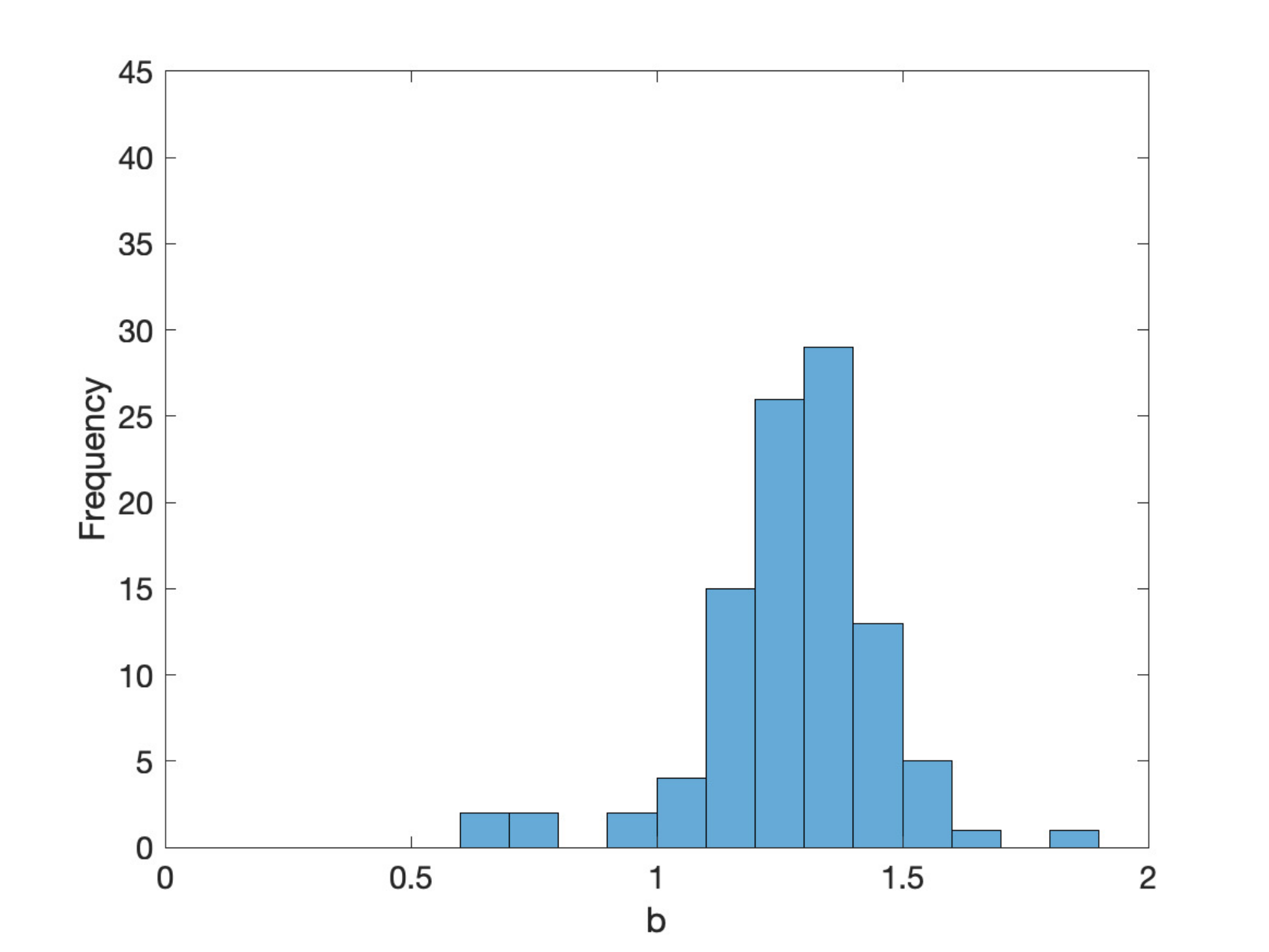}}
	\subfigure[$M=250$]{\includegraphics[width = .3\linewidth]{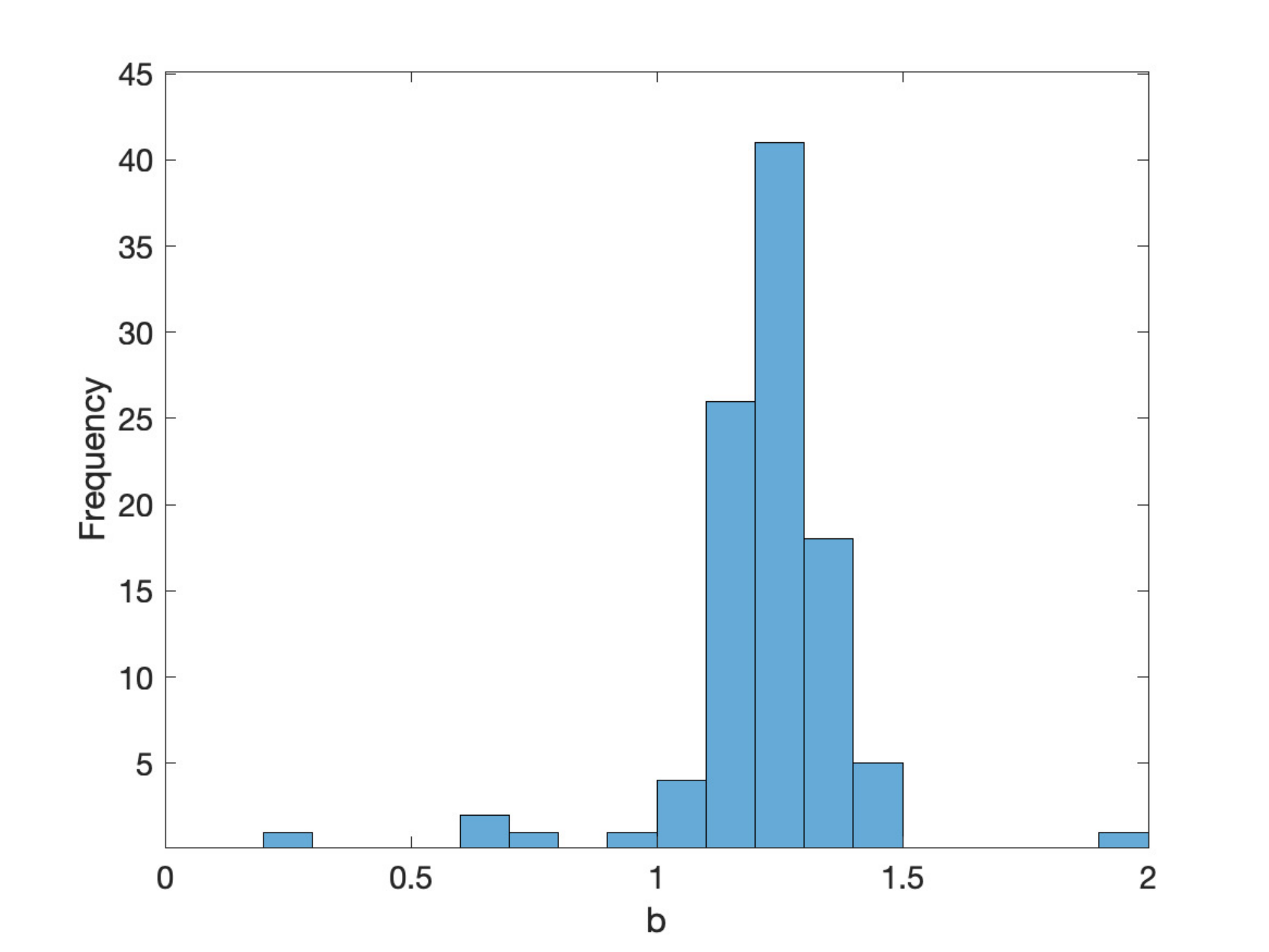}}
	\caption{Histogram of estimated $b$ values from 100 trials of SGD for various values of $M$, with data generated from $b=1.25$, $a=.25$, $\eta=5$}
	\label{fig:bHist}
\end{figure}

\begin{figure}[h]
 	\centering
 	%\subfigure[]
 	\subfigure[$M=50$]{\includegraphics[width = .3\linewidth]{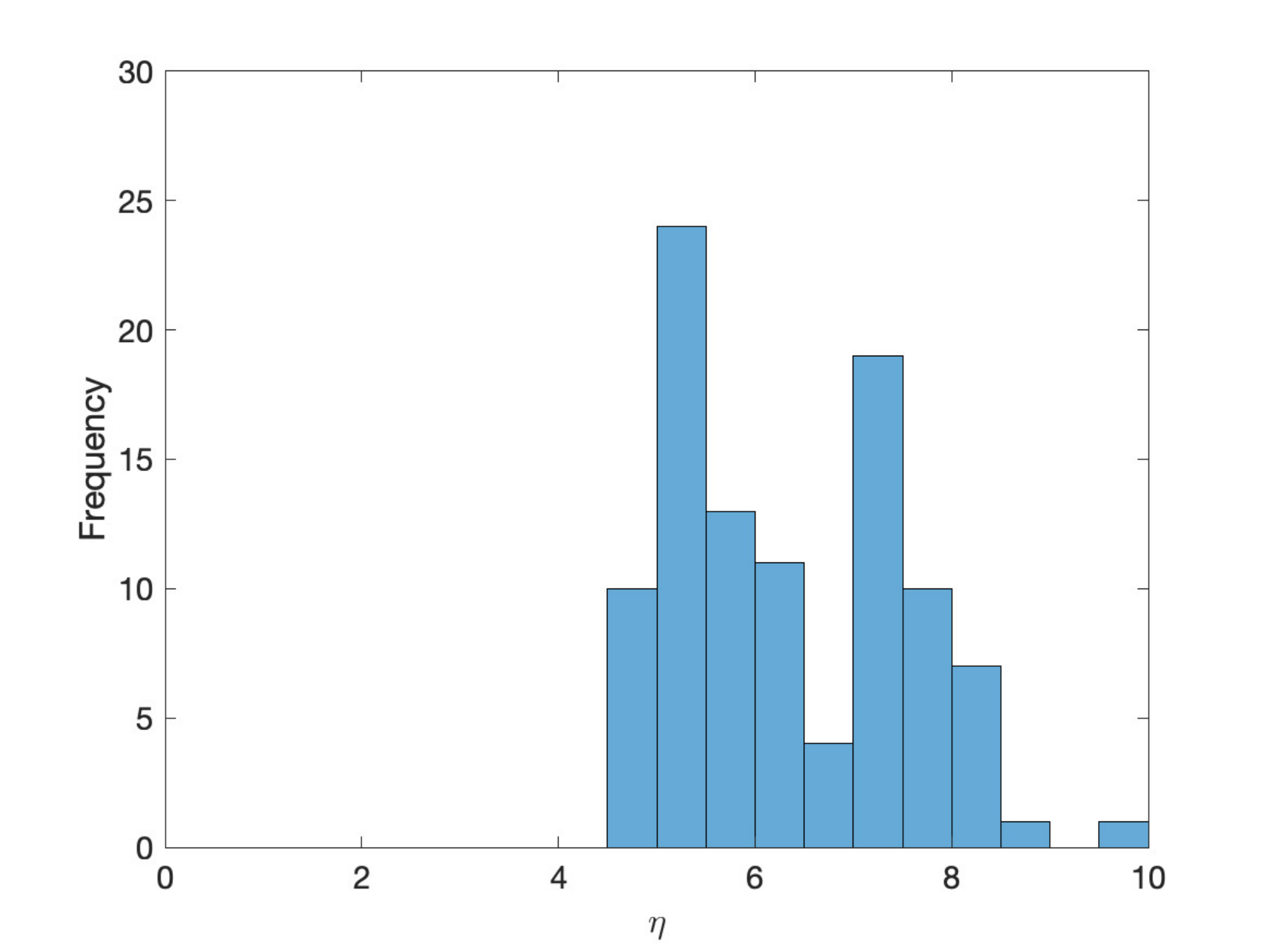}}
	\subfigure[$M=100$]{\includegraphics[width = .3\linewidth]{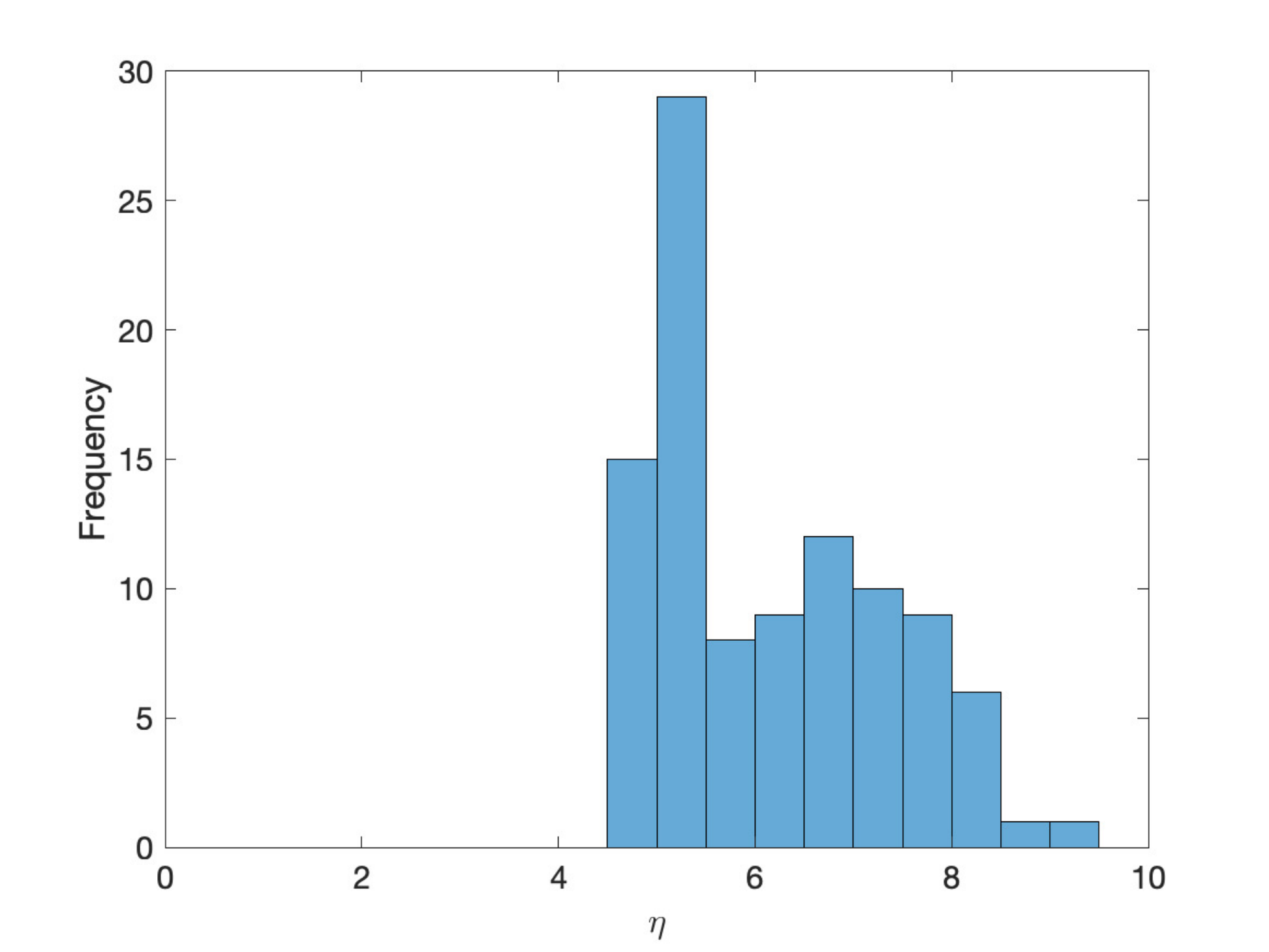}}
	\subfigure[$M=250$]{\includegraphics[width = .3\linewidth]{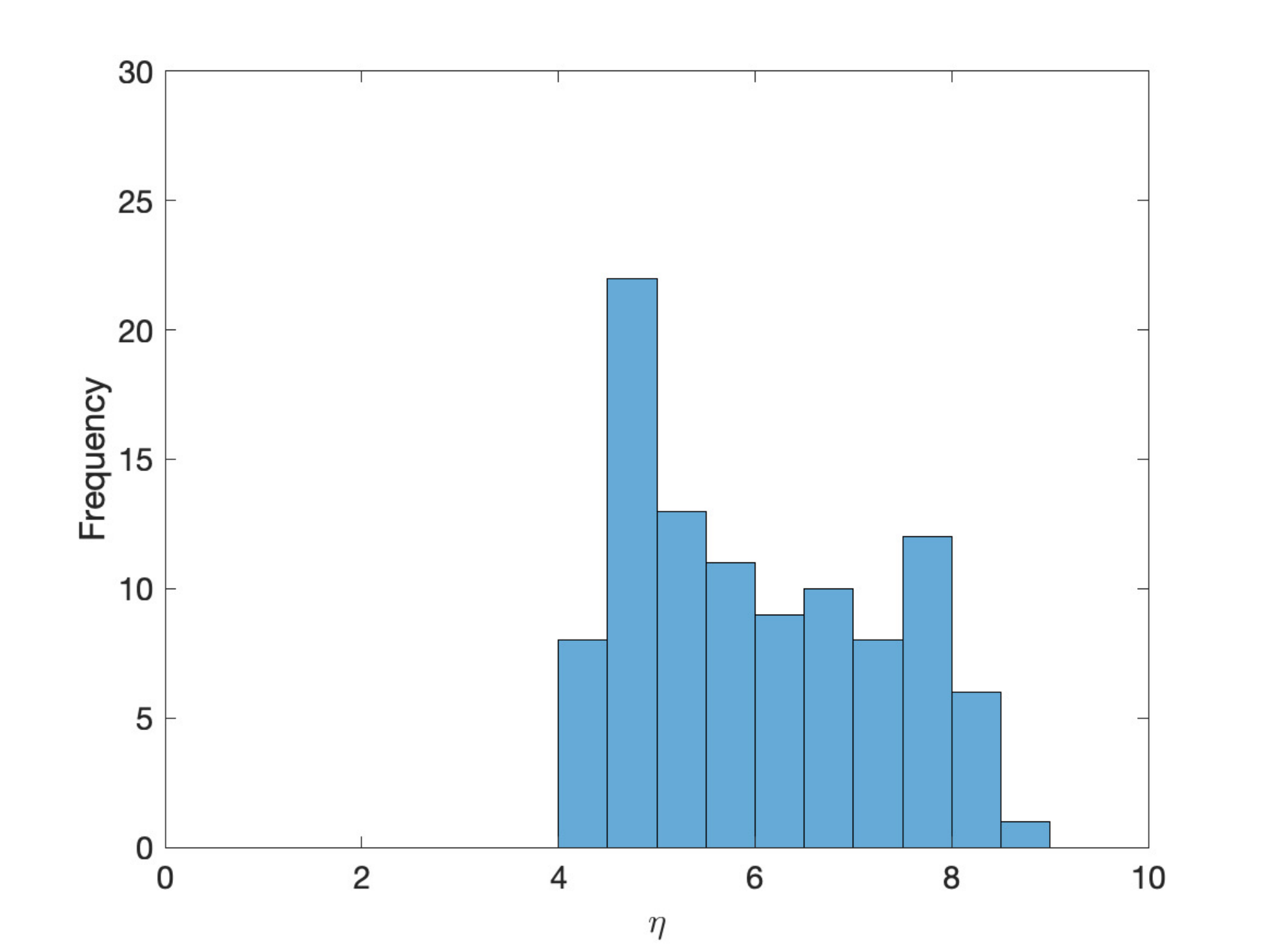}}
 	%\subfigure[Negative log-likelihood function]{\includegraphics[width = .25 \linewidth, height = 1.25in]{Histogram/OneParamEta/d250eta5.jpg}}
 	\caption{Histogram of estimated $\eta$ values from 100 trials of SGD for various values of $M$,with data generated from $b=1$, $a=.25$, $\eta=5$}
 	\label{fig:etaHist}
 \end{figure}

 In the interest of determining how often the algorithm converges to the correct parameters as well as the effect of the size of the data set, we completed 100 trials of maximum likelihood estimation to fit $b$ to data sets generated from $b=1.25$ for three different sizes of data sets, $M = 50$, $M = 100$, and $M  = 250$. The results are displayed in Figure \ref{fig:bHist}. A large majority of trials fall close to the true parameter value $1.25$, with fewer falling further away. This pattern gets more pronounced as the size of the data set increases. Figure \ref{fig:etaHist} demonstrates the analogous simulation for fitting $\eta$, and there is wider range of predicted parameters. Thus, we find that the log-likelihood function for $\eta$ is not as sensitive to parameter changes, particularly for parameters larger than the true parameter. As discussed above, the negative log-likelihood function for $\eta$ could lend insight to this. See Figure \ref{fig:fitetaLL}. The gradient is much shallower for values above $\eta = 5$ and there is a small difference in log-likelihood values  for $\eta$ values between five and eight.

\subsection{Data fitting with multiple parameters and multiple groups}

In this section, we fit both one and two parameters to the general multi-species model. As we increase the number of parameters and number of groups, we can no longer efficiently generate the log-likelihood function;
 thus, we must depend on stochastic gradient descent to minimize the log-likelihood function.  
 
\begin{figure}[h]
	\centering
	\noindent
	\begin{minipage}[c]{0.62\textwidth}
		\subfigure[Iterations of SGD and corresponding estimated parameter (top) and negative log-likelihood value (bottom) ]{\includegraphics[width=\linewidth]{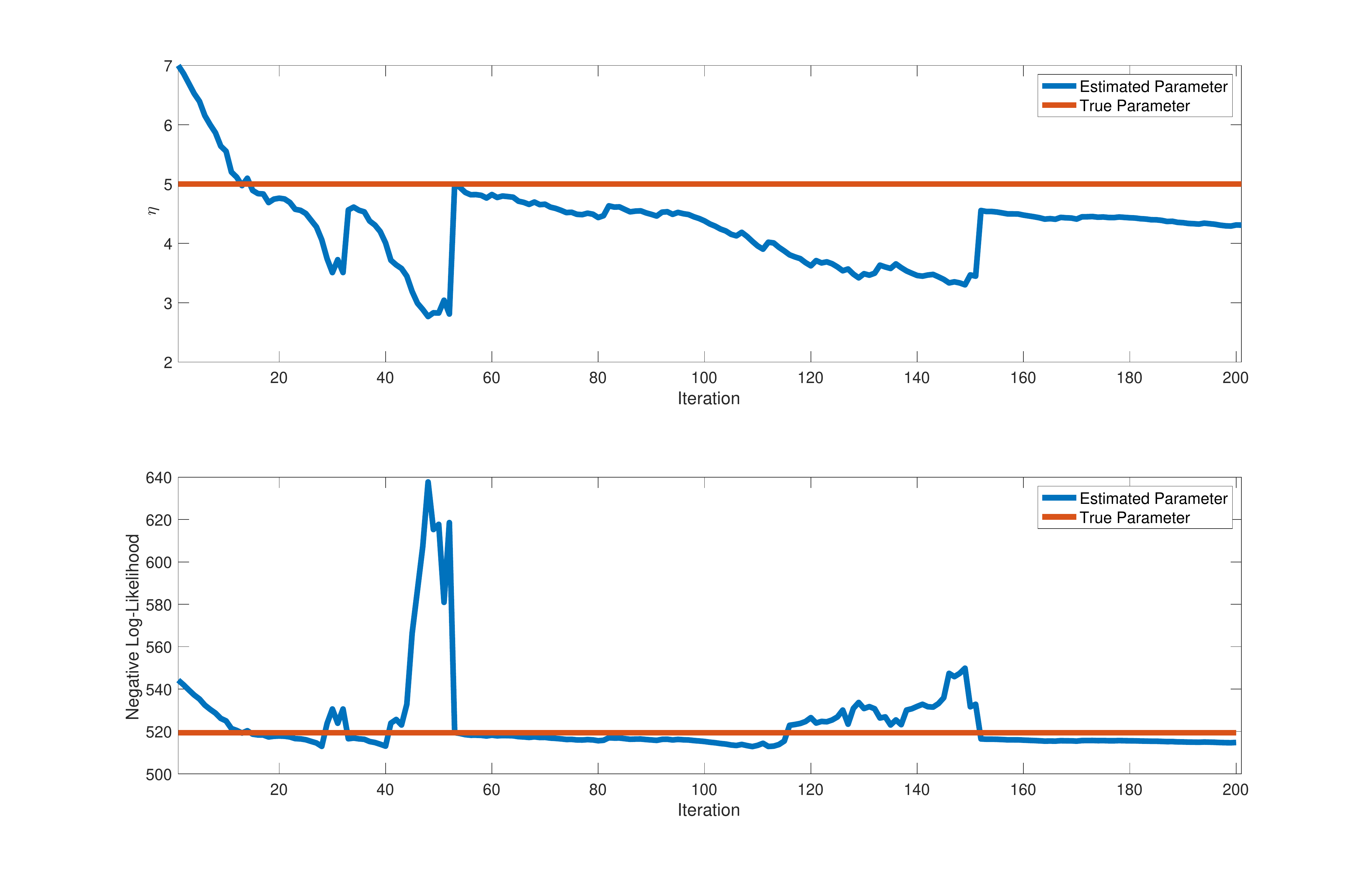}\label{fig:fitetan2SGD}}\\
\end{minipage} % no space if you would like to put them side by side
\begin{minipage}[c]{0.22\textwidth}
		\subfigure[ $\eta = 4.3$]{\includegraphics[width=\linewidth]{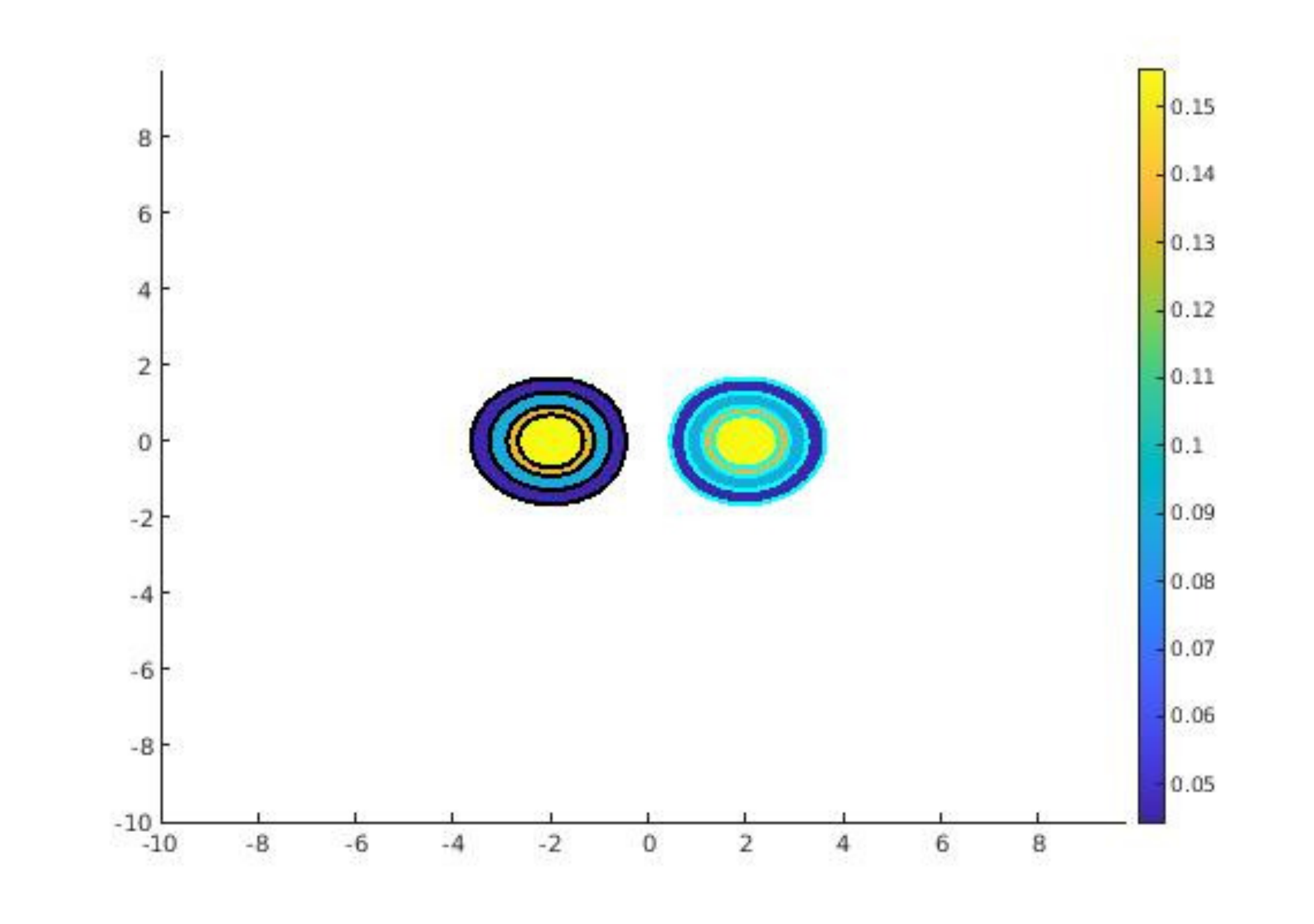}\label{fig:fitetan21}}\\
		\subfigure[$\eta=5$]{\includegraphics[width=\linewidth]{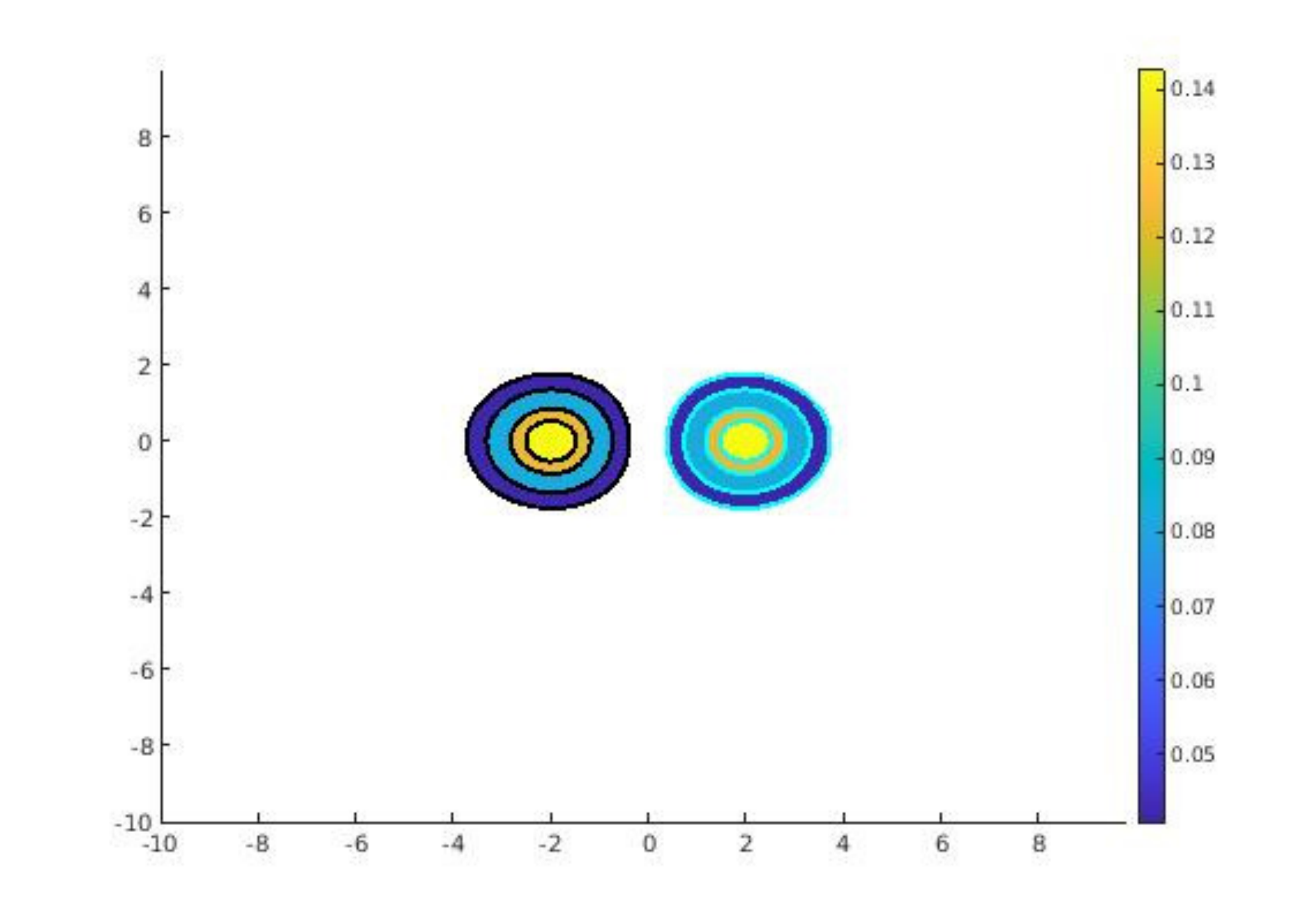}\label{fig:fitetan22}}
\end{minipage}
	\caption{(a) SGD minimizing the negative log-likelihood function to estimate $\eta$ using data generated with $M = 100$, $n=2$, $b=1$, $a = .25$, and $\eta=5$, and (b-c) the equilibrium solutions to system \eqref{eq:System} using the estimated parameter and the parameter used to generate the data, respectively }
	\label{fig:fitetan2}
\end{figure}

 In Figure \ref{fig:fitetan2SGD}, we estimate $\eta$ from data generated with $\eta = 5$, $a=.25$, and $N=2$. In this case, the algorithm does not converge to the parameter the data was generated with, but converges to a parameter value that has a lower log-likelihood value. It is worth looking at the predicted territories from both of these parameters. Figure \ref{fig:fitetan21} shows the equilibrium solution found using the parameters predicted from SGD, and Figure \ref{fig:fitetan2SGD} shows the equilibrium solution found using the parameters used to generate the data. The territories are slightly different, but the territory boundaries and the maximum values are close. The algorithm still predicts a similar territory to the territory that the data came from. %% not done here

\begin{figure}[h]
	\centering
	\noindent
\begin{minipage}[c]{0.45\linewidth}
	\subfigure[Iterations of SGD and corresponding estimated parameters (top, middle) and negative log-likelihood value (bottom) ]{\includegraphics[width=\linewidth]{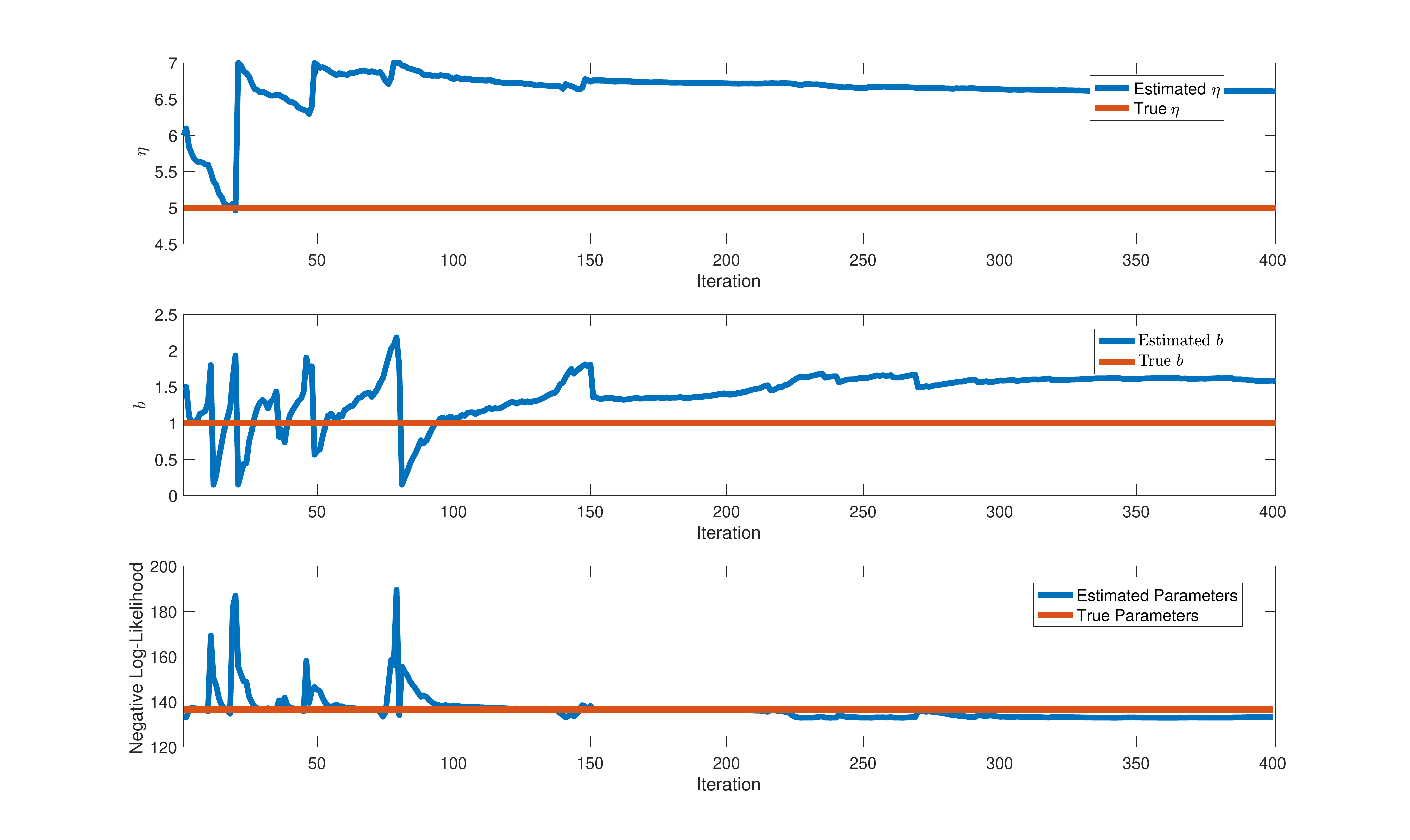}\label{fig:SGDnofit1}}
\end{minipage} % no space if you would like to put them side by side
\hspace{1pt}
\begin{minipage}[c]{0.45\linewidth}
	\subfigure[Iterations of SGD and corresponding estimated parameters (top, middle) and negative log-likelihood value (bottom) ]{\includegraphics[width=\linewidth]{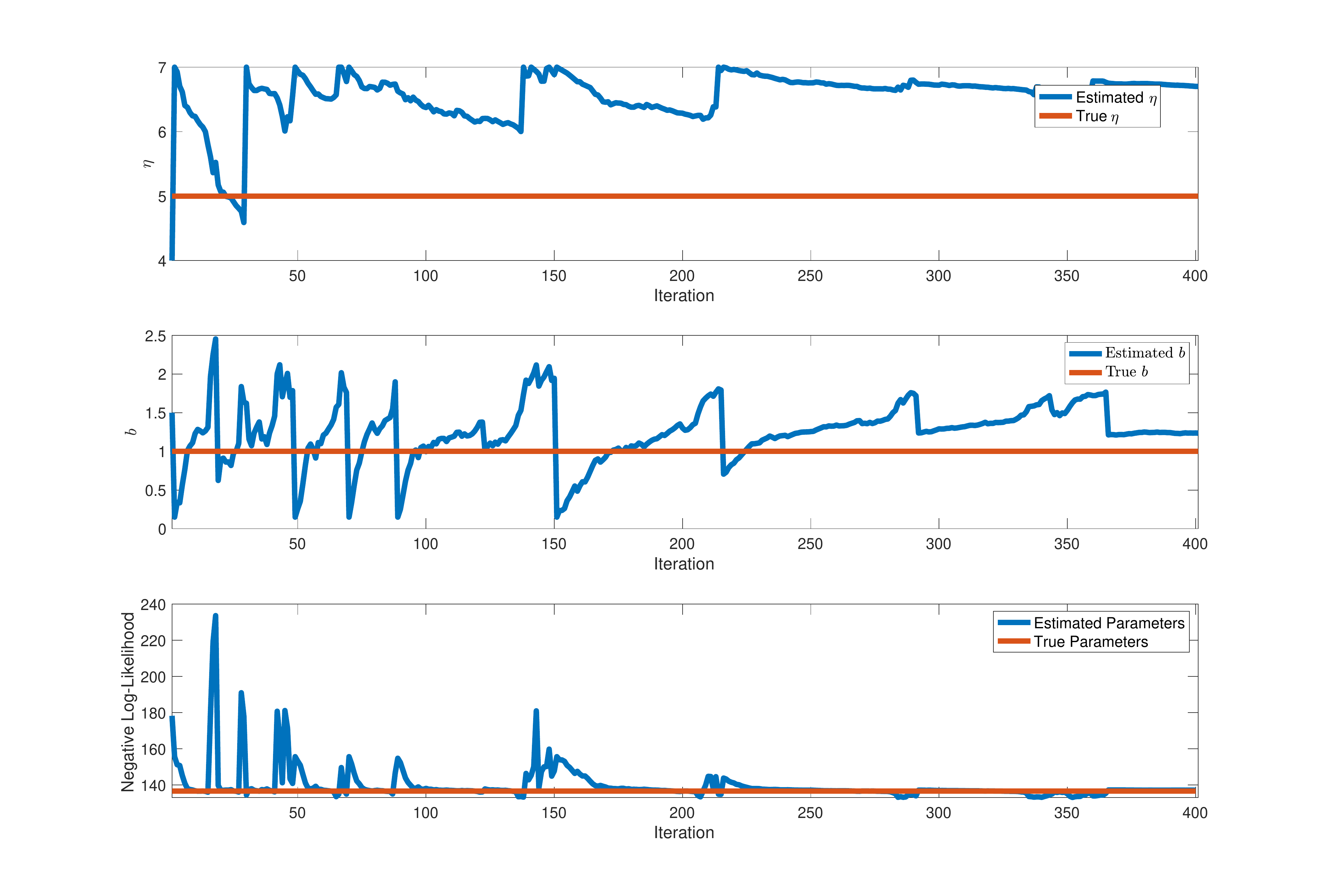}\label{fig:SGDnofit2}}\\
\end{minipage}
\subfigure[ $b=1$, $\eta = 5$]{\includegraphics[width=.3\linewidth]{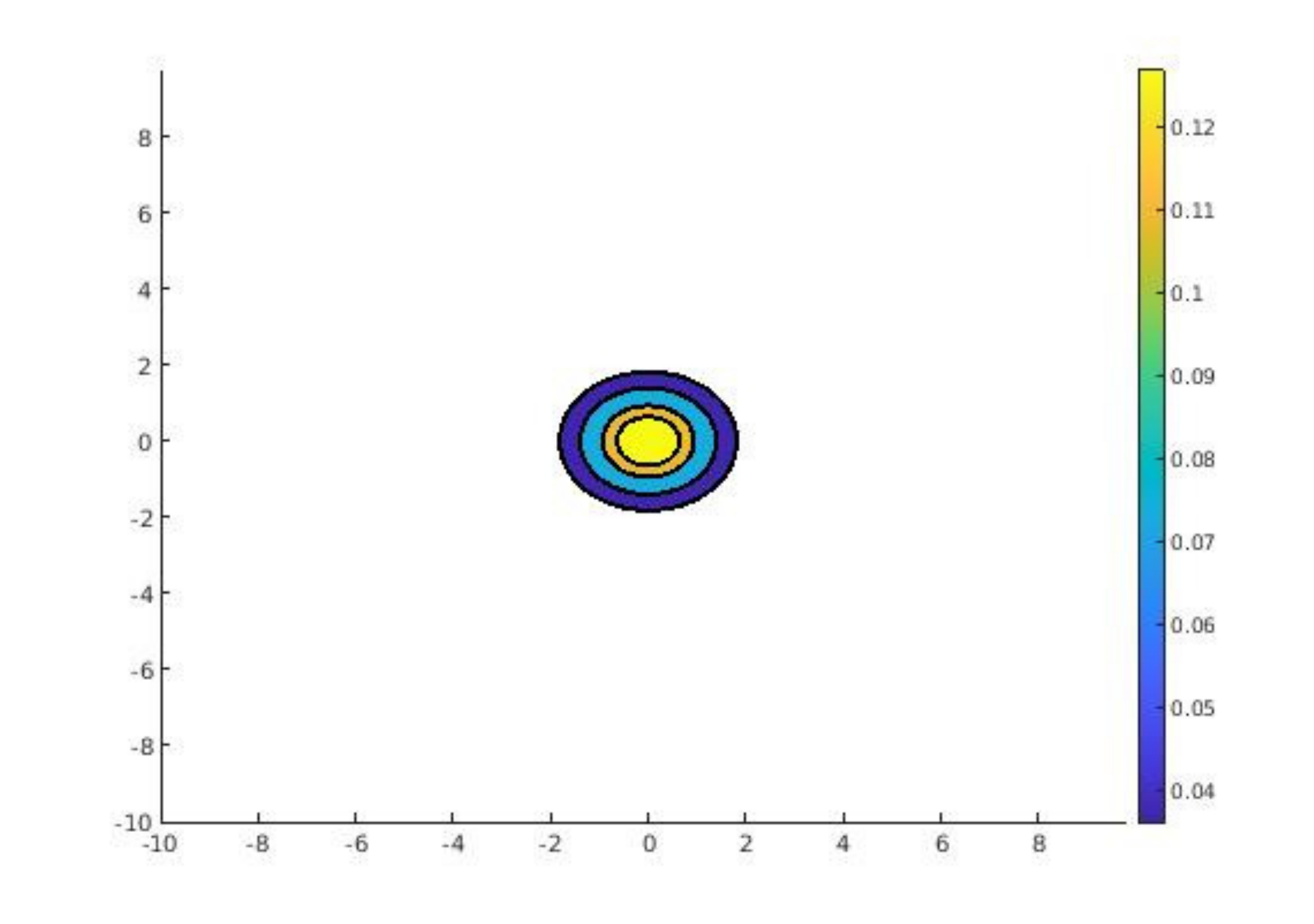}\label{fig:nofitTrue}}
\subfigure[$b=1.6$, $\eta = 6.6$]{\includegraphics[width=.3\linewidth]{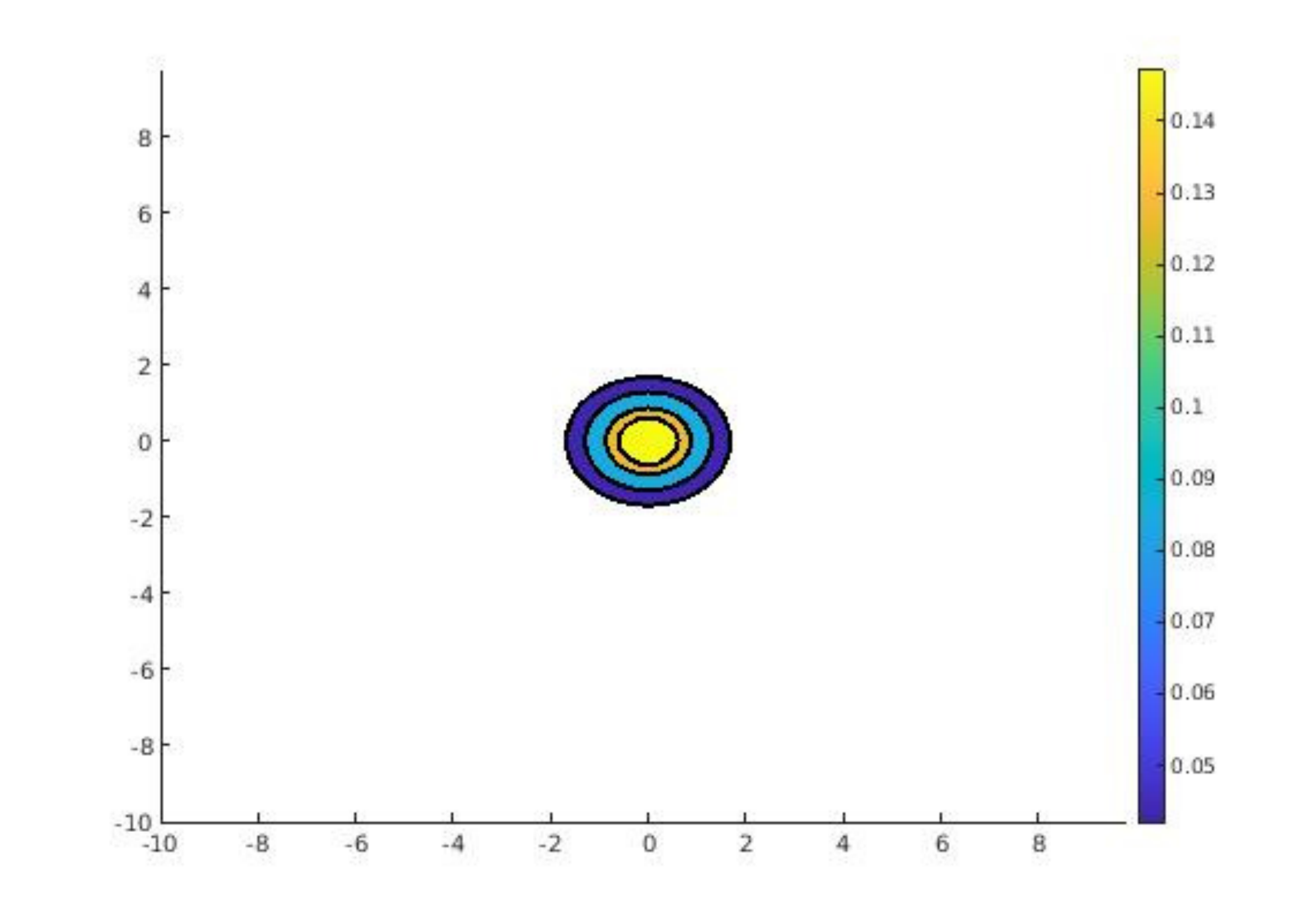}\label{fig:nofit1}}
\subfigure[$b=1.25$, $\eta = 6.7$]{\includegraphics[width=.3\linewidth]{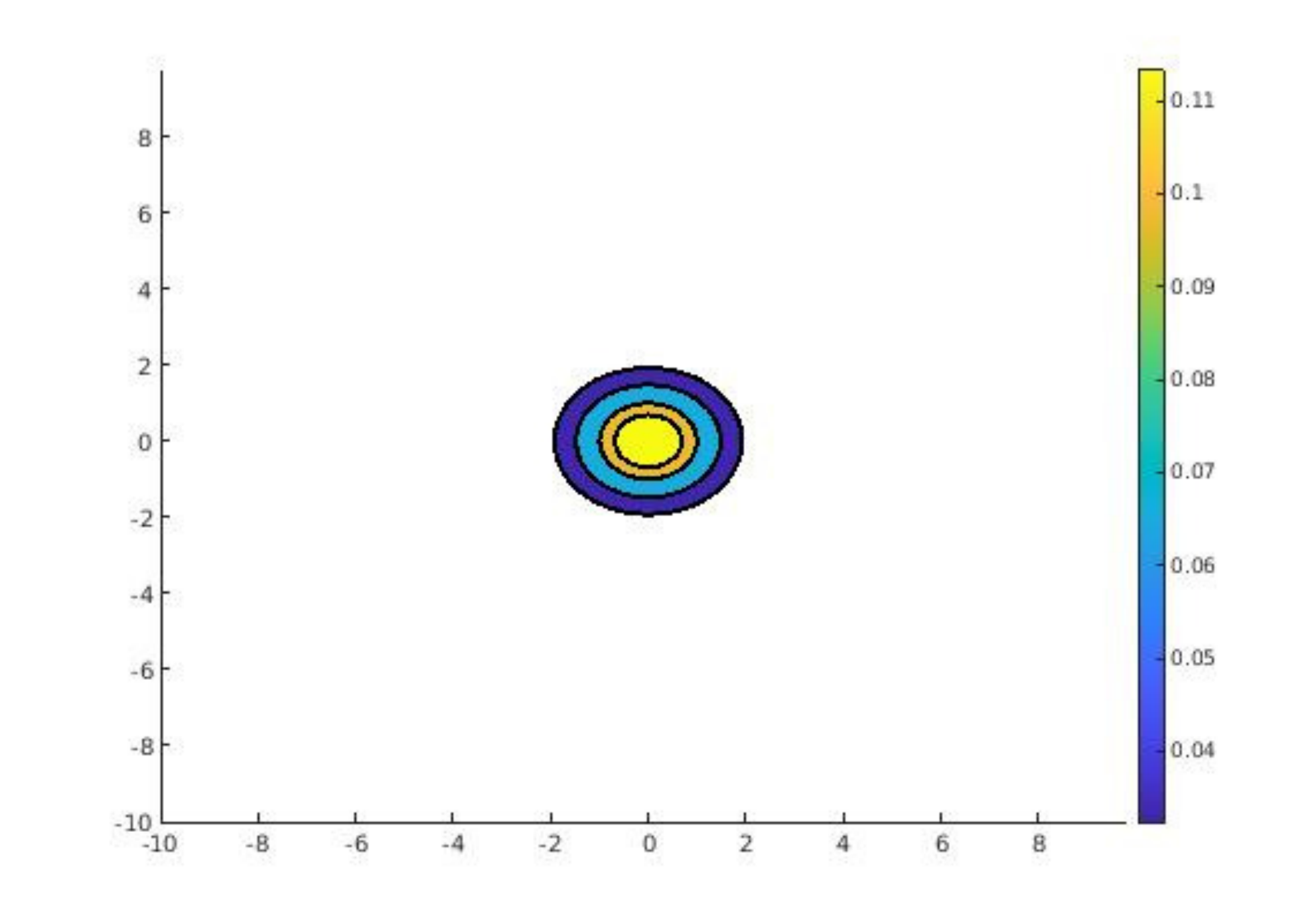}\label{fig:nofit2}}
	\caption{ (a-b) Two trials of SGD minimizing the negative log-likelihood function to estimate $\eta$ and $b$ using data generated with $\eta = 5$, $b = 1$, $a=.25$, and $M=100$, (c) equilibrium solutions to system \eqref{eq:System} with the true parameters, and (d-e) the predicted parameters}
	\label{fig:nofit}
\end{figure}

When we fit both $b$ and $\eta$ to the model, we are interested in the ratio of the parameters. This is due to the fact that as $b$ increases, a territory is more aggregated, but as $\eta$ increases, the territory spreads out. Therefore, overestimating or underestimating both parameters balance out and leads to a similar territory.
%There are cases where stochastic gradient descent does not converge to the parameters that were used to generate the data, but to parameters with a similar or lower log-likelihood value. 
In Figure \ref{fig:nofit}, we fit the model to two different data sets generated with $\eta=5$ and $b=1.5$, where $\eta/b \approx 3.333$. 
%There are two trials of stochastic gradient descent with different sets of data generated with $b=1$ and $\eta = 5$ 
SGD was used to fit both data sets to the model, and the results are shown in in Figure \ref{fig:SGDnofit1} and Figure \ref{fig:SGDnofit2}. In Figure \ref{fig:SGDnofit1}, the algorithm converges approximately to $\eta = 6.6$ and $b = 1.6$, where $\eta/b = 4.125$. In Figure \ref{fig:SGDnofit2}, the algorithm converges approximately to $\eta = 6.7$ and $b=1.25$, where $\eta/b=5.36$. The ratios the algorithm found are larger than the ratio of parameters the data was generated with, however they have a similar log-likelihood value to the true parameters. The equilibrium solutions to system \eqref{eq:System} with the parameters used to generate data and the two sets of estimated parameters are shown in Figures \ref{fig:nofitTrue}, \ref{fig:nofit1}, and \ref{fig:nofit2}.  These three sets of parameters predict similar territories, and all have a maximum value between $.11$ and $.14$. The equilibrium solution that is most different from the equilibrium solution found using the parameters the data was generated from is that in Figure \ref{fig:nofit2}, which has the ratio of parameters, $\eta/b$, furthest away from the true ratio. 
%While the algorithm converges to different parameters, the parameters found still predict a similar territory. 

\begin{figure}[h]
	\centering
	\subfigure[$M=50$]{\includegraphics[width = .3\linewidth]{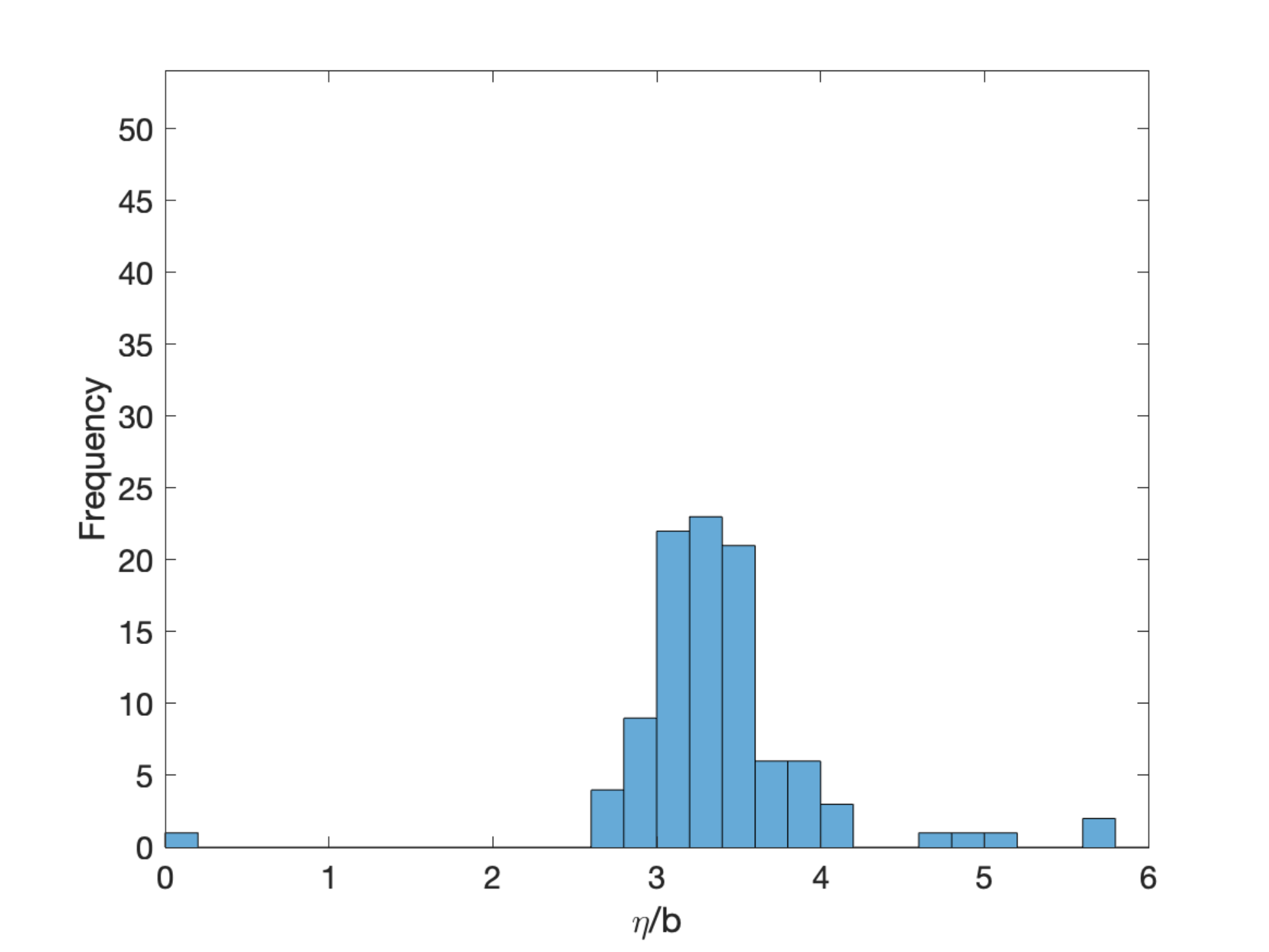}}
	\subfigure[$M=100$]{\includegraphics[width = .3\linewidth]{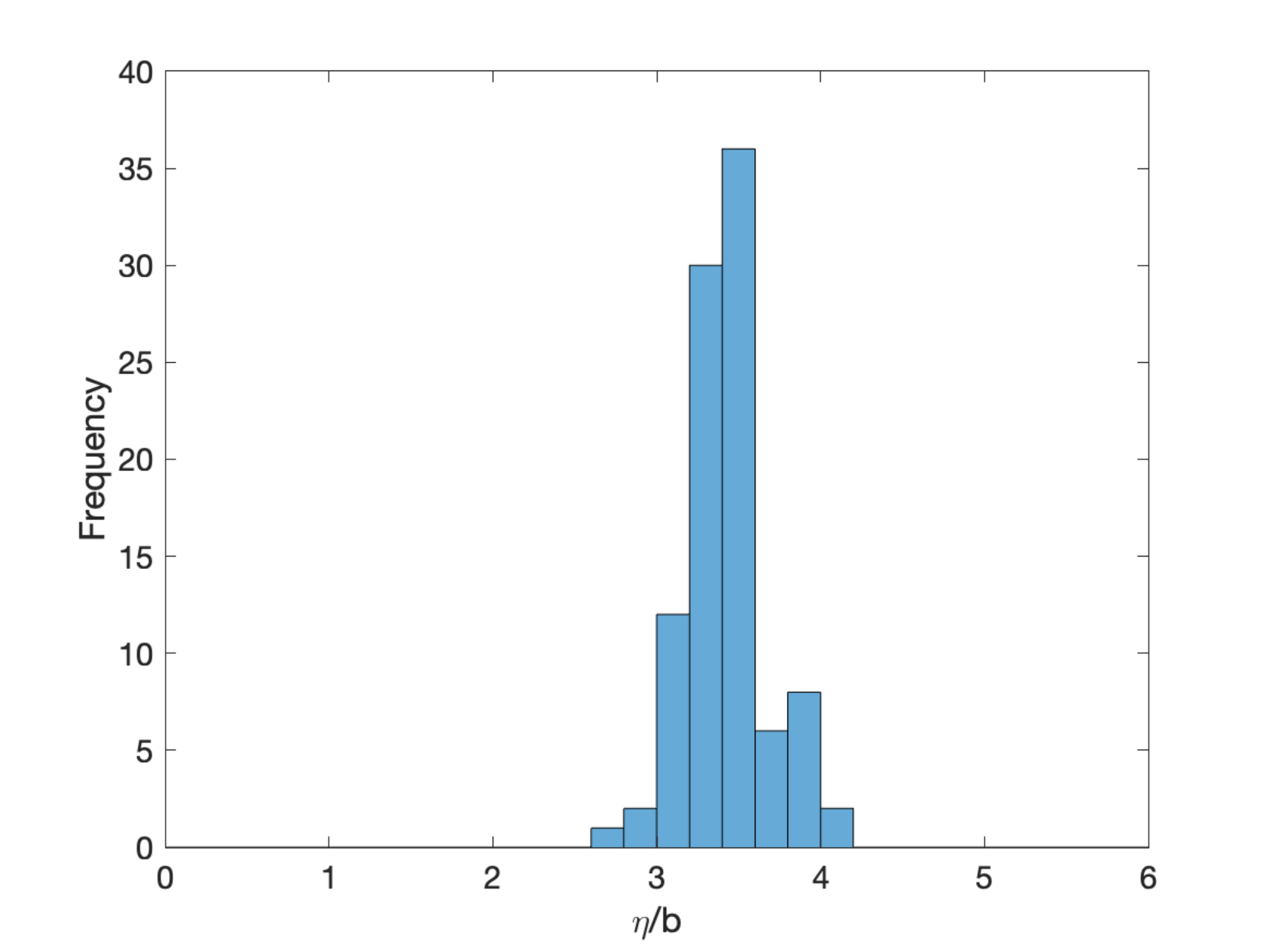}}
	\subfigure[$M=250$]{\includegraphics[width = .3\linewidth]{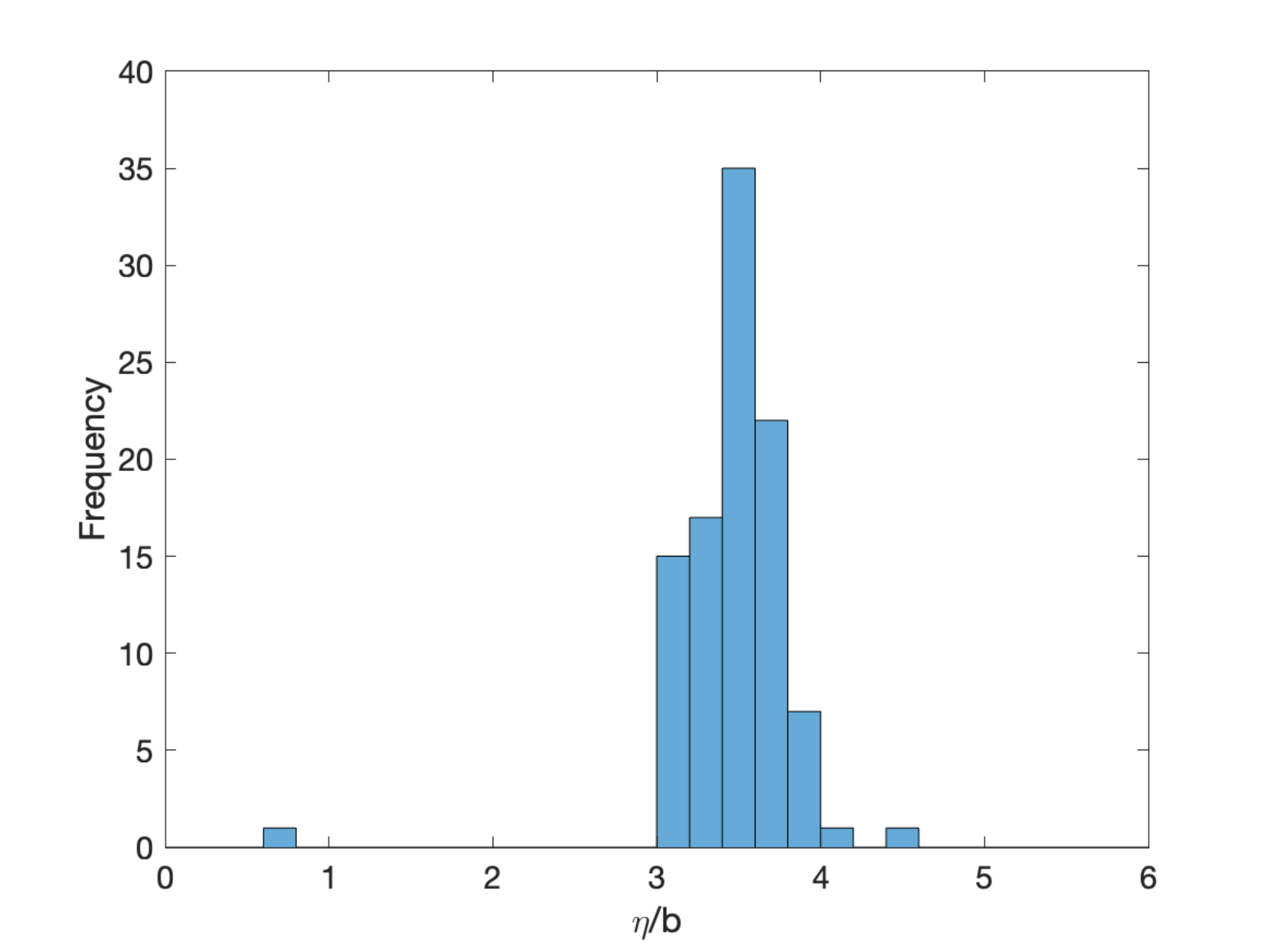}}
	\caption{Histogram of estimated $\eta/b$ values from 100 trials of SGD for various values of $M$, with data generated from $a=.25$, $\eta/b = 3 \frac{1}{3}$}
	\label{fig:RatioHist}
\end{figure}

Because the algorithm can converge to ratios of parameters that are different than the true parameter, we investigate how often the algorithm converges close to the true ratio, as well as the influence of the size of the data set. We completed 100 trials maximum likelihood estimation to fit $\eta$ and $b$ to the model for three sizes of data sets, $M = 50$, $M = 100$, and $M = 250$. The data was generated with $\eta = 5$, $a = .25$, and $b = 1.5$. The ratio of parameters the data was generated with is $\eta/b = 3 \frac{1}{3}$. The results in Figure \ref{fig:RatioHist} show the majority of trials of SGD fall at or near this ratio, with fewer falling further away. This pattern gets more pronounced as we increase the size of the data set from $M = 50$ to $M = 100$ or $M = 250$. Note that as the data set gets larger, the values seem to be skewed to the correct ratio, or larger. This fits with our observation that the values for $\eta$ at and above the true parameter value have similar log-likelihood values, so $\eta/b$ would be skewed above the true ratio as well.

\subsection{Fitting multiple models to synthetic data}

\begin{table}[h]
	\centering
	\begin{tabular}{|c|c|c c| c c| c c| }
		\hline
		  && \multicolumn{2}{|c|}{M = 50} & \multicolumn{2}{|c|}{M = 100} & \multicolumn{2}{|c|}{M = 250}\\
		\hline
		 model & trial& $\eta /b$& $-\ell \ell$  &  $\eta /b$ &$-\ell \ell$ &  $\eta /b$ & $-\ell \ell$\\
		\hline
		 \multirow{3}{8em}{$K = Be^{-\sqrt{\frac{3}{2}}|\textbf{x}|}$, $a = 1$}& 1& 4.50 & 268.4164 & 5.64 & 512.1422 &5.01 & 1307.3\\
		 &2& 3.85 & 254.5265 &5.07 & 525.5458 & 4.33& 1299.3\\
	      &3& 4.88 & 264.0203 & 5.57& 540.3379 &4.13 & 1283.9\\
		\hline
		 \multirow{3}{8em}{$K = Be^{-\sqrt{\frac{3}{2}}|\textbf{x}|}$, $a = 0$}&1 & 7.10 & 293.9026 &5.09 & 554.6673 & 5.24& 1410.7\\
		&2&  5.33 & 278.9786 &6.36 & 594.6089&5.34 &1412.5\\
		&3& 6.50 & 285.3712 & 4.99& 557.1079& 4.15&1291.0\\
		\hline
	\multirow{3}{8em}{$K_1 = 0$, $a = 1$, $K_2 =Be^{-\sqrt{\frac{3}{2}}|\textbf{x}|}$} &1& 3.86 & 299.8061 &3.87 & 588.3867 &5.06 &1522.6\\
		&2&  4.47 & 291.7603 &4.30 &604.2973 & 7.98& 1524.3\\
		&3& 4.07 & 302.2332 & 3.35& 592.4909&2.20 &1564.1\\
		\hline
		 \multirow{3}{8em}{$K_1 =Be^{-\sqrt{\frac{3}{2}}|\textbf{x}|}$, $K_2 = 0$, $a=1$}&1 & 7.40 & 296.1490 & 7.09& 579.9042&8.15 & 1492.5\\
		&2&  7.51& 290.4746&  7.95& 602.7097&7.10 &1486.2\\
		&3& 7.96 & 300.5563 & 6.84& 596.8582&7.14& 1483.4\\	
		\hline
	\end{tabular}
	\caption{ The predicted parameters and associated negative log-likelihood values for various models fitted to data generated from equilibrium solutions to system \eqref{eq:System} with $a= 1$, $b = 1.25$, $\eta =5$, and $n=2$, $K$ equal to the Laplace potential. The first, second, and third rows for each model were fit to the same data set .}
\label{T:fitmodels}
\end{table}

Finally, we explore model selection by fitting various models to data generated from equilibrium solutions to system \eqref{eq:System} and determining which model produces the lowest negative log-likelihood value. We repeat this process for three different sizes of data sets, $ M = 50$, $M=100$, and $M=250$. The data was generated with parameters $b=1.25$, $\eta = 5$, $a =.25$, $U$ Gaussian, and $N=2$. We fit parameters $b$ and $\eta$ to the original system, the system without segregation ($K_2 = 0$), the system without aggregation ($K_1 =0$), and to the system with $U=0$. The ratio of parameters predicted for each model and the resulting negative log-likelihood value are shown in Table \ref{T:fitmodels}. In each trial, the model used to generate the data was selected due to having the lowest negative log-likelihood value. In fact, in each trial, the model used to generate the data had the lowest log-likelihood value, the system without the environment potential had the second lowest value, then the system without segregation, and finally the system without aggregation.

\section{Discussion}\label{sec:disc}
Nonlocal mechanistic models can be a more realistic way to model many species that take nonlocal information into account when forming territories. However, nonlocal mechanistic models come at a higher computational cost than their local counterparts. When incorporating data into these models, they must be solved many times as we move through the parameter space, and often for a large system of equations (multiple species). Thus, we must solve the system efficiently. In this work, we use a nonlocal mechanstic model to describe territory formation of species that use nonlocal information. We efficiently solve the system using spectral methods and investigate computation times for various parameter regimes and number of species. We also consider the effect of the parameters and various terms in the system on the solutions. We find that when the diffusion and aggregation is well-balanced, the equilibrium solutions can be found more quickly. Additionally, in some cases, the steepness of the slope of the aggregation potential has a larger effect on the territory of a species than the variance of the aggregation potential.

With the ability to efficiently solve the system, we generate synthetic data and estimate the parameters used to generate the data via maximum likelihood estimation. The parameters are resurrected in cases with one parameter and one species, one parameter and two species, as well as for two parameters. By completing many trials, we are able to discuss how well this method estimates the true parameters. We find that increasing the size of the data set decreases the range of the predicted parameters and increases the frequency in which we predict the correct parameters. Finally, we perform model selection on data generated from the model, and in each case, the model the data was generated with was selected. 

Our main takeaway is that spectral methods can be an efficient way to solve this nonlocal system, and minimizing the negative log-likelihood function via stochastic gradient descent can be a viable way to incorporate data into the model. One limitation to this method is the necessity for periodic boundary conditions which are often not physical; we discuss ways in which we remedied these effects. In future work, it is of interest to incorporate meerkat location data into the model as well as environment data, performing model selection.

 {\bf Acknowledgments}: The authors would like to thank Mark Hoefer for his helpful ideas and expertise.
Both authors were partially funded by the NSF DMS 1909638 and NSF DMS 2042413.

{\bf Data Availability}: The method of generating datasets during the study are available from the corresponding author on reasonable request.

{\bf Statements and Declarations: }
The authors declare that they have no conflict of interest.

%\subsection{Details on reference citations}\label{subsec7}
%
%Standard \LaTeX\ permits only numerical citations. To support both numerical and author-year citations this template uses \verb+natbib+ \LaTeX\ package. For style guidance please refer to the template user manual.
%
%Here is an example for \verb+\citep{...}+: \citep{bib1}. Another example for \verb+\citepp{...}+: \citepp{bib2}. For author-year citation mode, \verb+\citep{...}+ prints Jones et al. (1990) and \verb+\citepp{...}+ prints (Jones et al., 1990).
%
%All cited bib entries are printed at the end of this article: \citep{bib3}, \cite{bib4}, \cite{bib5}, \cite{bib6}, \cite{bib7}, \cite{bib8}, \cite{bib9}, \cite{bib10}, \cite{bib11} and \cite{bib12}.

%%===========================================================================================%%
%% If you are submitting to one of the Nature Portfolio journals, using the eJP submission   %%
%% system, please include the references within the manuscript file itself. You may do this  %%
%% by copying the reference list from your .bbl file, paste it into the main manuscript .tex %%
%% file, and delete the associated \verb+\bibliography+ commands.                            %%
%%===========================================================================================%%

\bibliography{library2}% common bib file
%% if required, the content of .bbl file can be included here once bbl is generated
%%\input sn-article.bbl

%% Default %%
%%\input sn-sample-bib.tex%

\end{document}